\documentclass[11pt]{article}
\usepackage{graphicx,psfrag,amsmath,epsfig,amssymb,epsf,cite,
			latexsym,mathrsfs,subfigure,url,stfloats,array,dsfont}

\newcommand{\dotle} {\mbox{$\:\stackrel{\centerdot}{\le}\:$}}

\newcommand{\calA}{{\cal A}}
\newcommand{\calB}{{\cal B}}
\newcommand{\calC}{{\cal C}}
\newcommand{\calD}{{\cal D}}
\newcommand{\calE}{{\cal E}}
\newcommand{\calF}{{\cal F}}
\newcommand{\calG}{{\cal G}}
\newcommand{\calI}{{\cal I}}
\newcommand{\calK}{{\cal K}}
\newcommand{\calM}{{\cal M}}
\newcommand{\calP}{{\cal P}}
\newcommand{\calQ}{{\cal Q}}
\newcommand{\calS}{{\cal S}}
\newcommand{\calT}{{\cal T}}

\newcommand{\calV}{{\cal V}}
\newcommand{\calW}{{\cal W}}
\newcommand{\calX}{{\cal X}}
\newcommand{\calY}{{\cal Y}}
\newcommand{\calZ}{{\cal Z}}

\newcommand{\scrM}{\mathscr{M}}
\newcommand{\scrP}{\mathscr{P}}
\newcommand{\scrW}{\mathscr{W}}

\newcommand{\sfA}{\mathsf{A}}
\newcommand{\sfB}{\mathsf{B}}
\newcommand{\sfK}{\mathsf{K}}

\newcommand{\bs}{{\mathbf s}}
\newcommand{\bS}{{\mathbf S}}

\newcommand{\bV}{{\mathbf V}}
\newcommand{\bw}{{\mathbf w}}
\newcommand{\bW}{{\mathbf W}}
\newcommand{\bx}{{\mathbf x}}
\newcommand{\bX}{{\mathbf X}}
\newcommand{\by}{{\mathbf y}}
\newcommand{\bY}{{\mathbf Y}}
\newcommand{\bz}{{\mathbf z}}

\newcommand{\tp}{\tilde{p}}

\newcommand{\tby}{\tilde{\mathbf y}}

\newcommand{\oI}{{\overset{\circ}{I}}}

\def\eE{{\Bbb E}}

\def\rR{{\Bbb R}}

\def\one{\mathrm{one}}
\def\all{\mathrm{all}}
\def\simple{\mathrm{simple}}
\def\fair{\mathrm{fair}}
\def\good{\mathrm{good}}
\def\bad{\mathrm{bad}}
\def\memoryless{\mathrm{memoryless}}
\def\nom{\mathrm{nom}}
\def\psp{\mathrm{psp}}
\def\FP{\mathrm{FP}}

\newtheorem{definition}{Definition}[section]

\newtheorem{theorem}{Theorem}[section]
\newtheorem{lemma}[theorem]{Lemma}
\newtheorem{proposition}[theorem]{Proposition}
\newtheorem{corollary}[theorem]{Corollary}
\renewcommand{\theequation}{\arabic{section}.\arabic{equation}}
\begin{document}

\title{Universal Fingerprinting: \\
Capacity and Random-Coding Exponents}

\author{Pierre~Moulin 
\thanks{The author is with the ECE Department, the Coordinated Science Laboratory,
and the Beckman Institute at the University of Illinois at Urbana-Champaign, Urbana, IL 61801, USA.
Email: {\tt moulin@ifp.uiuc.edu}. This work was supported by NSF under grants CCR 03-25924,
CCF 06-35137 and CCF 07-29061. A 5-page version of this paper was presented at ISIT
in Toronto, July~2008. The current manuscript was submitted for publication on January~24, 2008
and revised on December~9, 2008, June~9, 2009, January~24, 2010, December 10, 2010, and May 24, 2011.}}

\maketitle
\begin{abstract}
This paper studies fingerprinting (traitor tracing) games in which the number
of colluders and the collusion channel are unknown. The fingerprints are embedded into
host sequences representing signals to be protected and
provide the receiver with the capability to trace back pirated copies
to the colluders. The colluders and the fingerprint embedder are
subject to signal fidelity constraints. Our problem setup unifies the signal-distortion
and Boneh-Shaw formulations of fingerprinting. The fundamental tradeoffs
between fingerprint codelength, number of users, number of colluders, 
fidelity constraints, and decoding reliability are then determined.

Several bounds on fingerprinting capacity have been presented in recent literature.
This paper derives exact capacity formulas and presents a new randomized fingerprinting
scheme with the following properties:
(1) the encoder and receiver assume a nominal coalition size but do not need to know
	the actual coalition size and the collusion channel;
(2) a tunable parameter $\Delta$ trades off false-positive
	and false-negative error exponents;
(3) the receiver provides a reliability metric for its decision; and
(4) the scheme is capacity-achieving when the false-positive exponent $\Delta$
	tends to zero and the nominal coalition size coincides with the actual coalition size.

A fundamental component of the new scheme is the use of a ``time-sharing'' randomized sequence.
The decoder is a {\em maximum penalized mutual information decoder}, where the significance
of each candidate coalition is assessed relative to a threshold, and the penalty
is proportional to the coalition size. A much simpler {\em threshold decoder}
that satisfies properties (1)---(3) above but not (4) is also given.\\[-0.1in]
\end{abstract}


\newpage

\section{Introduction}

Digital fingerprinting ({\em a.k.a.} traitor tracing)
is essentially a multiuser version of watermarking. A covertext --- such as
image, video, audio, text, or software --- is to be distributed to many users.
Prior to distribution, each user is assigned a fingerprint that is embedded
into the covertext. In a collusion attack, a coalition of users combine
their marked copies, creating a pirated copy that contains only weak traces
of their fingerprints. The pirated copy is subject to a fidelity requirement
relative to the coalition's copies. The fidelity requirement may take the form
of a {\em distortion constraint}, which is a natural model for media fingerprinting
applications \cite{Moulin02,Moulin02b,Moulin03,Somekh05,Somekh07,Wang06,Moulin07b};
or it may take the form of Boneh and Shaw's {\em marking assumption},
which is a popular model for software fingerprinting \cite{Boneh95,Tardos03,Barg08}.
To trace the forgery back to the coalition members, one needs
a fingerprinting scheme that can reliably identify the colluders' fingerprints
from the pirated copy. 

The fingerprinting problem presents two key challenges.
\begin{enumerate}
\item The number of colluders may be large, which makes
	it easier for the colluders to mount a strong attack.
	The difficulty of the decoding problem is compounded by the fact that
	{\em the number of colluders and the collusion channel are unknown to
	the encoder and decoder.}
\item There are two fundamental types of error events, namely {\em false positives},
	by which innocent users are wrongly accused, and {\em false negatives}, by which
	one or more colluders escape detection. For legal reasons, a maximum admissible value
	for the false-positive error probability should be specified.
\end{enumerate}
This paper proposes a mathematical model that satisfies these requirements
and derives the corresponding information-theoretic performance limits.
Prior art on related formulations of the fingerprinting problem is reviewed below.

The basic performance metric is capacity, which is defined with respect to a class of
collusion channels.  A multiuser data hiding problem was analyzed by
Moulin and O'Sullivan \cite[Sec.~8]{Moulin03}, and capacity expressions were obtained
assuming a compound class of memoryless channels, expected-distortion constraints
for the distributor and the coalition, and noncooperating, single-user decoders.
Despite clear mathematical similarities, this setup is quite different
from the one adopted in more recent fingerprinting papers. Somekh-Baruch and Merhav
\cite{Somekh05,Somekh07} studied a fingerprinting problem with a known number of
colluders and explored connections with the problem
of coding for the multiple-access channel (MAC).
The notion of false positives does not appear in their problem formulation.
Lower bounds on capacity were obtained assuming almost-sure distortion constraints
between the pirated copy and one \cite{Somekh05} or all \cite{Somekh07}
of the coalition's copies. The lower bounds on capacity correspond to a restrictive
encoding strategy, namely random constant-composition codes without time-sharing.

Other bounds on capacity and connections between MACs and fingerprinting under
the Boneh-Shaw assumption have been recently studied by Anthapadmanabhan {\em et al.}
\cite{Barg08}. The covertext is degenerate, and side information does not appear
in the information-theoretic formulation of this problem. 

In order to cope with unknown collusion channels and unknown number of colluders,
a special kind of universal decoder should be designed, where universality holds
not only with respect to some set of channels, but also with respect to an unknown
number of inputs. An early version of this idea in the context of the so-called
{\em random MAC} was introduced by Plotnik and Satt \cite{Plotnik91}.
In the context of fingerprinting, a tunable parameter should trade off the two
fundamental types of error probability. When the number of colluders is unknown,
two extreme instances of this tradeoff are to accuse {\em all} users or {\em none of them}.

While fingerprinting capacity is a fundamental measure of the ability
of any scheme to resist colluders, it only guarantees that the error probabilities
vanish if the codes are ``long enough''. Error exponents provide
a finer description of system performance. They provide estimates of the necessary
length of a fingerprinting code that can withstand a specified number of colluders,
given target false-positive and false-negative error probabilities.
This is especially valuable in any legal system where the reliability
of accusations should be assessed.

Besides capacity and error-exponent formulas, the information-theoretic analysis
sheds light about the structure of optimal codes. Particularly relevant
in this respect is a random coding scheme by Tardos \cite{Tardos03}, which
uses an auxiliary random sequence for encoding fingerprints. While
his scheme is presented at an algorithmic level (and no optimization was involved
in its construction), in our game-theoretic setting the auxiliary random variable
appears fundamentally as part of a randomized strategy in an information-theoretic
game whose payoff function is nonconcave with respect to the maximizing variable
(the fingerprint distribution).

Another issue that can be resolved in our game-theoretic setting
is the optimality of coalition strategies that are invariant to
permutations of the colluders. While one may heuristically expect
that such strategies are optimal, a proof of this property is established in
this paper. The approach used in previous papers was to {\em assume}
that coalitions employ such strategies, but often no performance guarantee
is given if the colluders employ asymmetric strategies.

Finally, in the aforementioned paper by Tardos \cite{Tardos03} and in the signal processing
literature, several simple algorithms have been proposed to detect colluders, involving
computing some correlation score between pirated copy and users' fingerprints, and setting up
a detection threshold. We study the limits of such strategies and compare them with joint
decoding strategies.

\subsection{Organization of This Paper}

As indicated by the bibliographic references, probabilistic analyses of digital
fingerprinting have been reported both in the information theory literature
and in the theoretical computer science literature. While the results derived
in this paper are put in the context of related information-theoretic work,
especially multiple-access channels, this paper is nevertheless intended to be
accessible to a broader community of readers that are trained in probability theory
and statistics. The main tools used in our derivations are the method of types
\cite{Csiszar81,Csiszar98} for analyzing random-coding schemes, Fano's lemma
for deriving upper bounds on capacity, sphere-packing methods, and elementary
properties of information-theoretic functionals.

A mathematical statement of our generic fingerprinting problem is given in
Sec.~\ref{sec:problem statement}, together with the definitions of codes,
collusion channels, error probabilities, capacity, and error exponents.
Our first main results are fingerprinting capacity theorems.
They are stated in Sec.~\ref{sec:C}.

The next two sections present the new random coding scheme and the resulting error exponents.
Sec.~\ref{sec:simple} presents a simple but suboptimal decoder that compares
empirical mutual information scores between received data and individual fingerprints,
and outputs a guilty decision whenever the score exceeds a certain tunable threshold.
This suboptimal decoder is closely related to strategies used in the signal
processing literature and in \cite{Tardos03}. For simplicity of the exposition,
the scheme and results are presented in the setup with degenerate side information,
which is directly applicable to the Boneh-Shaw problem.
Sec.~\ref{sec:joint} introduces and analyzes a more elaborate joint decoder that assigns
a penalized empirical equivocation score to candidate coalitions and selects the coalition
with the lowest score. The penalty is proportional to coalition size. 
The joint decoder is capacity-achieving. 

Sec.~\ref{sec:memoryless} outlines an extension to the problem where
the collusion channel is memoryless. The proofs of the main results appear
in Secs~\ref{Sec:Converse-all}---\ref{Sec:JointProof},
and the paper concludes in Sec.~\ref{sec:conclusion}.

\subsection{Notation}
\label{sec:notation}

We use uppercase letters for random variables, lowercase letters
for their individual values, calligraphic letters for finite alphabets,
and boldface letters for sequences. Given an integer $K$, we use the special
symbol $\sfK$ for the set $\{1,2,\cdots,K\}$. We denote by $\calM^\star$
the set of sequences of arbitrary length (including 0) whose elements
are in $\calM$.
The probability mass function (p.m.f.) of a random variable $X \in
\calX$ is denoted by $p_X=\{p_X(x),\,x \in \calX\}$. The variational
distance between two p.m.f's $p$ and $q$ over $\calX$ is denoted by
$d_V(p,q) = \sum_{x\in\calX} |p(x)-q(x)|$. The entropy
of a random variable $X$ is denoted by $H(X)$, and the mutual
information between two random variables $X$ and $Y$ is denoted by
$I(X;Y)=H(X)-H(X|Y)$. Should the dependency on the underlying p.m.f.'s
be explicit, we write the p.m.f.'s as subscripts, e.g., $H_{p_X}(X)$
and $I_{p_X \,p_{Y|X}}(X;Y)$. The Kullback-Leibler divergence
between two p.m.f.'s $p$ and $q$ is denoted by $D(p||q)$, and the
conditional Kullback-Leibler divergence of $p_{Y|X}$ and $q_{Y|X}$
given $p_X$ is denoted by
$D(p_{Y|X}||q_{Y|X}|p_X)=D(p_{Y|X}\,p_X||q_{Y|X}\,p_X)$.
All logarithms are in base 2 unless specified otherwise.

Given a sequence $\bx \in \calX^N$, denote by $p_\bx$ its type, or empirical p.m.f.
over the finite alphabet $\calX$. Denote by $T_\bx$ the type class associated with
$p_\bx$, i.e., the set of all sequences of type $p_\bx$. Likewise, $p_{\bx\by}$ denotes
the joint type of a pair of sequences $(\bx, \by) \in \calX^N \times \calY^N$,
and $T_{\bx\by}$ the associated joint type class.
The conditional type $p_{\by|\bx}$ of a pair of sequences ($\bx, \by$) is defined
by $p_{\bx\by}(x,y)/p_{\bx}(x)$ for all $x \in \calX$ such that $p_{\bx}(x) > 0$.
The conditional type class $T_{\by|\bx}$ given $\bx$, is
the set of all sequences $\tilde{\by}$ such that $(\bx,
\tilde{\by}) \in T_{\bx\by}$. We denote by $H(\bx)$ the empirical
entropy of the p.m.f. $p_{\bx}$, by $H(\by|\bx)$ the empirical conditional entropy,
and by $I(\bx;\by)$ the empirical mutual information for the joint p.m.f. $p_{\bx\by}$. 
Recall that the number of types and conditional types is polynomial in $N$
and that \cite{Csiszar81}
\begin{eqnarray}
   (N+1)^{- |\calX|} \,2^{NH(\bx)} \le & |T_{\bx}| & \le 2^{NH(\bx)} ,	\label{eq:type-size1} \\
   (N+1)^{- |\calX| \,|\calY|} \,2^{NH(\by|\bx)} 
	\le & |T_{\by|\bx}| & \le 2^{NH(\by|\bx)} .							\label{eq:type-size2}
\end{eqnarray}

We use the calligraphic fonts $\scrP_X$ and $\scrP^{[N]}_X$ to represent the set
of all p.m.f.'s and all empirical p.m.f.'s for length-$N$ sequences, respectively,
on the alphabet $\calX$. Likewise, $\scrP_{Y|X}$ and $\scrP^{[N]}_{Y|X}$ denote the
set of all conditional p.m.f.'s and all empirical conditional p.m.f.'s on
the alphabet $\calY$. The special symbol $\scrW_K$ will be used
to denote the feasible set of collusion channels $p_{Y|X_1, \cdots, X_K}$
that can be selected by a size-$K$ coalition.

Mathematical expectation is denoted by the symbol $\eE$.
The shorthands $a_N \doteq b_N$ and $a_N \dotle b_N$
denote asymptotic relations in the exponential scale, respectively
$\lim_{N \to \infty}\frac{1}{N}\log \frac{a_N}{b_N}=0$ and
$\limsup_{N \to \infty} \frac{1}{N}\log \frac{a_N}{b_N}\le 0$.
We define $|t|^+ \triangleq \max(t,0)$, and $\exp_2(t)\triangleq 2^t$.
The indicator function of a set $\calA$ is denoted by $\mathds1{\{x \in \calA\}}$.
The symbol $\calA \setminus \calB$ is used to denote the relative complement
(or set-theoretic difference) of set $\calB$ in set $\calA$.
(Note that $\calB$ is generally not a subset of $\calA$.)
Finally, we adopt the notational convention that the minimum of a function
over an empty set is $+\infty$, and the maximum is 0.

\section{Problem Statement and Basic Definitions}
\label{sec:problem statement}

\subsection{Overview}
\label{sec:overview}


\begin{figure}[hbt]
\begin{center}
\includegraphics[width=13cm]{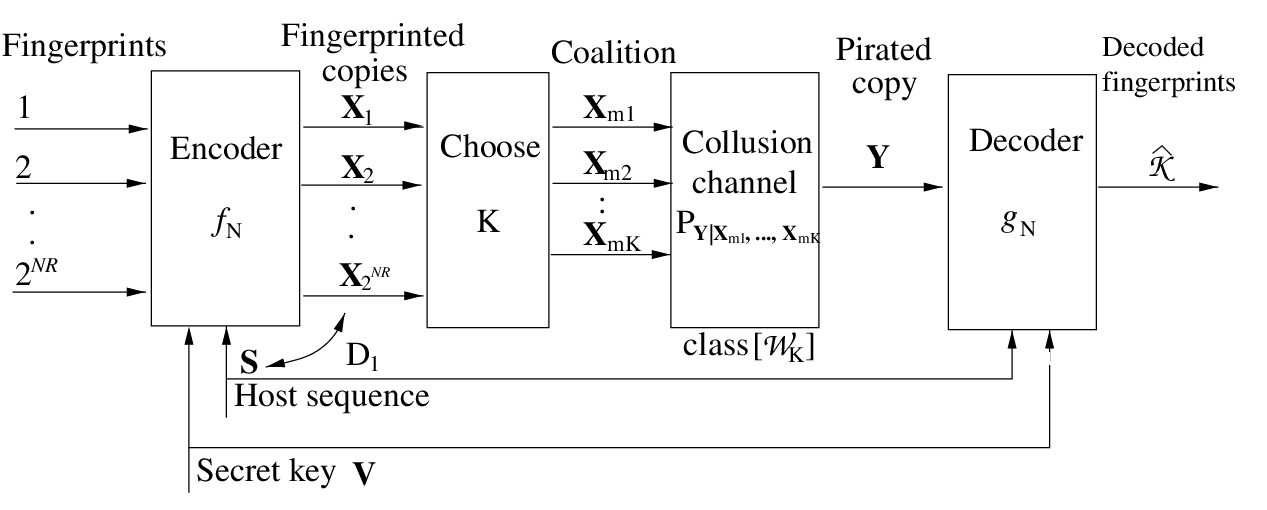}
\end{center}
\caption{Model for fingerprinting game, using randomized code $(f_N,g_N)$.
	In the Boneh-Shaw setup, the host sequence $\bS$ is degenerate and
	there is no distortion constraint ($D_1$). The class $\scrW_K$ characterizes
	the fidelity constraint on the collusion channel. The encoder and decoder know
	neither $K$ nor the collusion channel.}
\label{fig:Public FP}
\end{figure}

Our model for digital fingerprinting is diagrammed in Fig.~\ref{fig:Public FP}.
Let $\calS$, $\calX$, and $\calY$ be three finite alphabets. 
The covertext sequence $\bS=(S_1, \cdots, S_N) \in \calS^N$ consists of
$N$ independent and identically distributed (i.i.d.) samples drawn
from a p.m.f. $p_S(s)$, $s \in \calS$. A secret key $V$ taking values 
in an alphabet $\calV_N$, whose cardinality potentially grows with $N$,
is shared between encoder and decoder, and not publicly revealed.
The key $V$ is a random variable independent of $\bS$.
There are $2^{NR}$ users, each of which
receives a fingerprinted copy:
\begin{equation}
  \bX_m = f_N(\bS,V, m), \quad 1 \le m \le 2^{NR},
\end{equation}
where $f_N: \calS^N \times \calV_N \times \{ 1, \cdots, 2^{NR} \} \to \calX^N$
is the encoding function, and $m$ is the index of the user.
The fidelity requirement between $\bS$ and $\bX_m$ is expressed
via a distortion constraint. Let $d~:~\calS \times \calX \to \rR^+$
be the distortion measure and $d^N(\bs,\bx) = \frac{1}{N} \sum_{i=1}^N d(s_i,x_i)$
the extension of this measure to length-$N$ sequences.
The code $f_N$ is subject to the distortion constraint
\begin{equation}
   d^N(\bs,\bx_m) \le D_1 , \quad 1 \le m \le 2^{NR} .
\label{eq:D1}
\end{equation}

Let $\calK \triangleq \{m_1,\,m_2\,,\cdots,\,m_K\}$ be a coalition of $K$ users;
no constraints are imposed on the formation of coalitions.
The coalition uses its copies $\bX_{\calK} \triangleq \{\bX_m , \,m \in \calK\}$
to produce a pirated copy $\bY \in \calY^N$. Without loss of generality, we assume
that $\bY$ is generated stochastically according to a conditional p.m.f.
$p_{\bY|\bX_{\calK}}$ called the {\bf collusion channel}.
This includes deterministic mappings as a special case.
A fidelity constraint is imposed on $p_{\bY|\bX_{\calK}}$ to ensure that $\bY$
is ``close'' to the fingerprinted copies $\bX_m , \,m \in \calK$.
This constraint may take the form of a distortion constraint
(analogously to (\ref{eq:D1})), or alternatively, a constraint that
will be referred to as the Boneh-Shaw constraint.
The formulation of these constraints is detailed below and results
in the definition of a feasible set $\scrW_K(p_{\bx_\calK})$ 
for the conditional type $p_{\by|\bx_\calK}$.

The encoder and decoder assume a nominal coalition size $K_{\nom}$ but
know neither $K$ nor $p_{\bY|\bX_{\calK}}$ selected by
the $K$ colluders \footnote{
   If $K_{\nom} = K$, our random coding scheme of Sec.~\ref{sec:joint}
	is capacity-achieving.}.
The decoder has access to the pirated copy $\bY$, the host $\bS$,
and the secret key $V$. It produces an estimate
\begin{equation}
   \hat{\calK} = g_N(\bY,\bS,V)
\label{eq:detect all}
\end{equation}
of the coalition. Success can be defined as catching one colluder
or catching all colluders, the latter task being seemingly much more difficult.
An admissible decoder output is the empty set, $\hat{\calK} = \emptyset$,
reflecting the possibility that the signal submitted to the decoder
is unrelated to the fingerprints. If this possibility was not allowed,
an innocent user would be accused. Another good reason to allow 
$\hat{\calK} = \emptyset$ is simply that reliable detection is impossible
when there are too many colluders, and the constraint on the probability of
false positives would be violated if $\hat{\calK} = \emptyset$ was not an option.

\subsection{Randomized Fingerprinting Codes}

The formal definition of a fingerprinting code is as follows.
\begin{definition}
A {\bf randomized rate-$R$ length-$N$ fingerprinting code} $(f_N, g_N)$ with
embedding distortion $D_1$ is a pair of encoder mapping 
$f_N~:~\calS^N \times \calV_N \times \{1, 2, \cdots, \lceil 2^{NR} \rceil \} \to \calX^N$
and decoder mapping $g_N~:~\calY^N \times \calS^N \times \calV_N
\to \{1, 2, \cdots, \lceil 2^{NR} \rceil \}^\star$.
\label{def:deterministic-code}
\end{definition}

Many kinds of randomization are possible. In the most general setting,
the key space $\calV_N$ can grow superexponentially with $N$. For fingerprinting, three
kinds of randomization seem to be fundamental, each serving a different purpose.
All three kinds can be combined.
The first one is randomized permutation of the letters $\{1,2,\cdots,N\}$
to cope with channels with arbitrary memory, similarly to \cite{Moulin07}.
\begin{definition}
A {\bf randomly modulated} (RM) fingerprinting code is a randomized fingerprinting code
defined via permutations of a prototype $(\tilde{f}_N,\tilde{g}_N)$.
The code is of the form
\begin{eqnarray}
   \bx_m = \tilde{f}_N^{\pi}(\bs,w,m) & \triangleq & \pi^{-1} \tilde{f}_N (\pi \bs, w,m)
																\nonumber \\
	 \tilde{g}_N^{\pi}(\by,\bs,w) & \triangleq & \tilde{g}_N(\pi \by, \pi \bs,w)
\label{eq:RM}
\end{eqnarray}
where $\pi$ is chosen uniformly from the set of all $N!$ permutations
of the letters $\{1,2,\cdots,N\}$ and is not revealed publicly.
The sequence $\pi \bx_m$ is obtained by applying $\pi$ to the elements of $\bx_m$.
The secret key is $V=(\pi,W)$, where $W$ is independent of $\pi$.
\label{def:code-RM}
\end{definition}

The second kind of randomization is uniform permutations of the $2^{NR}$ fingerprint
assignments, to equalize error probabilities over all possible coalitions
\cite{Moulin07b,Barg08}.
\newpage

\begin{definition}
A {\bf randomly permuted} (RP) fingerprinting code is a randomized fingerprinting code
defined via permutations of a prototype $(\tilde{f}_N,\tilde{g}_N)$.
The code is of the form
\begin{eqnarray}
   \bx_m = \tilde{f}_N^{\pi}(\bs,w,m) & \triangleq & \tilde{f}_N (\bs, w, \pi^{-1}(m))
																\nonumber \\
   \tilde{g}_N^{\pi}(\by,\bs,w) & \triangleq & \pi \left( \tilde{g}_N(\by,\bs,w) \right)
\label{eq:RP}
\end{eqnarray}
where $\pi$ is chosen uniformly from the set of all $2^{NR}!$ permutations
of the user indices $\{1,2,\cdots,2^{NR}\}$ and is not revealed publicly.
The secret key is $V=(\pi,W)$, where $W$ is independent of $\pi$. 
In (\ref{eq:RP}), we have used the shorthand
$\pi(\hat{\calK}) \triangleq \{ \pi(m), \,m \in \hat{\calK}\}$.
\label{def:code-RP}
\end{definition}


The third kind of randomization arises via an auxiliary ``time-sharing''
random sequence. This strategy was not used in \cite{Somekh05,Somekh07,Barg08}
but a remarkable example was developed by Tardos \cite{Tardos03}.
For binary alphabets $\calS$, $\calX$, and $\calY$, i.i.d. random variables
$W_i \in (0,1), \,1 \le i \le N$, are generated, and next the fingerprint
letters $X_i(m)$ are generated as independent Bernoulli\,($W_i$) random variables.
Here $V = \{ W_i, \,1 \le i \le N\}$ is the secret key shared by encoder and decoder.

Given an embedding distortion $D_1$ and a size--$K$ coalition using a collusion channel
from class $\scrW_K$, there corresponds a capacity $C(D_1,\scrW_K)$ which is the supremum
over $(f_N,g_N)$ of all achievable $R$, under a prescribed error criterion.

\subsection{Collusion Channels}

First we define some basic terminology for MACs with $K$ inputs,
common input alphabet $\calX$, and output alphabet $\calY$. 
Recall that $\sfK = \{1,2,\cdots,K\}$ and let $X_\sfK = \{X_1, \cdots, X_K\}$.
Given a conditional p.m.f. $p_{Y|X_{\sfK}}$, consider the permuted conditional p.m.f.
\begin{equation}
  p_{Y|X_{\pi(\sfK)}}(y|x_1, \cdots, x_K)
	\triangleq p_{Y|X_{\sfK}}(y|x_{\pi(1)}, \cdots, x_{\pi(K)})
\label{eq:fair-pyxK}
\end{equation}
where $\pi$ is any permutation of the $K$ inputs.
We say that $p_{Y|X_{\sfK}}$ is permutation-invariant if
\[ p_{Y|X_{\pi(\sfK)}} = p_{Y|X_{\sfK}} , \quad \forall \pi . \]
A subset $\scrW_K$ of $\scrP_{Y|X_{\sfK}}$ is said to be permutation-invariant if
\[ p_{Y|X_\sfK} \in \scrW_K
	\;\Rightarrow\;p_{Y|X_{\pi(\sfK)}} \in \scrW_K , \quad \forall \pi . \]
In general, not all elements of such $\scrW_K$ are permutation-invariant.
The subset of permutation-invariant $\scrW_K$ that consists of permutation-invariant
conditional p.m.f.'s will be denoted by
\begin{equation}
   \scrW_K^{\fair} = \left\{ p_{Y|X_{\sfK}} \in \scrW_K
	~:~ p_{Y|X_{\pi(\sfK)}} = p_{Y|X_{\sfK}} , \;\forall \pi \right\} .
\label{eq:WK-fair}
\end{equation}
Finally, if $\scrW_K$ is permutation-invariant and convex, the permutation-averaged
conditional p.m.f. $\frac{1}{K!} \sum_{\pi} p_{Y|X_{\pi(\sfK)}}$ is also in $\scrW_K$
and is permutation-invariant by construction.

In the fingerprinting problem, the conditional type $p_{\by|\bx_{\calK}}
\in \scrP_{Y|X_{\calK}}^{[N]}$ is a random variable whose conditional distribution
given $\bx_{\calK}$ depends on the collusion channel $p_{\bY|\bX_{\calK}}$. 
Our fidelity constraint on the coalition is of the general form
\begin{equation}
  Pr[p_{\by|\bx_{\calK}} \in \scrW_K(p_{\bx_{\calK}})] = 1, 
\label{eq:WK}
\end{equation}
where for each $p_{\bx_{\calK}}$, $\scrW_K(p_{\bx_{\calK}})$
is a {\bf convex, permutation-invariant} subset of $\scrP_{Y|X_{\calK}}$.
That is, the empirical conditional p.m.f. of the pirated copy given the marked
copies is restricted. The choice of the feasible set $\scrW_K(p_{\bx_{\calK}})$ depends
on the application, as elaborated below. The explicit dependency of $\scrW_K$ on
$p_{\bx_{\calK}}$ will sometimes be omitted to simplify notation.
Note that assuming $\scrW_K$ is permutation-invariant does not imply that
$p_{\by|\bx_{\calK}}$ actually selected by the coalition is permutation-invariant.
Finally, it is assumed that the set-valued mapping $\scrW_K(p)$ is defined for
$p \in \scrP_{X^K}$ and is uniformly continuous in the variational distance,
in the sense that for every $\epsilon > 0$, there exists $\delta > 0$ such that
\begin{eqnarray}
   \lefteqn{\forall p_{X_\sfK}, p_{X_\sfK}' \in \scrP_{X^K} \;\mathrm{s.t.}\; 
	d_V(p_{X_\sfK}, p_{X_\sfK}') < \delta :} \nonumber \\
	& & \max_{p_{Y|X_\sfK} \in \scrW_K(p_{X_\sfK})} \min_{p_{Y|X_\sfK}' \in \scrW_K(p_{X_\sfK}')}
		d_V(p_{Y|X_\sfK} \,p_{X_\sfK}, \,p_{Y|X_\sfK}' \,p_{X_\sfK}') < \epsilon .
\label{eq:WK-continuous}
\end{eqnarray}

The model (\ref{eq:WK}) can be used to impose hard distortion constraints
on the coalition or to enforce the Boneh-Shaw marking assumption when $\calX = \calY$.

\begin{enumerate}

\item {\bf Distortion Constraints.} 
Consider the following variation on the constraints used in \cite{Moulin03,Somekh05,Somekh07}.
Define a {\em permutation-invariant} estimator $f~:\calX^K \to \calS$ which produces
an estimate $\hat{S}=f(X_{\calK})$ of the host signal sample based
on the corresponding marked samples. \footnote{
	A permutation-invariant estimator depends on the samples $\{X_k ,\,k\in\calK\}$
	only via their empirical distribution on $\calX$.}
The estimator could be, e.g., a maximum-likelihood estimator.
Then

\begin{equation}
   \scrW_K(p_{\bx_{\calK}}) = \left\{ p_{Y|X_{\calK}}
	~:~ \sum_{x_{\calK},y} \,p_{\bx_{\calK}}(x_{\calK})
	\,p_{Y|X_{\calK}}(y|x_{\calK}) \,d_2(f(x_{\calK}),y) \le D_2 \right\}
\label{eq:expected-D2}
\end{equation}
where $d_2 ~:~ \calS \times \calY \to \rR^+$ is the coalition's distortion function,
and $D_2$ is the maximum allowed distortion.
The constraint (\ref{eq:WK}) may be equivalently written as 
\begin{equation}
  Pr \left[ d_2^N(f(\bx_{\calK}),\by) = \frac{1}{N} \sum_{t=1}^N d_2(f(x_{\calK,t}),y_t)
	\le D_2 \right] = 1 .
\label{eq:as-D2}
\end{equation}

\item {\bf Interleaving Attack.}
Here each colluder contributes $N/K$ samples to the forgery
-- taken at arbitrary positions. The class $\scrW_K$ is a singleton:
\begin{equation}
   p_{Y|X_{\calK}}(y|x_{\calK}) = \frac{1}{K} \sum_{k \in \calK} \,\mathds1_{\{y=x_k\}} .
\label{eq:interleaving}
\end{equation}

\item {\bf Boneh-Shaw Marking Assumption.}
Assume $\calX = \calY$ and $\scrW_K$ is the set of conditional p.m.f.'s that satisfy
\begin{equation}
    x_1=\cdots=x_K \quad \Rightarrow \quad y = x_1 .
\label{eq:d-boneh}
\end{equation}
Then the constraint (\ref{eq:WK}) enforces the Boneh-Shaw {\em marking assumption}:
the colluders are not allowed to modify their samples at any location where these samples
agree. Thus $y_t = x_{m_1,t}$ at any position $1 \le t \le N$ such that
$x_{m_1,t} = \cdots = x_{m_K,t}$. Note that
$\scrW_K$ does not depend on $p_{\bx_\calK}$ and that the interleaving attack
(\ref{eq:interleaving}) satisfies the Boneh-Shaw condition. 

\end{enumerate}

\subsection{Strongly Exchangeable Collusion Channels}

Recall the definition of RM codes in (\ref{eq:RM}); a dual notion applies to
collusion channels. For any $p_{\bY|\bX_{\calK}}$ and permutation
$\pi$ of $\{1,2,\cdots,N\}$, define the permuted channel
$p_{\bY|\bX_{\calK}}^{\pi}(\by|\bx_{\calK})
\triangleq p_{\bY|\bX_{\calK}}(\pi \by| \pi \bx_{\calK})$. Then we have
\begin{definition} \cite{Somekh05}
A {\bf strongly exchangeable collusion channel} $p_{\bY|\bX_{\calK}}$
is a channel such that $p_{\bY|\bX_{\calK}}^{\pi}(\by|\bx_{\calK})$
is independent of $\pi$, for every $(\bx_{\calK},\by)$.
\label{def:exch}
\end{definition}

A strongly exchangeable collusion channel is defined by a probability assignment
$Pr[T_{\by|\bx_{\calK}}]$ on the conditional type classes. The distribution of $\bY$
conditioned on $\bY \in T_{\by|\bx_{\calK}}$ is uniform:
\begin{equation}
   p_{\bY|\bX_{\calK}}(\tby|\bx_{\calK})
	= \frac{Pr[T_{\by|\bx_{\calK}}]}{|T_{\by|\bx_{\calK}}|} ,
		\quad \forall \tby \in T_{\by|\bx_{\calK}} .
\label{eq:pYXS}
\end{equation}
In Sec.~\ref{sec:Pe} we show that for RM codes $(f_N,g_N)$, it is sufficient to consider
strongly exchangeable collusion channels to derive worst-case error probabilities.
Moreover, in the error probability calculations for random codes it will be sufficient
to use the trivial upper bound 
\begin{equation}
   Pr[T_{\by|\bx_{\calK}}] \le \mathds1\{ p_{\by|\bx_{\calK}} \in \scrW_K(p_{\bx_\calK}) \} .
\label{eq:bound1}
\end{equation}

\subsection{Fair Coalitions}
\label{sec:fair}

Two notions of fairness for coalitions will be useful.
Denote by $\pi$ a permutation of $\{1,2,\cdots,K\}$.
\begin{definition}
   The collusion channel $p_{\bY|\bX_{\calK}}$ is {\bf permutation-invariant} if
 \begin{equation}
   p_{\bY|\bX_{\calK}}(\by|\bx_{m_1}, \cdots, \bx_{m_K})
	= p_{\bY|\bX_{\calK}}(\by|\bx_{\pi(m_1)}, \cdots, \bx_{\pi(m_K)}) ,
		\quad \forall \pi .
\label{eq:fair-pYXK}
\end{equation}
\label{def:fair}
\end{definition}
For instance, if $\calX=\calY$ and $K=2$, the collusion channel
\begin{equation}
   p_{\bY|\bX_1 \bX_2}(\by|\bx_1,\bx_2) = \frac{1}{2} \left[
      \mathds1\{\by=\bx_1\} + \mathds1\{\by=\bx_2\} \right]
\label{eq:pYXK-pi}
\end{equation}
is permutation-invariant. Given $\bx_1, \bx_2$, there are two equally likely
choices for the pirated copy, namely $\by=\bx_1$ and $\by=\bx_2$.
Note that one colluder carries full risk and the other one zero risk.
A stronger definition of fairness (which will not be needed in this paper)
would require some kind of ergodic behavior of the inputs and output
of the collusion channel.

\begin{definition}
   The collusion channel $p_{\bY|\bX_{\calK}}$ is {\bf first-order fair} if
   $Pr[p_{\by|\bx_{\calK}} \in \scrW_K^{\fair}(p_{\bx_\calK})]=1$.
\label{def:1-fair}
\end{definition}
For any first-order fair collusion channel, the conditional type $p_{\by|\bx_{\calK}}$
is invariant to permutations of the colluders, with probability 1.
For instance, if $\calX=\calY$ and $K=2$, any collusion channel $p_{\bY|\bX_{\calK}}$
resulting in the conditional type $p_{\by|\bx_1 \bx_2}(y|x_1,x_2) = \frac{1}{2}
\left[ \mathds1\{y=x_1\} + \mathds1\{y=x_2\} \right]$ is first-order fair.
This is an interleaving attack in which each colluder contributes
exactly $N/2$ samples (in any order) to the pirated copy. 

A first-order fair collusion channel is not necessarily permutation-invariant,
and vice-versa. Further, if a collusion channel is first-order fair and strongly
exchangeable, then it is also permutation-invariant. However the converse is not true. 
For instance the collusion chanel of (\ref{eq:pYXK-pi}) is permutation-invariant
and strongly exchangeable but not first-order fair because the conditional type
$p_{\by|\bx_{\calK}}(y|x_1,x_2)$ is given by either $\mathds1\{y=x_1\}$ or 
$\mathds1\{y=x_2\}$, neither of which is permutation-invariant.

\subsection{Error Probabilities}
\label{sec:Pe}

Let $\calK$ be the coalition and $\hat{\calK} = g_N(\bY,\bS,V)$ the decoder's
output. There are several error probabilities of interest:
the probability of {\bf false positives} (one or more innocent users are accused):
\begin{equation}
   P_{\FP}(f_N,g_N,p_{\bY|\bX_{\calK}}) = Pr[\hat{\calK} \setminus \calK \ne \emptyset] ,
\label{eq:P-FP}
\end{equation}
the probability of missed detection for a specific coalition member $m \in \calK$:
\[
   P_{e,m}(f_N,g_N,p_{\bY|\bX_{\calK}}) = Pr[m \notin \hat{\calK}] ,
\]
the probability of failing to catch a single colluder:
\begin{equation}
   P_e^{\one}(f_N,g_N,p_{\bY|\bX_{\calK}}) = Pr[\hat{\calK} \cap \calK = \emptyset] ,
\label{eq:P-one}
\end{equation}
and the probability of failing to catch the full coalition: 
\begin{equation}
  P_e^{\all}(f_N,g_N,p_{\bY|\bX_{\calK}}) = Pr[\calK \not\subseteq \hat{\calK}] .
\label{eq:P-all}
\end{equation}
The error criteria (\ref{eq:P-one}) and (\ref{eq:P-all}) will be referred to as
the {\bf detect-one} and {\bf detect-all} criteria, respectively.

The above error probabilities may be written in the explicit form
\begin{equation}
   P_e(f_N,g_N,p_{\bY|\bX_{\calK}}) = \sum_{v,\bs,\bx_{\calK},\by} p_V(v) \,p_S^N(\bs)
			\left( \prod_{m\in\calK} \mathds1\{\bx_m = f_N(\bs,v,m)\} \right)
			p_{\bY|\bX_{\calK}}(\by|\bx_{\calK}) \,\mathds1\{\calE\}
\label{eq:Pe}
\end{equation}
where the error event $\calE$ is given by
$\calE_{\FP} = \{g_N(\by,\bs,v) \setminus \calK \ne \emptyset\}$,
or $\calE^{\one} = \{g_N(\by,\bs,v) \cap \calK = \emptyset\}$, or
$\calE^{\all} = \{\calK \not\subseteq g_N(\by,\bs,v)\}$, when $P_e$ is given by
(\ref{eq:P-FP}), (\ref{eq:P-one}), and (\ref{eq:P-all}), respectively.
The worst-case probability is given by
\[ P_e(f_N,g_N,\scrW_K) = \max_{p_{\bY|\bX_{\calK}}} \,P_e(f_N,g_N,p_{\bY|\bX_{\calK}}) \]
where the maximum is over all feasible collusion channels, i.e., such that
(\ref{eq:WK}) holds.

{\bf Maximum vs average error probability.}
The error probabilities (\ref{eq:P-FP})---(\ref{eq:P-all}) generally depend on $\calK$.
Prop.~\ref{prop:RP} below states that (a) in order to make them independent of $\calK$
and provide guarantees on error probability for any coalition, one may use RP codes,
and (b) random permutations of fingerprint assignments cannot increase the average
error probability of any code. 
Let $(\tilde{f}_N,\tilde{g}_N)$ be an arbitrary code and $(f_N,g_N)$
the RP code of (\ref{eq:RP}), obtained using $(\tilde{f}_N,\tilde{g}_N)$
as a prototype. Let $p_{\bY|\bX_{\calK}}$ be an arbitrary collusion
channel when coalition $\calK$ is in effect. Given any other coalition
$\calK' = \pi(\calK)$ of the same size, let $p_{\bY|\bX_{\calK'}}$
be the corresponding collusion channel, obtained by applying (\ref{eq:fair-pyxK}),
where $\pi$ is now a permutation of $\{1,\cdots, 2^{NR}\}$.
\begin{proposition}
For any code $\tilde{f}_N,\tilde{g}_N$ and collusion channel $p_{\bY|\bX_{\calK}}$,
we have
\begin{eqnarray}
  \forall \calK' ~:\quad P_e(f_N,g_N,p_{\bY|\bX_{\calK'}})
	= P_e(f_N,g_N,p_{\bY|\bX_{\calK}})
	\le \max_{\calK'} P_e(\tilde{f}_N,\tilde{g}_N,p_{\bY|\bX_{\calK'}})
\label{eq:RP-equilibrium}
\end{eqnarray}
where $(f_N,g_N)$ is the RP code of (\ref{eq:RP}),
and $P_e$ denotes any of the error probability criteria
(\ref{eq:P-FP}), (\ref{eq:P-one}), and (\ref{eq:P-all}).
\label{prop:RP}
\end{proposition}
{\em Proof}.
First consider the detect-one error criterion of (\ref{eq:P-one}): an error arises if 
$g_N(\bY,\bS,V) \cap \calK = \emptyset$. 
Given a RP fingerprinting code with prototype $(\tilde{f}_N, \tilde{g}_N)$ and
permutation parameter $\pi$, the detect-one error probability when coalition $\calK$
is in effect is given by
\begin{eqnarray}
   P_e^{\one}(f_N,g_N,p_{\bY|\bX_{\calK}})
	& = & Pr[g_N(\bY,\bS,V) \cap \calK = \emptyset] \nonumber \\
	& = & Pr[\tilde{g}_N^\pi(\bY,\bS,W) \cap \calK = \emptyset] \nonumber \\
	& = & Pr[\pi\left(\tilde{g}_N(\bY,\bS,W)\right) \cap \calK = \emptyset] \nonumber \\
	& = & Pr[\tilde{g}_N(\bY,\bS,W) \cap \pi^{-1}(\calK) = \emptyset] \nonumber \\
	& = & \eE_{\bY,\bS,W} \frac{1}{2^{NR}!} \underbrace{\sum_\pi \mathds1
			\{\tilde{g}_N(\bY,\bS,W) \cap \pi^{-1}(\calK) = \emptyset\} }_
			{\mathrm{independent~of~}\calK}
\label{eq:Pe1-RP}
\end{eqnarray}
which is independent of $\calK$, by virtue of the uniform distribution on $\pi$.
The derivation for the detect-all and the false-positive error probabilities
is analogous to (\ref{eq:Pe1-RP}). This establishes the first equality in
(\ref{eq:RP-equilibrium}). The inequality is proved similarly.
\hfill $\Box$

{\bf False Positives {\em vs} False Negatives.}
The tradeoff between false positives and false negatives is central
to statistical detection theory (the Neyman-Pearson problem)
and list decoding \cite{Forney68}. Note that in the classical formulation of list
decoding \cite[p.~166]{Gallager68}, an error is declared only if the message sent
does not appear on the decoder's output list. The false-negative error exponent increases
with list size and approaches the sphere packing exponent if the list size
is allowed to grow subexponentially with $N$.
This classical formulation does not include a cost for ``false positives''.

\subsection{Strongly Exchangeable Collusion Channels}

Prop.~\ref{prop:exch} below states that randomly modulated codes (Def.~\ref{def:code-RM})
and strongly exchangeable channels (Def.~\ref{def:exch}) satisfy a certain equilibrium
property: neither the fingerprint embedder nor the coalition has interest in deviating
from those strategies.
Let $(\tilde{f}_N,\tilde{g}_N)$ be an arbitrary code and $(f_N,g_N)$
the RM code of (\ref{eq:RM}), obtained using $\tilde{f}_N,\tilde{g}_N$
as a prototype. Given any feasible collusion channel $p_{\bY|\bX_{\calK}}$, denote by
\begin{equation}
   \overline{p}_{\bY|\bX_{\calK}}(\by|\bx_{\calK})
	= \frac{1}{N!} \sum_{\pi} p_{\bY|\bX_{\calK}}(\pi\by|\pi\bx_{\calK})
\label{eq:pYX-permuted1}
\end{equation}
the permutation-averaged channel, which is feasible and strongly exchangeable.
\begin{proposition}
For any code $\tilde{f}_N,\tilde{g}_N$ and collusion channel $p_{\bY|\bX_{\calK}}$,
we have
\begin{eqnarray}
  P_e(f_N,g_N,p_{\bY|\bX_{\calK}})
    & = & P_e(f_N,g_N,\overline{p}_{\bY|\bX_{\calK}}) \nonumber \\
    & = & P_e(\tilde{f}_N,\tilde{g}_N,\overline{p}_{\bY|\bX_{\calK}})
			\le \max_\pi P_e(\tilde{f}_N,\tilde{g}_N,p_{\bY|\bX_{\calK}}^\pi)
\label{eq:RM-equilibrium}
\end{eqnarray}
where $(f_N,g_N)$ is the RM code of (\ref{eq:RM})
and $P_e$ denotes any of the error probability criteria
(\ref{eq:P-FP}), (\ref{eq:P-one}), and (\ref{eq:P-all}).
\label{prop:exch}
\end{proposition}

{\em Proof.}
First consider the detect-one error criterion of (\ref{eq:P-one}): an error arises if 
$\tilde{g}_N(\bY,\bS,V) \cap \calK = \emptyset$. 
For any fixed $\calK$, the detect-one error probability is an average
over all possible permutations  $\pi$ and the other random variables $V,\bS,\bY$:
\begin{eqnarray}
   \lefteqn{P_e^{\one}(f_N,g_N,p_{\bY|\bX_{\calK}})} \nonumber \\
	& \stackrel{(a)}{=} & \frac{1}{N!} \sum_{\pi} \sum_{w,\bs,\bx_{\calK},\by} p_W(w) \,p_S^N(\bs)
			\left( \prod_{m\in\calK} \mathds1\{\pi\bx_m = \tilde{f}_N(\pi\bs,w,m)\} \right)
			p_{\bY|\bX_{\calK}}(\by|\bx_{\calK}) \nonumber \\
	& & \qquad \times \mathds1\{\tilde{g}_N(\pi\by,\pi\bs,w) \cap \calK = \emptyset \} \nonumber \\
	& \stackrel{(b)}{=} & \frac{1}{N!} \sum_{\pi} \sum_{w,\bs,\bx_{\calK},\by}
			p_W(w) \,p_S^N(\pi^{-1}\bs)
			\left( \prod_{m\in\calK} \mathds1\{\bx_m = \tilde{f}_N(\bs,w,m)\} \right)
			p_{\bY|\bX_{\calK}}(\pi^{-1}\by|\pi^{-1}\bx_{\calK}) \nonumber \\
	& & \qquad \times \mathds1\{\tilde{g}_N(\by,\bs,w) \cap \calK = \emptyset \} \nonumber \\
	& \stackrel{(c)}{=} & \sum_{w,\bs,\bx_{\calK},\by} p_W(w) \,p_S^N(\bs)
			\left( \prod_{m\in\calK} \mathds1\{\bx_m = \tilde{f}_N(\bs,w,m)\} \right)
			\left( \frac{1}{N!} \sum_{\pi} p_{\bY|\bX_{\calK}}
			(\pi^{-1}\by|\pi^{-1}\bx_{\calK}) \right) \nonumber \\
	& & \qquad \times \mathds1\{\tilde{g}_N(\by,\bs,w) \cap \calK = \emptyset \} \nonumber \\
	& = & \sum_{w,\bs,\bx_{\calK},\by} p_W(w) \,p_S^N(\bs)
			\left( \prod_{m\in\calK} \mathds1\{\bx_m = \tilde{f}_N(\bs,w,m)\} \right)
			\left( \frac{1}{N!} \sum_{\pi} p_{\bY|\bX_{\calK}}^{\pi}
			(\by|\bx_{\calK}) \right) \nonumber \\
	& & \qquad \times \mathds1\{\tilde{g}_N(\by,\bs,w) \cap \calK = \emptyset \} \nonumber \\
	& = & \sum_{w,\bs,\bx_{\calK},\by} p_W(w) \,p_S^N(\bs)
			\left( \prod_{m\in\calK} \mathds1\{\bx_m = \tilde{f}_N(\bs,w,m)\} \right)
			\overline{p}_{\bY|\bX_{\calK}}(\by|\bx_{\calK}) 
			\;\mathds1\{\tilde{g}_N(\by,\bs,w) \cap \calK = \emptyset \} \nonumber \\
	& = & P_e^{\one}(f_N,g_N,\overline{p}_{\bY|\bX_{\calK}})
\label{eq:Pe1-RM}
\end{eqnarray}
where (a) holds by definition of the RM code, (b) is obtained by applying the change of variables
$\bz \leftarrow \pi\bz$ to the sequences $\bs, \bx_{\calK}, \by$,
and (c) the fact that $p_S^N(\bs) = p_S^N(\pi\bs)$.
The derivation for the detect-all and the false-positive error probabilities
is analogous to (\ref{eq:Pe1-RM}). This establishes the first equality in
(\ref{eq:RM-equilibrium}). The second equality and the inequality are proved similarly.
\hfill $\Box$

\subsection{Risk for Fair Coalitions}

The maximum and the minimum of the error probabilities $P_{e,m}(\calK), \,m \in \calK$,
will be useful. The maximum value,
\begin{equation}
   \overline{P}_e(f_N,g_N,p_{\bY|\bX_{\calK}})
	= \max_{m \in \calK} P_{e,m}(f_N,g_N,p_{\bY|\bX_{\calK}}) ,
\label{eq:Pe-max}
\end{equation}
is the conventional error criterion for information transmission.
However, the minimum value, 
\begin{equation}
   \underline{P}_e(f_N,g_N,p_{\bY|\bX_{\calK}})
	= \min_{m \in \calK} P_{e,m}(f_N,g_N,p_{\bY|\bX_{\calK}}) ,
\label{eq:Pe-min}
\end{equation}
is more relevant to the coalition
because it represents the risk of their most vulnerable member.
Note that
\[ P_e^{\one}(f_N,g_N,p_{\bY|\bX_{\calK}}) \le \underline{P}_e(f_N,g_N,p_{\bY|\bX_{\calK}})
	\le \overline{P}_e(f_N,g_N,p_{\bY|\bX_{\calK}})
	\le P_e^{\all}(f_N,g_N,p_{\bY|\bX_{\calK}}) .
\]
While it is conceivable that some colluders could be tricked or coerced
into taking a higher risk than others, such strategy is not secure because
the whole coalition would be at risk if some of its members,
especially the vulnerable ones, are caught.
The proof of the following proposition is elementary.
\begin{proposition}
For randomly permuted codes (Def.~\ref{def:code-RP}),
if the collusion channel $p_{\bY|\bX_{\calK}}$ is permutation-invariant,
then all colluders incur the same risk:
\[ \underline{P}_e(f_N,g_N,p_{\bY|\bX_{\calK}})
	= \overline{P}_e(f_N,g_N,p_{\bY|\bX_{\calK}}) .
\]
\label{prop:prop:fair}
\end{proposition}

The proof of the following proposition is omitted because it is similar
to that of Prop.~\ref{prop:exch}. 
Assuming the fingerprint distributor uses RP codes, it follows from Prop.~\ref{prop:fair}
that permutation-invariant collusion channels are optimal for the colluders
under the detect-one error criterion.
\begin{proposition}
For randomly permuted codes, the maximum of the error probability criteria
(\ref{eq:P-FP}) and (\ref{eq:P-one}) is achieved by a
permutation-invariant collusion channel ((\ref{eq:fair-pYXK})) under the detect-one criterion.
\label{prop:fair}
\end{proposition}

Taken together with Prop.~\ref{prop:RP} on optimality of randomly-permuted
fingerprinting codes, Prop.~\ref{prop:fair} implies an equilibrium property: neither
the fingerprint embedder nor the coalition has interest in deviating from these symmetric
strategies, under the detect-one criterion.

\subsection{Capacity}

Having defined the detect-one and detect-all error criteria
and feasible classes of codes and collusion channels,
we may now define the corresponding notions of fingerprinting capacity.

\begin{definition}
A rate $R$ is achievable for embedding distortion $D_1$, collusion class $\scrW_K$,
and {\bf detect-one} criterion if there exists a sequence of $(N, \lceil 2^{NR} \rceil )$
randomized codes $(f_N, g_N)$ with maximum embedding distortion $D_1$, such that both
$P_e^{\one}(f_N,g_N,\scrW_K)$ and $P_{\FP}(f_N,g_N,\scrW_K)$ vanish as $N \to \infty$.
\label{def:detect-one}
\end{definition}

\begin{definition}
A rate $R$ is achievable for embedding distortion $D_1$, collusion class $\scrW_K$,
and {\bf detect-all} criterion if there exists a sequence of $(N, \lceil 2^{NR} \rceil )$
randomized codes $(f_N, g_N)$ with maximum embedding distortion $D_1$, such that both
$P_e^{\all}(f_N,g_N,\scrW_K)$ and $P_{\FP}(f_N,g_N,\scrW_K)$ vanish as $N \to \infty$.
\label{def:detect-all}
\end{definition}

\begin{definition}
Fingerprinting capacities $C^{\one}(D_1,\scrW_K)$ and $C^{\all}(D_1,\scrW_K)$ 
are the suprema of all achievable rates with respect to the detect-one and
detect-all criteria, respectively.
\label{def:C}
\end{definition}

We have $C^{\all}(D_1,\scrW_K) \le C^{\one}(D_1,\scrW_K)$
because an error event for the detect-one problem is also an error event
for the detect-all problem.

\subsection{Random-Coding Exponents}

For a sequence of randomized codes $(f_N, g_N)$, the error exponents are defined as
\[ E(R,D_1,\scrW_K) = \liminf_{N \to \infty} \left[ - \frac{1}{N} 
	\log P_e(f_N, g_N, \scrW_K) \right] \]
where $E$ represents the random coding exponent
$E_{\FP}$, $E^{\one}$, or $E^{\all}$.
Moreover, $E^{\all}(R,D_1,\scrW_K) \le E^{\one}(R,D_1,\scrW_K)$
because an error event for the detect-one problem is also an error event
for the detect-all problem. We have $E^{\all}=0$ if the class $\scrW_K$
includes channels in which one colluder can ``stay out,'' i.e.,
not contribute to the pirated copy.

\begin{figure}[h!]
\begin{center}
\includegraphics[width=9cm]{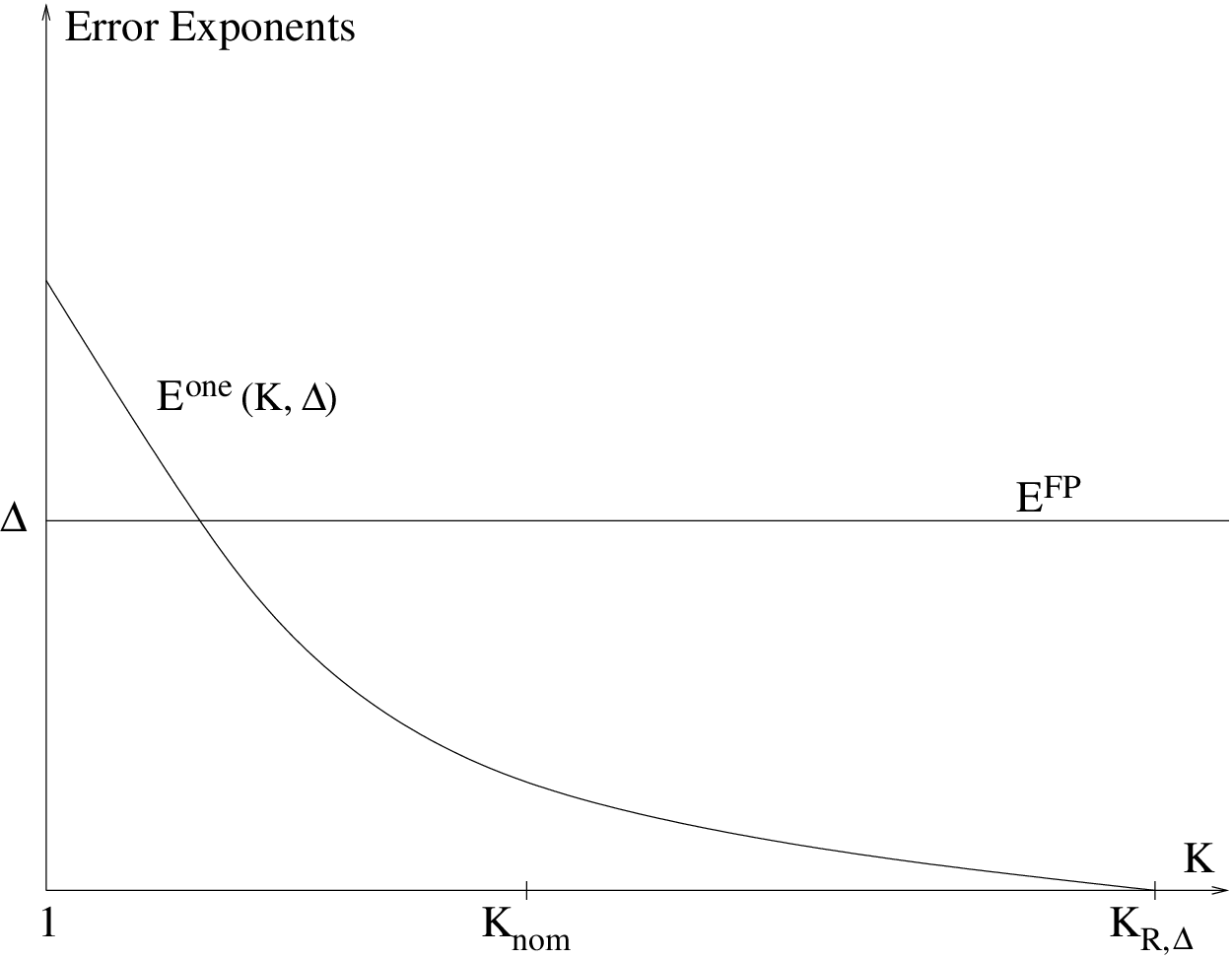}
\end{center}
\caption{False-positive and false-negative error exponents, as a function of
coalition size $K$, for fixed values of $R$ and $\Delta$.}
\label{fig:exponents}
\end{figure}

Fig.~\ref{fig:exponents} gives a preview of $E^{\one}$ and $E_{\FP}$ 
for our random coding scheme, viewed as a function of the number
$K$ of colluders. The false-positive exponent $E_{\FP}$ is 
equal to $\Delta$, for any value of $K$. The false-negative
exponent $E^{\one}$ decreases with $K$, up to some maximum value
$K_{R,\Delta}$ where it becomes zero. The decoder outputs $\hat{\calK}=\emptyset$
with high probability, and therefore reliable decoding of
any colluder is impossible, for any $K \ge K_{R,\Delta}$.

\begin{figure}[h!]
\begin{center}
\includegraphics[width=9cm]{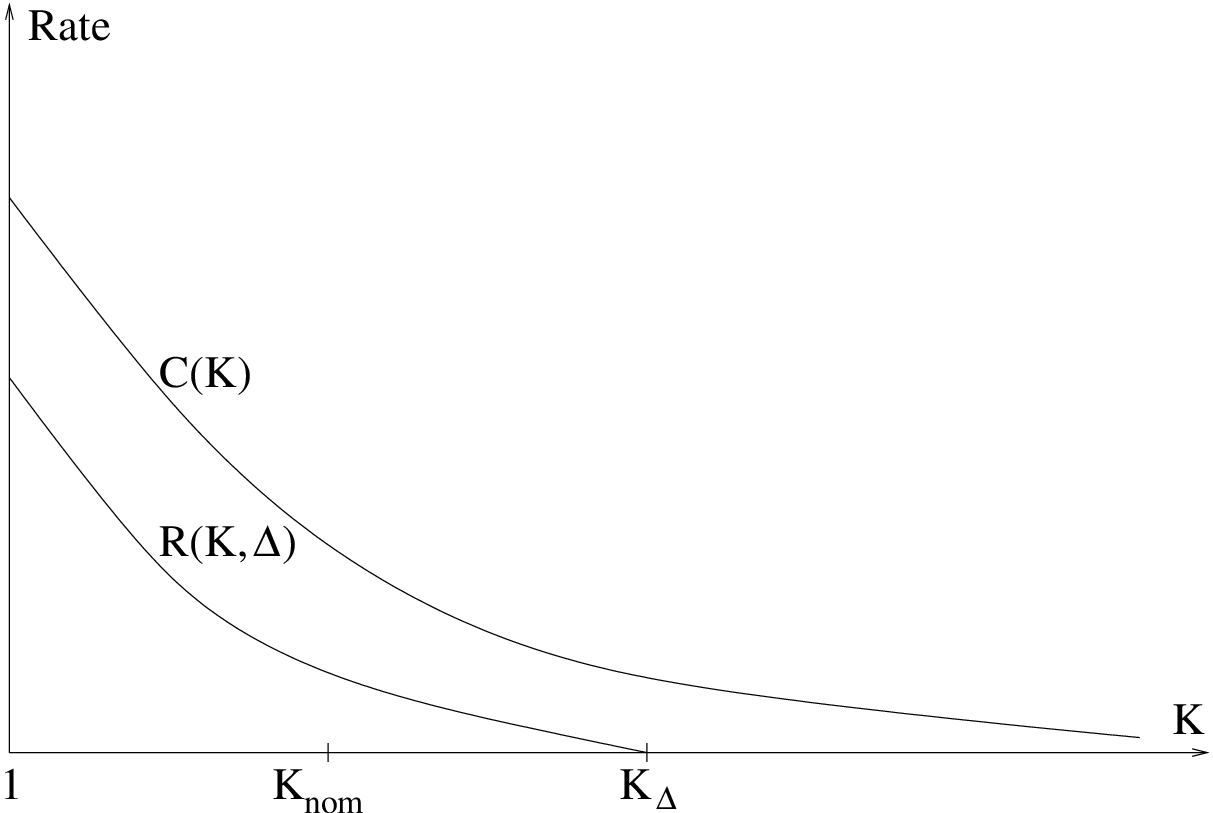}
\end{center}
\caption{Capacity $C$ and achievable rate $R$ (for false-positive error exponent
	equal to $\Delta$), as a function of coalition size $K$.}
\label{fig:rate}
\end{figure}

Fig.~\ref{fig:rate} illustrates the maximum rate $R(K,\Delta)$
that can be accommodated by the random coding scheme, for fixed $\Delta$.
This rate decreases with $K$ and becomes zero for $K \ge K_{\Delta}$.
If $\Delta \downarrow 0$, the rate curve $R(K,\Delta)$ tends to
the capacity function $C(K)$. Note that $C(K)$ vanishes as $K \to \infty$
but is generally positive for any finite $K$; in this case,
$\lim_{\Delta \to 0} K_{\Delta} = \infty$.

\subsection{Memoryless Collusion Channels}
\label{sec:memoryless-1}

As an alternative to the collusion channels subject to the hard constraint
$Pr[p_{\by|\bx_{\calK}} \in \scrW_K(p_{\bx_{\calK}})] = 1$, we may consider
memoryless collusion channels:
\begin{equation}
   p_{\bY|\bX_{\calK}}(\by|\bx_{\calK})
		= \prod_{t=1}^N p_{Y|X_{\calK}}(y_t|x_{\calK,t})
\label{eq:memoryless}
\end{equation}
where $p_{Y|X_{\calK}} \in \scrW_K(p_{\bx_{\calK}})$,
viewed as a {\em compound class} of channels \cite{Csiszar81}.
As we shall see there is a strong link between
the two problems in the form of Lemma~\ref{lem:memoryless} which is used
to establish our converse theorems; also see Sec.~\ref{sec:memoryless}.

\section{Fingerprinting Capacity}
\label{sec:C}
\setcounter{equation}{0}

This section presents fingerprinting capacity formulas under the detect-one
and detect-all error criteria. 
To put these results in context, let us first recall related results for MACs.
In the absence of side information, the capacity region of the MAC was determined
by Ahlswede \cite{Ahlswede71} and Liao \cite{Liao72}. This region is also
achievable for the random MAC \cite{Plotnik91}. For the MAC with common side
information at the transmitter and receiver, some very general capacity formulas
were derived by Das and Narayan \cite{Das02} under the assumption that $\bS$
is an ergodic process. In some special cases these formulas can be single-letterized. 
For fingerprinting with i.i.d. $\bS$ and coalition size equal to 2,
bounds on capacity were derived in \cite{Somekh05,Somekh07}.
Thus the presence of the side information $\bS$ causes difficulties in
deriving single-letter capacity formulas for both MAC and fingerprinting
problems.

The proof of the converse under the detect-all criterion is based
on the standard Fano inequality. Surprisingly, Fano's inequality
does not seem to be the right tool to prove the converse under the
detect-one criterion \cite{Barg08a}. A strong converse was presented
in \cite{Barg08}, but the resulting upper bound on capacity is loose. 
The direction we have pursued is based on explicit sphere-packing arguments,
specifically the fact that typical sets for $\bY$ given the embedded fingerprints
cannot have too much statistical overlap, otherwise reliable decoding
is impossible. The tools used here are different from those used for classical
problems such as the single-user discrete memoryless channel
\cite[pp.~173---176]{Gallager68} and the MAC \cite{Ahlswede82}. 
The use of a detect-one criterion requires a different machinery.
A simple technique is used to deal with codeword pairs whose self-information score
is well above average, and suffices to show that the error probability cannot
vanish for rates above capacity. We conjecture that a strong converse holds, namely:
for any rate above capacity,
\[ \lim_{N \to \infty} \min_{f_N,g_N} \max\{
	P_e^{\one}(f_N,g_N,\scrW_K), P_{\FP}(f_N,g_N,\scrW_K) \} = 1 .
\]
However, establishing this stronger result may require use of elaborate wringing techniques 
\cite{Ahlswede82}. Our lower bound on error probability does not tend to 1 as $N \to \infty$
because the bound (\ref{eq:Pc_good_sv_UB}) is likely loose.

\subsection{Mutual-Information Games}

The following lemma relates to Han's inequalities \cite{Han78} and will be useful
throughout this paper. Its proof appears in Appendix~\ref{Sec:Proof-Lemma-Fair}.\\
\begin{lemma}
Let $\sfK = \{1,2,\cdots,K\}$ and assume the distribution of $(X_{\sfK},Z)$
is invariant to permutations of $\sfK$.
Then for any nested sets $\sfA \subseteq \sfB \subseteq \sfK$, we have
\begin{eqnarray}
   \frac{1}{|\sfA|} H(X_{\sfA}|Z X_{\sfK\setminus\sfA})
	& \le & \frac{1}{|\sfB|} H(X_{\sfB}|Z X_{\sfK\setminus\sfB}) , \label{eq:HUY-fair} \\
   \frac{1}{|\sfA|} H(X_{\sfA}|Z)
	& \ge & \frac{1}{|\sfB|} H(X_{\sfB}|Z) .
\label{eq:HUS-fair}
\end{eqnarray}
Both inequalities hold with equality if $X_k, \,k\in\sfK$, are conditionally independent
given $Z$.
\label{lem:H-fair}
\end{lemma}

We will derive two simple formulas by application of this lemma.
First, applying (\ref{eq:HUY-fair}) with $Z=(Y,S,W)$ and (\ref{eq:HUS-fair}) with $Z=(S,W)$
and subtracting the first inequality from the second, we obtain 
\begin{equation}
   \frac{1}{|\sfA|} I(X_{\sfA};Y X_{\sfK\setminus\sfA}|SW)
	\ge \frac{1}{|\sfB|} I(X_{\sfB};YX_{\sfK\setminus\sfB}|SW) ,
	\quad \forall \sfA \subseteq \sfB \subseteq \sfK
\label{eq:I-fair}
\end{equation}
with equality if $X_k, \,k\in\sfK$, are conditionally independent given $Z$.
Second, for $X_k, \,k\in\sfK$ conditionally i.i.d. given $(S,W)$, we have
\begin{eqnarray}
  I(X_1;Y|S,W)
	& = & H(X_1|S,W) - H(X_1|Y,S,W) \nonumber \\
	& = & \frac{1}{K} \,H(X_{\sfK}|S,W) - H(X_1|Y,S,W) \nonumber \\
	& \le & \frac{1}{K} \,H(X_{\sfK}|S,W) - \frac{1}{K} \,H(X_{\sfK}|Y,S,W) \nonumber \\
	& = & \frac{1}{K} \,I(X_{\sfK};Y|S,W)
\label{eq:I2-fair}
\end{eqnarray}
where the inequality follows from (\ref{eq:HUS-fair}) with $Z=(Y,S,W)$.

Now consider an auxiliary random variable $W$ defined over an alphabet
$\calW = \{1, 2, \cdots, L\}$, and independent of $S$.
Define the set of conditional p.m.f.'s
\begin{eqnarray}
   \lefteqn{\scrP_{X_{\sfK} W|S}(p_S,L,D_1)} \nonumber \\
	& \triangleq & \left\{ p_{X_{\sfK} W|S} = p_W \prod_{k\in\sfK} p_{X_k|SW}
		~:~ p_{X_1|SW} = \cdots = p_{X_K|SW} , \; \eE d(S,X_1) \le D_1 \right\}
\label{eq:PXS-set}
\end{eqnarray}
and the functions
\begin{eqnarray}
	C_L^{\one}(D_1,\scrW_K)
	& = & \max_{p_{X_{\sfK} W|S} \in \scrP_{X_{\sfK} W|S}(p_S,L,D_1)}
		\;\min_{p_{Y|X_{\sfK}} \in \scrW_K^{\fair}(p_{X_{\sfK}})}
		\;\frac{1}{K} I(X_{\sfK};Y|S,W) 						\label{eq:CL-one} \\
	C_L^{\all}(D_1,\scrW_K)
	& = & \max_{p_{X_{\sfK} W|S} \in \scrP_{X_{\sfK} W|S}(p_S,L,D_1)}
		\;\min_{p_{Y|X_{\sfK}} \in \scrW_K(p_{X_{\sfK}})}
		\;\min_{\sfA \subseteq \sfK}
		\;\frac{1}{|\sfA|} I(X_{\sfA};Y|S,X_{\sfK\setminus\sfA},W) .
																\label{eq:CL-all}
\end{eqnarray}
Using the same derivation as in Lemma~2.1 of \cite{Moulin07}, it is easily
shown that $C_L^{\one}(D_1,\scrW_K)$ and $C_L^{\all}(D_1,\scrW_K)$
are nondecreasing functions of $L$ and converge to finite limits:
\begin{eqnarray}
   \widetilde{C}^{\one}(D_1,\scrW_K) & \triangleq & \lim_{L \to \infty} \;C_L^{\one}(D_1,\scrW_K)
																\label{eq:C-one} \\
   \widetilde{C}^{\all}(D_1,\scrW_K) & \triangleq & \lim_{L \to \infty} \;C_L^{\all}(D_1,\scrW_K) .
																\label{eq:C-all}
\end{eqnarray}
Moreover, the gap to each limit may be bounded by a polynomial function of $L$,
see \cite[Sec.~3.5]{Moulin07} for a similar derivation. The basic idea is to
discretize each $\scrW_K(p_{X_{\sfK}})$ to a fine grid of $\tilde{L}$ collusion channels.
By application of Caratheodory's theorem, the supremum of $C_L$ over $L$
is achieved by $L \le |\calS|\,|\calX| + \tilde{L}$. The gap between the
minimum of the cost function over $\scrW_K(p_{X_{\sfK}})$ and over its discrete approximation
can be bounded by $c \,\tilde{L}^{- |\calY|^{-1}\,|\calX|^{-K}}$ where $c$ is a constant.

Since $\scrW_K^{\fair}(p_{X_\sfK}) \subseteq \scrW_K(p_{X_\sfK})$, we have
$\widetilde{C}^{\all}(D_1,\scrW_K^{\fair}) \ge \widetilde{C}^{\all}(D_1,\scrW_K)$.
In fact the right side is zero if $\scrW_K(p_{X_{\sfK}})$
contains conditional p.m.f.'s $p_{Y|X_{\sfK}}$ such that $Y$ is independent of
one of the inputs $X_k, \,k\in\sfK$.

\begin{lemma}
For any $D_1$ and $\scrW_K$ we have
\begin{equation}
   \widetilde{C}^{\all}(D_1,\scrW_K) \le \widetilde{C}^{\one}(D_1,\scrW_K) .
\label{eq:C-all-vs-one}
\end{equation}
Equality holds for any class of fair collusion channels ($\scrW_K = \scrW_K^{\fair}$).
\label{lem:C-all-vs-one}
\end{lemma}
{\em Proof}:
Property (\ref{eq:C-all-vs-one}) follows from  (\ref{eq:CL-one})---(\ref{eq:C-all}) 
and the fact that $\scrW_K^{\fair}(p_{X_\sfK}) \subseteq \scrW_K(p_{X_\sfK})$.
Now consider $\scrW_K = \scrW_K^{\fair}$. Application of Property (\ref{eq:I-fair})
to any fair collusion channel yields
\[ \frac{1}{|\sfK|} I(X_{\sfK};Y|S,W)
	\le \frac{1}{|\sfA|} I(X_{\sfA};Y|X_{\sfK\setminus\sfA},S,W) ,
	\quad \forall \sfA \subseteq \sfK . \]
Hence the inner minimum in (\ref{eq:CL-all}) is achieved by $\sfA=\sfK$, and equality
holds in (\ref{eq:C-all-vs-one}).
\hfill $\Box$

\subsection{Capacity Theorems}

The following lemma will be used to prove Theorems \ref{thm:C-all}
and \ref{thm:C-one} below. Its proof is given in
Appendix~\ref{sec:Proof-Lemma-memoryless} and borrows ideas from
\cite[Theorem~3.7]{Moulin07}.

\begin{lemma}
Consider the compound family $\scrW_K(p_{\bx_{\sfK}})$ of memoryless channels
in (\ref{eq:memoryless}). Under both the detect-one and detect-all criteria,
the compound capacity for this problem is an upper bound on the capacity for
the main problem of (\ref{eq:WK}), in which $p_{\by|\bx_\sfK} \in \scrW_K(p_{\bx_{\sfK}})$
with probability 1.
\label{lem:memoryless}
\end{lemma}

We now give a direct coding theorem \ref{thm:C-achievable}
and two converse theorems \ref{thm:C-all} and \ref{thm:C-one}
pertaining to the detect-all and the detect-one criteria, respectively.
These theorems, combined with Lemma~\ref{lem:C-all-vs-one}, establish
the capacity theorem~\ref{thm:C}.

\begin{theorem}
Under the continuity assumption (\ref{eq:WK-continuous}),
all fingerprinting code rates below $\widetilde{C}^{\all}(D_1,\scrW_K)$
and $\widetilde{C}^{\one}(D_1,\scrW_K)$ are achievable
under the detect-all and the detect-one criteria, respectively.
\label{thm:C-achievable}
\end{theorem}

Theorem~\ref{thm:C-achievable} is a direct consequence of Theorem~\ref{thm:joint}(vi),
stated and proved later in this paper.

\begin{theorem}
When $\scrW_K$ is independent of $p_{\bx_{\sfK}}$, no fingerprinting code rate
$R$ exceeding $\widetilde{C}^{\all}(D_1,\scrW_K)$ is achievable under the detect-all
criterion. The same holds for the compound memoryless class of (\ref{eq:memoryless}).
\label{thm:C-all}
\end{theorem}

\begin{corollary}
Under the continuity assumption (\ref{eq:WK-continuous}),
when $\scrW_K$ depends on $p_{\bx_{\sfK}}$, the following holds. If the colluders
are constrained to select a fair collusion channel, then
$\scrW_K(p_{\bx_{\sfK}}) = \scrW_K^{\fair}(p_{\bx_{\sfK}})$, and no rate above
$\widetilde{C}^{\all}(D_1,\scrW_K^{\fair})$ is achievable under the detect-all criterion. 
\label{cor:C-all}
\end{corollary}

The proof of Theorem~\ref{thm:C-all} and Corollary~\ref{cor:C-all}
is given in Sec.~\ref{Sec:Converse-all}.

\begin{theorem}
When $\scrW_K$ is independent of $p_{\bx_{\sfK}}$, no fingerprinting code rate $R$
exceeding
\begin{eqnarray}
   \widetilde{C}^{\one}(D_1,\scrW_K^{\fair}) = \widetilde{C}^{\one}(D_1,\scrW_K)
\label{eq:C-one-fair}
\end{eqnarray}
is achievable under the detect-one criterion.
The same holds for the compound memoryless class of (\ref{eq:memoryless}).
\label{thm:C-one}
\end{theorem}


The proof of Theorem~\ref{thm:C-one} 
is given in Sec.~\ref{Sec:Converse-one}.

\begin{theorem}
Consider fingerprinting for coalitions of size at most $K$. Let $\scrW_K$
be the set of all conditional distributions $p_{Y|X_{\sfK}}$ (collusion
attacks) that can be selected by the coalition.
\begin{description}
\item[(a)] {\bf Detect-all case.}
Fingerprinting capacity is lower-bounded by $\widetilde{C}^{\all}(D_1,\scrW_K)$
given by (\ref{eq:C-all}). If in addition one of the following holds:
\begin{description}
\item[(i)] The set $\scrW_K$ of attacks available to the coalition is independent of 
	the joint type of the fingerprints $p_{\bx_{\sfK}}$ assigned to the coalition; or
\item[(ii)] For every $p_{\bx_{\sfK}}$, the set $\scrW_K(p_{\bx_{\sfK}})$ of attacks
	given the joint type $p_{\bx_{\sfK}}$ contains only permutation-invariant attacks
	($\scrW_K(p_{\bx_{\sfK}}) = \scrW_K^{\fair}(p_{\bx_{\sfK}})$),
\end{description}
then fingerprint capacity under the detect-all criterion is equal to
$\widetilde{C}^{\all}(D_1,\scrW_K)$.
\item[(b)] {\bf Detect-one case.}
Fingerprinting capacity is lower-bounded by $\widetilde{C}^{\one}(D_1,\scrW_K)$
given by (\ref{eq:C-one}). If in addition the set $\scrW_K$ of attacks available
to the coalition is independent of the joint type of the fingerprints $p_{\bx_{\sfK}}$
assigned to the coalition, then
\[
   \widetilde{C}^{\one}(D_1,\scrW_K) = \widetilde{C}^{\one}(D_1,\scrW_K^{\fair})
	= \widetilde{C}^{\all}(D_1,\scrW_K^{\fair}) ,
\]
and fingerprint capacity under the detect-one criterion is equal to this common value.
\end{description}
\label{thm:C}
\end{theorem}

The lower bounds on fingerprinting capacity derived in \cite{Somekh05,Somekh07}
are of the form (\ref{eq:CL-one}) with $L=1$, i.e., the auxiliary random variable $W$
is degenerate. Since the payoff function
$I_{p_S \,p_{X|S}^{\sfK} \,p_{Y|X_{\sfK}}}(X_{\sfK};Y|S)$ is generally
nonconcave with respect to $p_{X|S}$, a randomized strategy in which the variable
$p_{X|S}$ is randomized will generally outperform a deterministic strategy in
which $p_{X|S}$ is fixed. The auxiliary random variable $W$ plays the role
of selector of $p_{X|S}$ in this mutual-information game.

Apparently the benefits of this randomization can be dramatic for large $K$.
For the Boneh-Shaw problem, the value of the maxmin of (\ref{eq:CL-one}) with $L=1$
is $C_1^{\one}(D_1,\scrW_K) = K^{-1} \,2^{-(K-1)}$. However Tardos' scheme
\cite{Tardos03} uses $\calW = [0,1]$ and achieves a rate $O(K^{-2})$ which is 
therefore much larger than $C_1^{\one}(D_1,\scrW_K)$ for large $K$.
The rate of his code is necessarily a lower bound on $C^{\one}(D_1,\scrW_K)$.

\section{Simple Fingerprint Decoder}
\label{sec:simple}
\setcounter{equation}{0}

This section introduces our random coding scheme and a simple decoder that tests candidate
fingerprints one by one. This decoder is closely related to the correlation decoders that
have been used in Tardos' paper \cite{Tardos03} and in the signal processing literature.
(Such decoders evaluate a measure of correlation between the received
sequence and the individual fingerprints, and retain the fingerprints whose correlation
score is above a certain threshold.)
We derive error exponents for this scheme and establish maximum rates for reliable decoding.
These rates fall short of the fingerprinting capacities $C^{\all}(D_1,\scrW_K)$ and
$C^{\one}(D_1,\scrW_K)$ given by Theorem~\ref{thm:C-all} and \ref{thm:C-one}.
The derivations are given for the case without side information ($S=\emptyset$)
or distortion constraint ($D_1$) for the fingerprint distributor. This setup is 
directly applicable to the Boneh-Shaw model, and the derivations are much easier
to follow. This setup also contains several key ingredients of the error analysis for
the more elaborate joint fingerprint decoder of Sec.~\ref{sec:joint}.
In particular, the false-negative error exponents are determined by the worst conditional
type $T_{\by\bx_{\calK}|\bw}$.

\subsection{Codebook}

The scheme is designed to achieve a false-positive error exponent equal to $\Delta$
and assumes a {\em nominal value} $K_{\nom}$ for coalition size. (Reliable
decoding will generally be possible for $K > K_{\nom}$ though.) These parameters are used
to identify a joint type class $T_{\bw\bx}^*$ defined below (\ref{eq:Epsp-simple-N}).
An arbitrarily large $L$ is selected, defining an alphabet $\calW = \{1,2,\cdots,L\}$.
A random constant-composition code $\calC(\bw) = \{\bx_m, \,1 \le m \le 2^{NR}\}$
is generated for each $\bw \in T_{\bw}^*$ by drawing $2^{NR}$ sequences independently
and uniformly from the conditional type class $T_{\bx|\bw}^*$. 

\subsection{Encoding Scheme}

A sequence $\bW$ is drawn uniformly from the type class $T_{\bw}^*$ and shared
with the receiver. User $m$ is assigned codeword $\bx_m$ from $\calC(\bW)$,
for $1 \le m \le 2^{NR}$.

\subsection{Decoding Scheme}

The receiver makes an innocent/guilty decision on each user {\em independently of
the other users}, and there lies the simplicity but also the suboptimality
of this decoder. Specifically, the estimated coalition $\hat{\calK}$ is the collection
of all $m$ such that
\begin{equation}
  I(\bx_m;\by|\bw) > R+\Delta .
\label{eq:decision-simple}
\end{equation}
If no such $\hat{\calK}$ is found, the receiver outputs $\hat{\calK} = \emptyset$.
The users whose empirical mutual information score exceeds the threshold $R+\Delta$
are declared guilty.

\subsection{Error Exponents}

Theorem~\ref{thm:simple} below gives the false-positive and false-negative
error exponents for this coding scheme. These exponents are given in terms
of the functions defined below.

Define the set of conditional p.m.f.'s for $X_{\sfK}$ given $W$
whose conditional marginals are the same for all components of $X_{\sfK}$:
\[ \scrM(p_{X|W}) = \{ p_{X_{\sfK}|W} ~:~ p_{X_m|W} = p_{X|W} ,
	\,\forall m \in \sfK \} . \]
Denote by $\scrP_{XW}(L)$ the set of p.m.f.'s $p_{XW}$ defined over
$\calX \times \calW$.
Define for each $m\in\sfK$ the set of conditional p.m.f.'s
\begin{eqnarray}
   \scrP_{YX_{\sfK}|W}(p_{XW}, \scrW_K, R,L, m)
	& \triangleq  & \left\{ \tp_{YX_{\sfK}|W}\,: ~\tp_{X_{\sfK}|W} \in \scrM(p_{X|W}) ,
	\;\tp_{Y|X_{\sfK}} \in \scrW_K(\tilde{p}_{X_{\sfK}}) , \right. \nonumber \\
	& & \qquad \qquad \left. \;I_{\tp_{YX_{\sfK}|W} p_W}(X_m;Y|W) \le R \right\}
\label{eq:set-m}
\end{eqnarray}
and the {\em pseudo sphere packing exponent}
\begin{eqnarray}
	\tilde{E}_{\psp,m}(R,L,p_{XW},\scrW_K)
	& = & \min_{\tp_{YX_{\sfK}|W} \,\in \,\scrP_{YX_{\sfK}|W}(p_{XW}, \scrW_K, R,L, m)} 
		\;D(\tp_{YX_{\sfK}|W} \| \tp_{Y|X_{\sfK}} \,p_{X|W}^K \,|\,p_W) .
\label{eq:Epsp-m-simple}
\end{eqnarray}
The terminology {\em pseudo sphere-packing exponent} is used because despite its superficial
resemblance to a sphere-packing exponent \cite{Csiszar81}, (\ref{eq:Epsp-m-simple})
does not provide a fundamental asymptotic lower bound on error probability. 

Taking the maximum and minimum of $\tilde{E}_{\psp,m}$ above over $m\in\sfK$,
we respectively define
\begin{eqnarray}
   \overline{\tilde{E}}_{\psp}(R,L,p_{XW},\scrW_K)
		& = & \max_{m\in\sfK} \tilde{E}_{\psp,m}(R,L,p_{XW},\scrW_K) ,
															\label{eq:Emax-tilde-simple} \\
   \underline{\tilde{E}}_{\psp}(R,L,p_{XW},\scrW_K)
		& = & \min_{m\in\sfK} \tilde{E}_{\psp,m}(R,L,p_{XW},\scrW_K) .
															\label{eq:Emin-tilde-simple}
\end{eqnarray}
If these expressions are evaluated for the set $\scrW_K^{\fair}$ which is
permutation invariant,
then (\ref{eq:set-m}) and 
(\ref{eq:Epsp-m-simple}) are independent of $m \in \sfK$, and the expressions
(\ref{eq:Emax-tilde-simple}) and (\ref{eq:Emin-tilde-simple}) coincide. Define
\begin{equation}
   E_{\psp}(R,L,\scrW_K) = \max_{p_{XW} \in \scrP_{XW}(L)}
		\tilde{E}_{\psp,1}(R,L,p_{XW},\scrW_{K_{\nom}}^{\fair}) .
\label{eq:Epsp-simple}
\end{equation}
Denote by $p_{XW}^*$ the maximizer in (\ref{eq:Epsp-simple}),
which depends on $R$ and $\scrW_{K_{\nom}}^{\fair}$.
Finally, define
\begin{eqnarray}
   \overline{E}_{\psp}(R,L,\scrW_K) & = & 
	\overline{\tilde{E}}_{\psp}(R,L,p_{XW}^*,\scrW_K) ,	\label{eq:Emax-simple} \\
   \underline{E}_{\psp}(R,L,\scrW_K) & = & 
	\underline{\tilde{E}}_{\psp}(R,L,p_{XW}^*,\scrW_K) ,	\label{eq:Emin-simple}
\end{eqnarray}
where no fairness requirement is imposed on $\scrW_K$.

\begin{theorem}
The threshold decision rule (\ref{eq:decision-simple}) yields the following error exponents.
\begin{description}
\item[(i)] The false-positive error exponent is
	\begin{equation}
	   E_{\FP}(R,L,\scrW_K,\Delta) = \Delta .
	\label{eq:E-FP-simple}
	\end{equation}
\item[(ii)] The detect-one error exponent is
	\begin{equation}
	   E^{\one}(R,L,\scrW_K,\Delta) = \overline{E}_{\psp}(R+\Delta,L,\scrW_K) .
	\label{eq:E-one-simple}
	\end{equation}
\item[(iii)] The detect-all error exponent is
	\begin{equation}
	   E^{\all}(R,L,\scrW_K,\Delta) = \underline{E}_{\psp}(R+\Delta,L,\scrW_K) .
	\label{eq:E-all-simple}
	\end{equation}
\item[(iv)] A fair collusion strategy is optimal under the detect-one error
	criterion: $E^{\one}(R,L,\scrW_K,\Delta) = E^{\one}(R,L,\scrW_K^{\fair},\Delta)$.
\item[(v)] The detect-one and detect-all error exponents 
	are the same when the colluders restrict their choice to fair strategies:
	$E^{\one}(R,L,\scrW_K^{\fair},\Delta) = E^{\all}(R,L,\scrW_K^{\fair},\Delta)$.
\item[(vi)] For $K = K_{\nom}$, the supremum of all rates for which the detect-one
	error exponent of (\ref{eq:E-one-simple}) is positive is given by
	\begin{eqnarray}
		C^{\simple}(\scrW_K) & = & C^{\simple}(\scrW_K^{\fair}) \nonumber \\
		& = & \lim_{L \to \infty} \;\max_{p_{XW}  \in \scrP_{XW}(L)}
			\;\min_{p_{Y|X_{\sfK}}  \in \scrW_K^{\fair}(p_{X_{\sfK}})}
			\;I_{p_W\,p_{X|W}^{\sfK}\,p_{Y|X_{\sfK}}}(X_1;Y|W)
	\label{eq:C1}
	\end{eqnarray}
	and is achieved by letting $\Delta \to 0$ and $L \to \infty$.
\end{description}
\label{thm:simple}
\end{theorem}

{\bf Note.}
Applying (\ref{eq:I2-fair}) with $S = \emptyset$, we have
$I(X_1;Y|W) \le \frac{1}{K} \,I(X_{\sfK};Y|W)$ for any permutation-invariant
$p_{Y|X_{\sfK}}$. Since this inequality is generally strict, $C^{\simple}(\scrW_K)$
is generally lower than the fingerprinting capacity $C^{\one}(\scrW_K)$ of (\ref{eq:C-one}).
Hence the simple thresholding rule (\ref{eq:decision-simple}) is generally
not capacity-achieving.

\section{Joint Fingerprint Decoder}
\label{sec:joint}
\setcounter{equation}{0}

The encoder and joint decoder are presented in this section,
and the performance of the new scheme is analyzed.  
As in the previous section, the encoder  ensures a false-positive
error exponent $\Delta$ and assumes a {\em nominal value} $K_{\nom}$ for coalition size.
An arbitrarily large $L$ is selected, defining an alphabet $\calW = \{1,2,\cdots,L\}$.
A random constant-composition code $\calC(\bs,\bw) = \{\bx_m, \,1 \le m \le 2^{NR}\}$
is generated for each $\bs \in \calS^N$ and $\bw \in T_{\bw}^*$ by drawing $2^{NR}$ sequences
independently and uniformly from a conditional type class $T_{\bx|\bs\bw}^*$.
Both $T_{\bw}^*$ and $T_{\bx|\bs\bw}^*$ depend on $\Delta$ and $K_{\nom}$
as defined below (\ref{eq:Epsp-N}).
Prior to encoding, a sequence $\bW \in \calW^N$ is drawn independently of $\bS$
and uniformly from $T_{\bw}^*$, and shared with the receiver.
Next, user $m$ is assigned codeword $\bx_m \in \calC(\bS,\bW)$, for $1 \le m\le 2^{NR}$.

In terms of decoding, the fundamental improvement over the simple strategy of Sec.~\ref{sec:simple}
resides in the use of a joint decoding rule. Specifically, the decoder maximizes a penalized
empirical mutual information score over all possible coalitions of any size.
The penalty is proportional to the size of the coalition.

\subsection{Mutual Information of $k$ Random Variables}
\label{sec:MI}

Our fingerprint decoding scheme is based on the notion of
mutual information between $k$ random variables $X_1, \cdots, X_k$.
For $k=3$, this mutual information is defined as \cite[p.~57]{Csiszar81} \cite[p.~378]{Liu96}
\[ \oI(X_1;X_2;X_3) = H(X_1) + H(X_2) + H(X_3) - H(X_1,X_2,X_3) . \]
We use the symbol $\oI$ to distinguish it from the symbol $I$ for standard mutual
information between two random variables. Note the chain rule
\[ \oI(X_1;X_2;X_3) = I(X_1; X_2 X_3) + I(X_2;X_3) . \]

The mutual information between $k$ random variables $X_1, \cdots, X_k$
is similarly defined as the sum of their individual entropies minus
their joint entropy \cite[p.~57]{Csiszar81}
or equivalently, the divergence between their joint distribution
and the product of their marginals:
\begin{eqnarray}
   \oI(X_1;\cdots;X_k) & = & H(X_1) + \cdots + H(X_k) - H(X_1,\cdots,X_k)
												\label{eq:MI-k} \\
					 & = & D(p_{X_1 \cdots X_k} \| p_{X_1} \cdots p_{X_k}) . \nonumber
\end{eqnarray}
Note the following properties, including the chain rules (P3) and (P4):
\begin{description}
\item[(P1)] The mutual information (\ref{eq:MI-k}) is symmetric in its arguments;
\item[(P2)] $\oI(X_1;X_2) = I(X_1;X_2)$;
\item[(P3)] $\oI(X_1;\cdots;X_k) = I(X_1;X_2 \cdots X_k) + \oI(X_2;\cdots;X_k) 
		= \sum_{i=1}^{k-1} I(X_i;X_{i+1} \cdots X_k)$;
\item[(P4)] $\oI(X_1;\cdots;X_k)
		= \oI(X_1;\cdots ;X_i;X_{i+1} \cdots X_k) + \oI(X_{i+1};\cdots;X_k)$
		for any $i \in \{1, 2, \cdots, k-2\}$;
\item[(P5)] $\oI(X_1;\cdots;X_k)
		= \sum_{i=1}^{k-1} H(X_i) - H(X_1 \cdots X_{k-1} \,|\,X_k)$.
\end{description}

Similarly to (\ref{eq:MI-k}), we define the empirical mutual information
$\oI(\bx_1;\cdots;\bx_k)$ between $k$ sequences $\bx_1, \cdots, \bx_k$, as 
the mutual information with respect to the joint type of $\bx_1, \cdots, \bx_k$.
Analogously to Property (P5), we have
\begin{equation}
   \oI(\bx_1; \cdots; \bx_k; \by) = \sum_{i=1}^k H(\bx_i) - H(\bx_1 \cdots \bx_k | \by) .
\label{eq:I-empirical}
\end{equation}
This leads to the following alternative interpretation of the minimum-equivocation
decoder of Liu and Hughes \cite{Liu96}.
If $\bx_1, \cdots, \bx_k$ are codewords from a constant-composition code $\calC$, then 
$H(\bx_i)$ is the same for all $i$, then the minimum-equivocation decoder
is equivalent to a maximum-mutual-information decoder:
\begin{equation}
   \min_{\bx_1 \cdots \bx_k \in \calC} H(\bx_1 \cdots \bx_k | \by)
	\quad \Leftrightarrow \quad \max_{\bx_1 \cdots \bx_k \in \calC} 
		\oI(\bx_1; \cdots; \bx_k; \by) .
\label{eq:equivalence}
\end{equation}
There is no similar interpretation when ordinary mutual information
$I(\bx_1 \cdots \bx_k; \by)$ is used \cite{Liu96}.
Liu and Hughes showed that the minimum-equivocation decoder
outperforms the ordinary maximum-mutual-information decoder
in terms of random-coding exponent.

\subsection{MPMI Criterion}
\label{sec:MPMI}

The restriction of $\bx_{\calM}$ to a subset $\calA$ of $\calM$ will be denoted
by $\bx_\calA = \{ \bx_m, \,m \in \calA\}$.
For disjoint sets $\calA = \{ m_1, \cdots, m_{|\calA|} \}$ and
$\calB = \{ m_{|\calA|+1}, \cdots, m_{|\calA|+|\calB|} \}$, we use the shorthand
\begin{equation}
   \oI(\bx_{\calA};\by\bx_{\calB}|\bs\bw) \triangleq
	\oI(\bx_{m_1}; \cdots; \bx_{m_{|\calA|}}; \by\bx_{\calB}|\bs\bw)
\label{eq:MPMI-shorthand}
\end{equation}
for the mutual information between the $|\calA|+1$ random variables
$\bx_{m_1}, \cdots, \bx_{m_{|\calA|}}$, and $(\by,\bx_{\calB})$,
conditioned on $(\bs,\bw)$.


Define the function
\begin{equation}
   MPMI(k) = \left\{ \begin{array}{ll} 0 & :~\mathrm{if~} k = 0 \\
	\underset{\bx_{\calK}\in\calC^k(\bs,\bw)}{\max}
	\left[\oI(\bx_{\calK};\by|\bs\bw) - k(R+\Delta) \right]
				& :~\mathrm{if~} k=1,2,\cdots
				\end{array} \right.
\label{eq:MPMI-k}
\end{equation}
where $k=|\calK|$ and
\vspace*{-0.1in}
\begin{equation}
   \oI(\bx_{\calK};\by|\bs\bw) = \oI(\bx_1;\cdots;\bx_k;\by|\bs\bw) 
	= kH(\bx|\bs\bw) - H(\bx_{\calK}|\,\by\bs\bw)
\label{eq:oI-decomp}
\end{equation}
is the mutual information between the $k+1$ sequences $\bx_1,\cdots,\bx_k,\by$,
conditioned on $(\bs,\bw)$, as defined in (\ref{eq:MPMI-shorthand}).
Again we stress that $\oI(\bx_1;\cdots;\bx_k;\by|\bs\bw)$ should not be confused with
the ordinary mutual information $I(\bx_1 \cdots \bx_k;\by|\bs\bw)$ between the $k$-tuple
$(\bx_1 , \cdots, \bx_k)$ and $\by$, conditioned on $(\bs,\bw)$.
Our joint fingerprint decoder is a {\bf Maximum Penalized Mutual Information} (MPMI) decoder:
\begin{equation}
   \max _{k \ge 0} MPMI(k) .
\label{eq:MPMI}
\end{equation}
In case of a tie, the largest value of $k$ is retained.
The decoder seeks the coalition size $k$ and the codewords $\{\bx_m, \,m\in\hat{\calK}\}$
in $\calC(\bs,\bw)$ that achieve the MPMI criterion above. The indices of these codewords
form the decoded coalition $\hat{\calK}$. If the maximizing $k$ in (\ref{eq:MPMI})
is zero, the receiver outputs $\hat{\calK} = \emptyset$.
Similarly to (\ref{eq:equivalence}), the MPMI decoder may equivalently
be interpreted as a Minimum Penalized Equivocation criterion.


\subsection{Properties}
\label{sec:joint-properties}

The following lemma shows that
1) each subset of the estimated coalition is significant, and 
2) any extension of the estimated coalition would fail a significance test. 
\begin{lemma}
Let $\hat{\calK}$ achieve the maximum in (\ref{eq:MPMI-k})
(\ref{eq:MPMI}). Then
\begin{equation}
  \forall \calA \subseteq \hat{\calK} ~:\quad
  \oI(\bx_{\calA};\by\bx_{\hat{\calK}\setminus\calA}\,|\bs\bw) > |\calA| (R+\Delta) .
\label{eq:property1}
\end{equation}
Moreover, for every $\calA$ disjoint with $\hat{\calK}$,
\begin{equation}
  \oI(\bx_{\calA};\by\bx_{\hat{\calK}}\,|\bs\bw) \le |\calA| (R+\Delta) .
\label{eq:property2}
\end{equation}
\label{lem:optimalK}
\end{lemma}
\vspace*{-0.3in}
{\em Proof}.
For any $\calA \subseteq \hat{\calK}$, we have
\begin{eqnarray*}
   \lefteqn{\oI(\bx_{\calA};
		\by\bx_{\hat{\calK}\setminus\calA}\,|\bs\bw) - |\calA| \,(R+\Delta)} \\
	& \stackrel{(a)}{=} & [\oI(\bx_{\hat{\calK}};\by\,|\bs\bw) - \hat{K} \,(R+\Delta)]
		- [\oI(\bx_{\hat{\calK}\setminus\calA};\by\,|\bs\bw) - (\hat{K}-|\calA|) \,(R+\Delta)] \\
	& \stackrel{(b)}{=} & MPMI(\hat{K})
		- [\oI(\bx_{\hat{\calK}\setminus\calA};\by\,|\bs\bw) - (\hat{K}-|\calA|) \,(R+\Delta)] \\
	& \ge & MPMI(\hat{K}) - MPMI(\hat{K}-|\calA|) \\
	& \stackrel{(c)}{\ge} & 0
\end{eqnarray*}
where (a) follows from the chain rule for $\oI$,
(b) holds because $\hat{\calK}$ achieves the maximum in (\ref{eq:MPMI-k}), and
(c) because $\hat{K}$ achieves the maximum in (\ref{eq:MPMI}).
This proves (\ref{eq:property1}).

To prove (\ref{eq:property2}), consider any $\calA$ disjoint with $\hat{\calK}$
and let $\calK' = \hat{\calK} \cup \calA$. We have
\begin{eqnarray*}
   \lefteqn{\oI(\bx_{\calA};\by\bx_{\hat{\calK}}\,|\bs\bw) - |\calA| \,(R+\Delta)} \\
	& \stackrel{(a)}{=} & [\oI(\bx_{\calK'};\by\,|\bs\bw) - K' \,(R+\Delta)]
		- [\oI(\bx_{\hat{\calK}};\by\,|\bs\bw) - \hat{K} \,(R+\Delta)] \\
	& \stackrel{(b)}{=} & [\oI(\bx_{\calK'};\by\,|\bs\bw) - K' \,(R+\Delta)] - MPMI(\hat{K}) \\
	& \le & MPMI(K') - MPMI(\hat{K}) \\
	& \stackrel{(c)}{\le} & 0 ,
\end{eqnarray*}
where (a), (b), (c) are justified in the same way as above.
This proves (\ref{eq:property2}).
\hfill $\Box$

{\bf Reliability metric.}
The score
\vspace*{-0.1in}
\[ \oI(\bx_{\hat{\calK}};\by\,|\bs\bw) - \hat{K}R
	> \hat{K} \Delta \]
represents a guilt index for the estimated coalition $\hat{\calK}$.
The larger this quantity is, the stronger the evidence
that the members of $\hat{\calK}$ are guilty.
Likewise,
\[ \oI(\bx_m;\by \bx_{\hat{\calK}\setminus\{m\}}\,|\bs\bw) - R > \Delta \]
is a guilt index for accused user $m \in \hat{\calK}$, and
\[ \oI(\bx_m;\by \bx_{\hat{\calK}}\,|\bs\bw) - R \le \Delta \]
is a guilt index for user $m \notin \hat{\calK}$.
The smaller this index is, the stronger the evidence that $m$ is innocent.

\subsection{Error Exponents}
\label{sec:joint-exp}

Theorem~\ref{thm:joint} below gives the false-positive and false-negative
error exponents for our coding scheme. These exponents are given in terms
of the functions defined below.

Recall $\scrP_{X_{\sfK} W|S}(p_S,L,D_1)$ defined in (\ref{eq:PXS-set}).
We similarly define
\begin{eqnarray*}
   \scrP_{X_{\sfK}|SW}(p_{SW},L,D_1)
	& \triangleq & \left\{ p_{X_{\sfK}|SW} = \prod_{k\in\sfK} p_{X_k|SW}
		~:~ p_{X_1|SW} = \cdots = p_{X_K|SW} , \; \eE d(S,X_1) \le D_1 \right\} .
\end{eqnarray*}
Define now the following set of conditional p.m.f.'s for $X_{\sfK}$ given $S,W$
whose conditional marginal p.m.f. $p_{X|SW}$ is the same for each $X_m, m \in \sfK$:
\[ \scrM(p_{X|SW}) = \{ p_{X_{\sfK}|SW} ~:~ p_{X_m|SW} = p_{X|SW} , \,\forall m \in \sfK \} . \]
Define for each $\sfA \subseteq \sfK$ the set of conditional p.m.f.'s
\begin{eqnarray}
   \lefteqn{\scrP_{YX_{\sfK}|SW}(p_W,\tp_{S|W}, p_{X|SW}, \scrW_K, R, L, \sfA)} \nonumber \\
	& \triangleq & \left\{ \tp_{YX_{\sfK}|SW}\,: ~\tp_{X_{\sfK}|SW} \in \scrM(p_{X|SW}) ,
						\;\tp_{Y|X_{\sfK}} \in \scrW_K(\tilde{p}_{X_{\sfK}}),
						\right. \nonumber \\
	& & \quad \left. \frac{1}{|\sfA|} \oI_{p_W\,\tp_{S|W} \,\tp_{YX_{\sfK}|SW}}
			(X_{\sfA};YX_{\sfK\setminus\sfA}|S,W) \le R \right\}
\label{eq:set-A}
\end{eqnarray}
and the {\em pseudo sphere packing exponent} 
\begin{eqnarray}
   \lefteqn{\tilde{E}_{\psp,\sfA}(R,L,p_W,\tp_{S|W},p_{X|SW},\scrW_K)} \nonumber \\
	& = & \min_{\tp_{YX_{\sfK}|SW} \in \scrP_{YX_{\sfK}|SW}
			(p_W, \tp_{S|W}, p_{X|SW}, \scrW_K, R, L, \sfA)}
			\;D(\tp_{YX_{\sfK}|SW}\,\tp_{S|W} \| \tp_{Y|X_{\sfK}} \,p_{X|SW}^K \,p_S \,|\,p_W) .
\label{eq:Epsp-A}
\end{eqnarray}
Taking the maximum \footnote{
   The property that $\sfK$ achieves $\max_{\sfA \subseteq \sfK} \tilde{E}_{\psp,\sfA}$
   is derived in the proof of Theorem~\ref{thm:joint}, Part (iv).
}
and the minimum of $\tilde{E}_{\psp,\sfA}$ above over all subsets
$\sfA \subseteq \sfK$, we define
\begin{eqnarray}
   \overline{\tilde{E}}_{\psp}(R,L,p_W,\tp_{S|W},p_{X|SW},\scrW_K)
		& = & \tilde{E}_{\psp,\sfK}(R,L,p_W,\tp_{S|W},p_{X|SW},\scrW_K) ,
																	\label{eq:Emax-tilde} \\
   \underline{\tilde{E}}_{\psp}(R,L,p_W,\tp_{S|W},p_{X|SW},\scrW_K)
		& = & \min_{\sfA \subseteq \sfK}
				\tilde{E}_{\psp,\sfA}(R,L,p_W,\tp_{S|W},p_{X|SW},\scrW_K) .
																	\label{eq:Emin-tilde}
\end{eqnarray}
Now define
\begin{eqnarray}
   E_{\psp}(R,L,D_1,\scrW_K) & = & \max_{p_W \in \scrP_W} \min_{\tp_{S|W} \in \scrP_{S|W}}
	\max_{p_{X|SW} \in \scrP_{X|SW}(p_W,\tp_{S|W},L,D_1)}			\nonumber \\
	& & \quad \tilde{E}_{\psp,\sfK}(R,L,p_W,\tp_{S|W},p_{X|SW},\scrW_{K_{\nom}}^{\fair}) .
\label{eq:Epsp}
\end{eqnarray}
Denote by $p_W^*$ and $p_{X|SW}^*$ the maximizers in (\ref{eq:Epsp}),
where the latter is to be viewed as a function of $\tp_{S|W}$. Also note that
both $p_W^*$ and $p_{X|SW}^*$ implicitly depend on $R$ and $\scrW_{K_{\nom}}^{\fair}$.
Finally, define
\begin{eqnarray}
   \overline{E}_{\psp}(R,L,D_1,\scrW_K) & = & \min_{\tp_{S|W} \in \scrP_{S|W}}
	\overline{\tilde{E}}_{\psp}(R,L,p_W^*,\tp_{S|W},p_{X|SW}^*,\scrW_K) ,
													\label{eq:Emax} \\
   \underline{E}_{\psp}(R,L,D_1,\scrW_K) & = & \min_{\tp_{S|W} \in \scrP_{S|W}}
	\underline{\tilde{E}}_{\psp}(R,L,p_W^*,\tp_{S|W},p_{X|SW}^*,\scrW_K) .
													\label{eq:Emin}
\end{eqnarray}
\newpage

\begin{theorem}
The decision rule (\ref{eq:MPMI}) yields the following error exponents.
\begin{description}
\item[(i)] The false-positive error exponent is
	\begin{equation}
	   E_{\FP}(R,D_1,\scrW_K,\Delta) = \Delta .
	\label{eq:E-FP}
	\end{equation}
\item[(ii)] The error exponent for the (false negative) probability that the decoder fails to catch
	all colluders (misses some of them) is
	\begin{equation}
	   E^{\all}(R,L,D_1,\scrW_K,\Delta) = \underline{E}_{\psp}(R+\Delta,L,D_1,\scrW_K) .
	\label{eq:E-all}
	\end{equation}
\item[(iii)] The error exponent for the (false negative) probability that the decoder
	fails to catch even one colluder (misses every single colluder) is
	\begin{equation}
	   E^{\one}(R,L,D_1,\scrW_K,\Delta) = \overline{E}_{\psp}(R+\Delta,L,D_1,\scrW_K) .
	\label{eq:E-one}
	\end{equation}
\item[(iv)] $E^{\one}(R,L,D_1,\scrW_K,\Delta) = E^{\one}(R,L,D_1,\scrW_K^{\fair},\Delta)$.
\item[(v)]  $E^{\all}(R,L,D_1,\scrW_K^{\fair},\Delta) = E^{\one}(R,L,D_1,\scrW_K^{\fair},\Delta)$.
\item[(vi)] If $K = K_{\nom}$, the supremum of all rates for which the error exponents
	of (\ref{eq:E-all}) and (\ref{eq:E-one}) are positive are $C^{\all}(D_1,\scrW_K)$
	and $C^{\one}(D_1,\scrW_K)$ of (\ref{eq:C-all}) and (\ref{eq:C-one}), respectively.
\end{description}
\label{thm:joint}
\end{theorem}

{\bf Note.}
The expressions (\ref{eq:E-all}) and (\ref{eq:E-one}) for
the false-negative error exponents may be viewed as sequences indexed by $L$.
As discussed below (\ref{eq:CL-all}) and in \cite[Sec.~3.5]{Moulin07},
one may show that these sequences are nondecreasing and converge to
finite limits at a polynomial rate.

\section{Error Exponents for Memoryless Collusion Channels}
\label{sec:memoryless}
\setcounter{equation}{0}

Consider the compound class (\ref{eq:memoryless}) of memoryless channels.
The theorems of Sec.~\ref{sec:C} showed that compound capacity is the same
as for the main problem of (\ref{eq:WK}).
We now outline how the derivation of the error exponents.

Retracing the steps of the proof of Theorem~\ref{thm:joint}, it may be seen that
the expressions (\ref{eq:E-FP}), (\ref{eq:E-all}) and (\ref{eq:E-one}) for the error exponents
remain valid, with two modifications. First, in (\ref{eq:set-A}),
the constraint $\tp_{Y|X_{\sfK}} \in \scrW_K$ is removed, and so
the resulting set $\scrP_{YX_{\sfK}|SW}^{\memoryless}$ is larger than
$\scrP_{YX_{\sfK}|SW}$ of (\ref{eq:set-A}). Second, the divergence cost function
\begin{equation}
  D(\tp_{YX_{\sfK}|SW}\,\tp_{S|W} \| \tp_{Y|X_{\sfK}} \,p_{X|SW}^K \,p_S \,|\,p_W)
\label{eq:div-1}
\end{equation}
in the expression (\ref{eq:Epsp-A}) for the pseudo sphere packing exponent
$\tilde{E}_{\psp,\sfA}$ is replaced by \footnote{
	This can be traced back to (\ref{eq:PrT}), where $p_{\by|\bx_{\calK}}$
	is now replaced with $p_{Y|X_{\calK}}$ in the asymptotic expression for
	the probability of the conditional type class $T_{\by\bx_{\calK}|\bs\bw}$.
	}
\begin{equation}
   \min_{p_{Y|X_{\sfK}} \in \scrW_K}
	\;D(\tp_{YX_{\sfK}|SW}\,\tp_{S|W} \| p_{Y|X_{\sfK}} \,p_{X|SW}^K \,p_S \,|\,p_W) ;
\label{eq:div-2}
\end{equation}
denote by $\tilde{E}_{\psp,\sfA}^{\memoryless}$ the corresponding pseudo sphere packing
exponent.

The divergences in (\ref{eq:div-1}) and (\ref{eq:div-2}) coincide
when $p_{Y|X_{\sfK}} = \tp_{Y|X_{\sfK}}$, thus (\ref{eq:div-2}) is upper-bounded
by (\ref{eq:div-1}). Since $p_{Y|X_{\sfK}} = \tp_{Y|X_{\sfK}}$ is feasible
for $\scrP_{YX_{\sfK}|SW}$ of (\ref{eq:set-A}), we conclude that
$\tilde{E}_{\psp,\sfA}^{\memoryless}$ $\le \tilde{E}_{\psp,\sfA}$ of (\ref{eq:Epsp-A}).
Hence the false-negative error exponents in the memoryless case are upper-bounded
by those of Theorem~\ref{thm:joint}. This phenomenon is similar to results
in \cite{Moulin07}: due to the use of RM codes, the colluders' optimal strategy
is a nearly-memoryless strategy, but they are precluded from using a truly memoryless
strategy because that would violate the hard constraint $p_{\by|\bx_{\sfK}} \in \scrW_K$.
In the memoryless case, the worst conditional type (which determines the false-negative
error exponents) might be such that $p_{\by|\bx_{\calK}} \notin \scrW_K$.

\section{Proof of Converse Under Detect-All Criterion}
\label{Sec:Converse-all}
\setcounter{equation}{0}

\subsection{Proof of Theorem~\ref{thm:C-all}}

The encoder generates marked copies $\bx_m = f_N(\bs,v,m)$ for
$1 \le m \le 2^{NR}$ and the decoder outputs an estimated coalition
$g_N(\by,\bs,v) \in \{ 1, \cdots, 2^{NR} \}^\star$.
By Lemma~\ref{lem:memoryless}, it suffices to prove the claim for the compound
class of memoryless channels $\scrW_K$ of (\ref{eq:memoryless}).
Let $K$ be the size of the coalition and $(f_N,g_N)$ a sequence of length-$N$,
rate-$R$ codes. We show that for any such sequence of
codes, reliable decoding of the fingerprints is possible only if
$R \le \widetilde{C}^{\all}(D_1,\scrW_K)$ under the detect-all criterion.

{\bf Step~1.}
A lower bound on error probability is obtained when a helper provides
some information to the decoder. Here the helper informs the decoder
that the coalition size is $K$. 
There are $\left( \begin{array}{c} 2^{NR} \\ K \end{array} \right) \le 2^{KNR}$
possible coalitions of size $K$.
We represent a coalition as $M_\sfK \triangleq \{ M_1, \cdots, M_K\}$, where
$M_k, \,k\in\sfK = \{1,2,\cdots,K\}$, are assumed to be drawn i.i.d. uniformly \footnote{
   Capacity could be higher if there were constraints on the formation of coalitions,
	for instance if the users form social networks \cite{Moulin09-ICASSP}.}
from $\{1, \cdots, 2^{NR}\}$.
We similarly write $\bX_k \triangleq \bx_{M_k}, \,k\in\sfK$,
and $\bX_\sfK \triangleq \{ \bX_1, \cdots, \bX_K\}$. The component of $\bX_\sfK$ at position
$t \in \{1,\cdots,N\}$ is denoted by $\bX_{\sfK,t} \triangleq \{X_{1t}, \cdots, X_{Kt}\}$.
Assuming memoryless collusion channel $p_{Y|X_\sfK} \in \scrW_K$ is in effect,
the joint p.m.f. of $(M_\sfK,\bS,V,\bX_\sfK,\bY)$ is given by
\begin{equation}
 p_{M_\sfK \bS V \bX_\sfK \bY} = p_S^N \,p_V \,\prod_{k\in\sfK} \left(
	p_{M_k} \,\mathds1\{\bX_k=f_N(\bS,V,M_k)\} \right) \,p_{Y|X_\sfK}^N .
\label{eq:jointProbate}
\end{equation}

Define the random variables $Q_t = \{V, \,S_j, j \ne t\} \in \calV_N \times \calS^{N-1}$
for $1 \le t \le N$. By assumption, $S_t$ and $Q_t$ are independent,
and $X_{kt}, \,k\in\sfK$, are conditionally i.i.d. given $(S_t,Q_t) = (\bS,V)$.
However, note that $X_{kt}, \,1 \le k \le K$, are generally conditionally {\em dependent}
given $(S_t, V)$ alone. The joint p.m.f. of $(S_t,Q_t,X_{\sfK,t},Y_t)$ is
\begin{equation}
   p_{S_t} p_{Q_t} \left( \prod_{1 \le k \le K} p_{X_{kt}|S_tQ_t} \right)
	\,p_{Y|X_{\sfK}} , \quad 1 \le t \le N
\label{eq:joint-t}
\end{equation}
where the conditional p.m.f. $p_{X_{kt}|S_tQ_t}$ is the same for all $k\in\sfK$.
Now define a time-sharing random variable $T$, uniformly distributed over
$\{1, \cdots, N\}$, and independent of the other random variables.
Let
\begin{eqnarray}
   X_{\sfK} & \triangleq & X_{\sfK,T} \in \calX^K, \quad
		Y \triangleq Y_T \in \calY, \quad S \triangleq S_T \in \calS, \nonumber \\
	W & \triangleq & (Q_T, T) \in \calW \triangleq \calV_N \times \calS^{N-1}
		\times \{1,\cdots,N\} .
\label{eq:converse-all-RVs}
\end{eqnarray}
By (\ref{eq:joint-t}) and (\ref{eq:converse-all-RVs}), the code $f_N$ and
the random variables $\bS,V,M_\sfK$ induce an empirical p.m.f.
$p_{X_{\sfK}}$ which can be viewed as a function of $f_N$.
The joint p.m.f. of $(S,W,X_{\sfK},Y)$ is 
\begin{equation}
   p_S \,p_W \left( \prod_{k\in\sfK} p_{X_k|SW} \right) \,p_{Y|X_{\sfK}}
\label{eq:jointate}
\end{equation}
where the conditional p.m.f. $p_{X_k|SW}$ is the same for all $k\in\sfK$.
Moreover
\[ D_1 \ge \eE \left[ \frac{1}{N} \sum_{t=1}^N d(S_t,X_{kt}) \right]
	= \eE \,d(S,X_k) , \quad k\in\sfK . \]
Hence $p_{X_{\sfK} W|S}$ belongs to the set $\scrP_{X_{\sfK} W|S}(p_S,L,D_1)$
of (\ref{eq:PXS-set}), with $L = |\calW| = N \times \,\calV_N \times \,|\calS|^N$.




{\bf Step~2.}
Our single-letter expressions are derived from the following inequality,
which is valid for all $\sfA \subseteq \sfK$ and $p_{Y|X_\sfK} \in \scrW_K$:
\begin{eqnarray}
  I(M_\sfA;\bY|\bS,V)
	& \stackrel{(a)}{=} & I(\bX_\sfA;\bY|\bS,V) \nonumber\\
	& = & I(\bX_\sfA;\bY|\bX_{\sfK\setminus\sfA},\bS,V)
			+ \underbrace{I(\bX_\sfA;\bX_{\sfK\setminus\sfA}|\bS,V)}_{=0} 
			- I(\bX_\sfA;\bX_{\sfK\setminus\sfA}|\bY,\bS,V) \nonumber\\
	& \stackrel{(b)}{\le} & I(\bX_\sfA;\bY|\bX_{\sfK\setminus\sfA},\bS,V) \nonumber\\
	& = & H(\bY|\bX_{\sfK\setminus\sfA},\bS,V) - H(\bY|\bX_{\sfK},\bS,V) \nonumber \\
	& \stackrel{(c)}{=} & H(\bY|\bX_{\sfK\setminus\sfA},\bS,V) - H(\bY|\bX_{\sfK}) \nonumber \\
	& \stackrel{(d)}{=} & \sum_{t=1}^N H(Y_t|Y^{t-1},\bX_{\sfK\setminus\sfA},\bS,V)
			- \sum_{t=1}^N H(Y_t|X_{\sfK,t}) \nonumber \\
	& \stackrel{(e)}{\le} & \sum_{t=1}^N H(Y_t|X_{\sfK\setminus\sfA,t},\bS,V)
			- \sum_{t=1}^N H(Y_t|X_{\sfK,t}) \nonumber \\
	& \stackrel{(f)}{=} & \sum_{t=1}^N H(Y_t|X_{\sfK\setminus\sfA,t},S_t,Q_t)
		- \sum_{t=1}^N H(Y_t|X_{\sfK,t},S_t,Q_t) \nonumber \\
	& = & \sum_{t=1}^N I(X_{\sfA,t};Y_t|X_{\sfK\setminus\sfA,t},S_t,Q_t) \nonumber \\
	& = & N \,I(X_{\sfA};Y|X_{\sfK\setminus\sfA},S,W)
\label{eq:I-sum}
\end{eqnarray}
where (a) is due to the data processing inequality and the fact that
$\bX_\sfA$ is a function of $(M_\sfA, \bS, V)$,
(b) holds because the codewords $\{\bX_k, \,1 \le k \le K\}$ are mutually independent
given $(\bS,V)$, 
(c) because $(\bS,V) \to \bX_{\sfK} \to \bY$ forms a Markov chain,
(d) is obtained using the chain rule for entropy and the fact that the
collusion channel is memoryless,
(e) holds because conditioning reduces entropy, and
(f) because $(\bS,V)=(S_t,Q_t) \to X_{\sfK,t} \to Y_t$ forms a Markov chain.

{\bf Step~3.}
Under collusion channel $p_{Y|X_\sfK} \in \scrW_K$, let
$P_e^{\all}(p_{Y|X_\sfK})=Pr[\hat{\calK} \ne \calK]$ be the decoding error probability
of the detect-all decoder.
The following inequalities hold for every subset $\sfA$ of $\sfK$
and for every $p_{Y|X_{\sfK}}$:
\begin{eqnarray}
   |\sfA| \,NR \stackrel{(a)}{=} H(M_{\sfA}) \stackrel{(b)}{=} H(M_{\sfA}|\bS,V)
		& = & H(M_{\sfA}|\bY,\bS,V) + I(M_{\sfA};\bY|\bS,V) \nonumber\\
		& \le & H(M_{\sfK}|\bY,\bS,V) + I(M_{\sfA};\bY|\bS,V) \nonumber\\
		& \stackrel{(c)}{\le} & 1 + P_e^{\all}(p_{Y|X_\sfK}) \cdot KNR + I(M_{\sfA};\bY|\bS,V)
\label{eq:fano-all}
\end{eqnarray}
where (a) holds because $M_{\sfA}$ is uniformly distributed over $\{1, \cdots, 2^{|\sfA|\,NR}\}$,
(b) because $M_{\sfA}$ and $(\bS,V)$ are independent,
and (c) because of Fano's inequality.

For the error probability $P_e^{\all}(p_{Y|X_\sfK})$ to vanish for each
$p_{Y|X_\sfK} \in \scrW_K$, we need
\begin{equation}
   R \le \liminf_{N \to \infty} \min_{p_{Y|X_\sfK} \in \scrW_K}
	\min_{\sfA \subseteq \sfK} \frac{1}{N |\sfA|} I(M_{\sfA};\bY|\bS,V).
\label{eq:KRate}
\end{equation}

We have
\begin{eqnarray}
   \lefteqn{\min_{p_{Y|X_\sfK} \in \scrW_K} \min_{\sfA \subseteq \sfK}
			\frac{1}{N |\sfA|} I(M_{\sfA};\bY|\bS,V)} \nonumber \\
	& \stackrel{(a)}{\le} & \min_{p_{Y|X_\sfK} \in \scrW_K} \min_{\sfA \subseteq \sfK}
			\frac{1}{|\sfA|} I(X_{\sfA};Y|X_{\sfK\setminus\sfA},S,W) \nonumber \\
	& \stackrel{(b)}{\le} & \max_{p_{X_{\sfK} W|S} \,\in \,\scrP_{X_{\sfK} W|S}(p_S,L(N),D_1)}
			\;\min_{p_{Y|X_\sfK} \in \scrW_K} \min_{\sfA \subseteq \sfK}
			\frac{1}{|\sfA|} \,I(X_{\sfA};Y|X_{\sfK\setminus\sfA},S,W) \nonumber \\
	& \le & \sup_{L \to \infty} \;\max_{p_{X_{\sfK} W|S} \,\in \,\scrP_{X_{\sfK} W|S}(p_S,L,D_1)}
			\;\min_{p_{Y|X_\sfK} \in \scrW_K} \min_{\sfA \subseteq \sfK}
			\frac{1}{|\sfA|} \,I(X_{\sfA};Y|X_{\sfK\setminus\sfA},S,W) \nonumber \\
	& \stackrel{(c)}{\le} & \lim_{L \to \infty}
			\;\max_{p_{X_{\sfK} W|S} \,\in \,\scrP_{X_{\sfK} W|S}(p_S,L,D_1)}
			\;\min_{p_{Y|X_\sfK} \in \scrW_K} \min_{\sfA \subseteq \sfK}
			\frac{1}{|\sfA|} \,I(X_{\sfA};Y|X_{\sfK\setminus\sfA},S,W)
\label{eq:I-converse-all}
\end{eqnarray}
where (a) is due to (\ref{eq:I-sum}),
(b) to the fact that $p_{X_{\sfK} W|S}$
given in (\ref{eq:jointate}) belongs to the set $\scrP_{X_{\sfK} W|S}(p_S,L,D_1)$
defined in (\ref{eq:PXS-set}), with $L=L(N)= N \times \,\calV_N \times \,|\calS|^N$,
and (c) because the supremand is nondecreasing in $L$.

Combining (\ref{eq:KRate}) and (\ref{eq:I-converse-all}), we obtain
\begin{eqnarray}
   R & \le & \lim_{L \to \infty} \;\max_{p_{X_{\sfK} W|S} \in \scrP_{X_{\sfK} W|S}(p_S,L,D_1)}
			\;\min_{p_{Y|X_\sfK} \in \scrW_K(p_{X_{\sfK}})} \min_{\sfA \subseteq \sfK}
			\frac{1}{|\sfA|} \;I(X_{\sfA};Y|X_{\sfK\setminus\sfA},S,W) \nonumber \\
	 & = & \lim_{L \to \infty} C_L^{\all}(D_1, \scrW_K) \nonumber \\
	 & = & \widetilde{C}^{\all}(D_1,\scrW_K)
\label{eq:Rmax-C-all}
\end{eqnarray}
which concludes the proof of Theorem~\ref{thm:C-all}.
\hfill $\Box$

\subsection{Proof of Corollary~\ref{cor:C-all}}

By assumption, here the coalition is fair and $\scrW_K = \scrW_K^{\fair}$ depends on
the joint type $p_{\bx_\calK}$ of the colluders' fingerprinted sequences. We denote
this joint type by $Z \in \calZ = \scrP_{X_\sfK}^{[N]}$ to make the notation
more compact. Note that $Z$ is a function of $(\bS, V, \calK)$
and that the cardinality of $\calZ$ is at most $(N+1)^{|\calX|^K}$.
Since the channel $p_{Y|X_\sfK}$ selected by the coalition may depend on $Z$,
we indicate this dependency explicitly by representing the channel
as $p_{Y|X_\sfK Z}$ and the set of feasible channels as
\begin{equation}
  \widetilde{\scrW}_K^{\fair} = \{ p_{Y|X_\sfK Z} ~:~ 
	p_{Y|X_\sfK, Z=z} \in \scrW_K^{\fair}(z) ,\;\forall z \in \calZ \} .
\label{eq:tildeWK}
\end{equation}
By Lemma~\ref{lem:memoryless}, it suffices to prove the claim for the compound
class of memoryless channels $\scrW_K^{\fair}(p_{\bx_{\calK}})$.

Define the set
\[ \scrP_{X_{\sfK} WS}(p_S,L,D_1) \triangleq \left\{ p_S \,p_{X_{\sfK} W|S} ~:~
	p_{X_{\sfK} W|S}\in \scrP_{X_{\sfK} W|S}(p_S,L,D_1) \right\} \]
and slice it into the following disjoint collection of sets:
\begin{equation}
   \forall z \in \calZ ~:\quad \scrP_{X_{\sfK} WS}(p_S,L,D_1,z)
	\triangleq \left\{ p_{X_{\sfK} WS}\in \scrP_{X_{\sfK} WS}(p_S,L,D_1)
		~:\; p_{X_\sfK} = z \right\} .
\label{eq:Pxkws-z}
\end{equation}

The error probability of the decoder is not increased if a helper reveals
the joint type $Z$. The entropy of $Z$ is at most
$\log|\calZ| \le |\calX|^K \log(N+1)$. Fano's inequality (\ref{eq:fano-all})
applied to $\sfA = \sfK$ becomes
\begin{eqnarray}
  KNR & = & H(M_{\sfK}|\bS,V) \nonumber \\
	& \le & H(M_{\sfK},Z|\bS,V) \nonumber \\
	& = & |\calX|^K \log(N+1) + H(M_{\sfK}|\bS,V,Z) \nonumber \\
	& \le & |\calX|^K \log(N+1) + 1
		+ P_e^{\all}(p_{Y|X_\sfK Z}) \cdot KNR + I(M_{\sfK};\bY|\bS,V,Z) .
\label{eq:fano-lambda}
\end{eqnarray}
Analogously to (\ref{eq:I-sum}), the following single-letter expression
holds for every $z\in\calZ$ and $p_{Y|X_{\sfK}} \in \scrW_K(z)$:
\begin{eqnarray}
  I(M_\sfK;\bY|\bS,V,Z=z)
	& = & I(\bX_\sfK;\bY|\bS,V,Z=z) \nonumber \\
	& = & H(\bY|\bS,V,Z=z) - H(\bY|\bX_\sfK,\bS,V,Z=z) \nonumber \\
	& \stackrel{(a)}{=} & H(\bY|\bS,V,Z=z) - H(\bY|\bX_\sfK,Z=z) \nonumber \\
	& \stackrel{(b)}{=} & \sum_{t=1}^N H(Y_t|Y^{t-1},\bS,V,Z=z) 
								- \sum_{t=1}^N H(Y_t|X_{\sfK,t},Z=z) \nonumber \\
	& \le & \sum_{t=1}^N H(Y_t|\bS,V,Z=z) - \sum_{t=1}^N H(Y_t|X_{\sfK,t},Z=z) \nonumber \\
	& \stackrel{(c)}{=} & \sum_{t=1}^N H(Y_t|S_t,Q_t,Z=z) - \sum_{t=1}^N H(Y_t|X_{\sfK,t},S_t,Q_t,Z=z) \nonumber \\
	& = & \sum_{t=1}^N I(X_{\sfK,t};Y_t | S_t,Q_t,Z=z) \nonumber \\
	& = & N \,I(X_{\sfK};Y|S,W,Z=z) \label{eq:I-sum-Z} \\
	& = & N I_{p_{X_\sfK WS|Z=z} \,p_{Y|X_\sfK}}(X_{\sfK};Y|S,W) \nonumber
\end{eqnarray}
where (a) holds because $(\bS,V) \to (\bX_\sfK, Z) \to \bY$ forms a Markov chain,
(b) because the collusion channel remains memoryless even when conditioned on $Z$, and (c)
because $(\bS,V) = (S_t,Q_t) \to (X_\sfK, Z) \to Y_t$ forms a Markov chain for each $1 \le t \le N$.

For the error probability $P_e^{\all}(p_{Y|X_\sfK Z})$
to vanish for each $p_{Y|X_\sfK Z} \in \widetilde{\scrW}_K^{\fair}$, we need
\begin{eqnarray}
   R & \stackrel{(a)}{\le} & \liminf_{N \to \infty} \min_{p_{Y|X_\sfK Z} \in \widetilde{\scrW}_K^{\fair}}
	\frac{1}{NK} I(M_{\sfK};\bY|\bS,V,Z) \nonumber \\
		& \stackrel{(b)}{\le} & \liminf_{N \to \infty} \min_{p_{Y|X_\sfK Z} \in \widetilde{\scrW}_K^{\fair}}
	\frac{1}{K} I_{p_{SWX_\sfK Z} \,p_{Y|X_\sfK Z}}(X_{\sfK};Y|S,W,Z) \nonumber \\
		& \le & \lim_{N \to \infty} \;\max_{p_Z \in \scrP_Z}
			\;\max_{\{p_{X_{\sfK} WS|Z=z} \in \scrP_{X_{\sfK} WS}(p_S,L(N),D_1,z)\}_{z\in\calZ}}
			\;\min_{\{p_{Y|X_\sfK} \in \scrW_K^{\fair}(z)\}_{z\in\calZ}} \nonumber \\
		& & \hspace*{0.5in} \frac{1}{K}
			\sum_{z\in\calZ} p_Z(z) \,I_{p_{X_\sfK WS|Z=z} \,p_{Y|X_\sfK}}(X_{\sfK};Y|S,W) \nonumber \\
		& \stackrel{(c)}{=} & \lim_{N \to \infty} \max_{z\in\calZ}
			\;\max_{p_{X_{\sfK} WS|Z=z} \in \scrP_{X_{\sfK} WS}(p_S,L(N),D_1,z)}
			\;\min_{p_{Y|X_\sfK} \in \scrW_K^{\fair}(z)} \,\frac{1}{K}
			I_{p_{X_\sfK WS|Z=z} \,p_{Y|X_\sfK}}(X_{\sfK};Y|S,W) \nonumber \\
		& = & \lim_{N \to \infty}
			\;\max_{p_{X_{\sfK} WS} \in \scrP_{X_{\sfK} WS}(p_S,L(N),D_1)}
			\;\min_{p_{Y|X_\sfK} \in \scrW_K^{\fair}(p_{X_\sfK})} 
			\,\frac{1}{K} I_{p_{X_\sfK WS} \,p_{Y|X_\sfK}}(X_{\sfK};Y|S,W) \nonumber \\
		& \le & \lim_{L \to \infty} \;\max_{p_{X_{\sfK} WS}
			\in \scrP_{X_{\sfK} WS}(p_S,L,D_1)} \min_{p_{Y|X_\sfK} \in \scrW_K^{\fair}(p_{X_\sfK})}
			\frac{1}{K} I(X_{\sfK};Y|S,W) \nonumber \\
		& =& \lim_{L \to \infty} \;\max_{p_{X_{\sfK} W|S}
			\in \scrP_{X_{\sfK} W|S}(p_S,L,D_1)} \min_{p_{Y|X_\sfK} \in \scrW_K^{\fair}(p_{X_\sfK})}
			\frac{1}{K} I(X_{\sfK};Y|S,W) \nonumber \\
		& = & \widetilde{C}^{\all}(D_1,\scrW_K^{\fair}) 
\label{eq:KRate-lambda}
\end{eqnarray}
where (a) follows from (\ref{eq:fano-lambda}),
(b) from (\ref{eq:I-sum-Z}), and (c) from the fact that in a game in which $Z$
is a variable chosen by the first player (here the embedder) but
known to all players (embedder, colluders, receiver), there can be
no advantage in randomizing $Z$, i.e., a deterministic choice of $Z$
suffices to achieve the value of the maxmin game. More formally, equality (c)
is a direct consequence of the following simple lemma, using the fingerprint
distributor's feasible set $\scrP_{X_{\sfK} WS}(p_S,L(N),D_1,z)$ in place of $\calF(z)$,
the colluders' feasible set $\scrW_K^{\fair}(z)$ in place of $\calG(z)$, and the conditional
mutual information $I(X_{\sfK};Y|S,W)$ as the payoff function $\phi$. 
This concludes the proof.
\hfill $\Box$
			
\begin{lemma}
Consider a discrete set $\calZ$ and two families of sets $\calF(z), \,z\in\calZ$
and $\calG(z), \,z\in\calZ$ indexed by the elements of $\calZ$.
Then the following game with payoff function $\phi$:
\begin{equation}
   V = \max_{p \in \scrP_Z} \max_{\{f_z \in \calF(z)\}_{z\in\calZ}} 
	\min_{\{g_z \in \calG(z)\}_{z\in\calZ}} \sum_{z\in\calZ} p(z) \phi(f_z,g_z)
\label{eq:V}
\end{equation}
admits a pure-strategy solution, i.e., the maximum over the p.m.f. $p \in \scrP_Z$
is achieved by deterministic $p$.
\label{lem:game}
\end{lemma}

{\em Proof}:
Write $f = \{f_z\}_{z\in\calZ}$ and $g = \{g_z\}_{z\in\calZ}$
where each $f_z \in \calF(z)$ and $g_z \in \calG(z)$.
For each $(p,f)$, let $g^*(p,f)$ achieve the minimum over $g$
of the function $\sum_{z\in\calZ} p(z) \phi(f_z,g_z)$. For each $p$, let $f^*(p)$
achieve the maximum over $f$ of the function $\sum_{z\in\calZ} p(z) \phi(f_z,g_z^*(p,f))$.
By inspection of (\ref{eq:V}), the following elementary properties hold
for each $p \in \scrP_Z$ and $z \in \calZ$:
\begin{itemize}
\item The minimizing $g_z^*$ depends on $(p,f)$ via $f_z$ only, and we denote this limited
	dependency explicitly by $g_z^*(f_z)$. The minimizer satisfies
	\begin{equation}
	\phi(f_z,g_z^*(f_z)) = \min_{g_z \in \calG(z)} \phi(f_z,g_z) .
	\label{eq:gz*}
	\end{equation}
\item The maximizing $f_z^*$ does not depend on $p$ and satisfies
	\begin{equation}
	\phi(f_z^*,g_z^*(f_z^*)) = \max_{f_z \in \calF(z)} \min_{g_z \in \calG(z)} \phi(f_z,g_z) .
	\label{eq:fz*}
	\end{equation}
\end{itemize}
Substituting (\ref{eq:fz*}) into (\ref{eq:V}), we obtain
\begin{eqnarray*}
   V & = & \max_{p \in \scrP_Z} \sum_{z\in\calZ} p(z) \phi(f_z^*,g_z^*(f_z^*)) \\
	& = & \max_{z\in\calZ} \phi(f_z^*,g_z^*(f_z^*)) \\
	& = & \max_{z\in\calZ} \max_{f_z \in \calF(z)} \min_{g_z \in \calG(z)} \phi(f_z,g_z)
\end{eqnarray*}
which proves the claim.
\hfill $\Box$


\section{Proof of Theorem~\ref{thm:C-one}: Converse Under Detect-One Criterion}
\label{Sec:Converse-one}
\setcounter{equation}{0}

By Lemma~\ref{lem:memoryless}, it suffices to prove the claim for the compound
class of memoryless channels $\scrW_K$.
Let $\calM_N = \{1,2,\cdots,2^{NR}\}$. 
For notational simplicity, assume two colluders ($K=2$). The proof extends
straightforwardly to larger coalitions. For the detect-one criterion,
it is sufficient to consider decoding rules that return {\em exactly} one user index,
i.e., the decoding rule is a mapping
\begin{equation}
  g_N ~:~ \calY^N \times \calS^N \times \calV_N \to \calM_N .
\label{eq:return-one}
\end{equation}
Indeed, consider momentarily a more general decoder that returns a list of accused users.
By definition of the detect-one and false-positive error criteria, correct decoding occurs
if and only if the list size $L \ge 1$ and {\bf all} users on the output list are guilty.
One can then construct a new decoder of the form (\ref{eq:return-one}) that returns an arbitrary
user if $L=0$ and an arbitrary element of the original size-$L$ list if $L \ge 1$.
The correct-decoding event for the original decoder is also a correct-decoding
event for the new decoder, and so the new decoder has {\em at least} the same probability
of correct decoding as the original decoder. \footnote{
	The new decoder performs better than the original one in the event that the list
	has size $L \ge 2$ and consists of a mix
	of guilty and innocent users (an error is then declared for the original decoder),
	and the list member selected by new decoder is guilty (a correct decision is made).}
In the following, we only consider decoding rules of the form (\ref{eq:return-one}).
	
Denote by $\calD_i(\bs,v)$ the decoding region for user $i$, i.e.,
\[ \by \in \calD_i(\bs,v) \quad \Leftrightarrow \quad g_N(\by,\bs,v) = i ,
	\quad \forall i \in \calM_N . \]
The decoding regions form a partition of $\calY^N$.
The average probability of correct decoding is given by
\begin{eqnarray}
  \lefteqn{P_c(f_N,g_N,p_{Y|X_1 X_2})} \nonumber \\
	&& \quad = Pr[g_N(\bY,\bS,V) \in \calK ] \nonumber \\
	&& \quad = \frac{1}{2^{2NR}} \sum_{i,j\in\calM_N}
			\,\sum_{\bs \in \calS^N} p_S^N(\bs) \sum_{v\in\calV_N} p_V(v)
			\,\sum_{\by \in \calD_i(\bs,v) \cup \calD_j(\bs,v)}
			p_{Y|X_1 X_2}^N(\by|\bx_i(\bs,v),\bx_j(\bs,v)) .
\label{eq:Pc}
\end{eqnarray}
Without loss of optimality we assume that
randomly modulated codes (Def.~\ref{def:code-RM}, Prop.~\ref{prop:exch}) are used.

The proof is organized along thirteen steps.
An arbitrarily small parameter $\delta > 0$ is chosen.
Step~1 defines for each $(\bs, v)$ a set of bad codewords that have exponentially
many neighbors within Hamming balls of radius $N\delta$ centered at these codewords.
The remaining codewords constitute the so-called good set.
Step~2 introduces a dense, nested family $\scrW_{K,\delta}^{\fair}$ of subsets
of $\scrW_K^{\fair}$ indexed by $\delta$ and consisting of ``nice channels''.
An equivalence is given between Hamming distance of two codewords and statistical
distinguishability of the output of any $p_{Y|X_1 X_2} \in \scrW_{K,\delta}^{\fair}$.
For clarity of the exposition we initially derive error probabilities assuming that
the good set is large and that both colluders are assigned codewords in the good set;
these assumptions are subsequently relaxed in Steps 10 and 11.
All the error probabilities up to that point are conditioned on $\bS,V$.
Step~3 introduces the basic random variables used in the proof. Step~4 does
(a) define a reference product conditional p.m.f for $\bY$ given $\bS, V$;
(b) associate a conditional self-information to each pair of codewords; and
(c) define a large set of codeword pairs whose conditional self-information
is within $\delta^2$ of their average value.
Step~5 defines a typical set for $\bY$ given $\bS,V$, and $\calK$.
Step~6 shows that typical sets for good codeword pairs have weak overlap.
Step~7 defines a collection of refined typical sets for $\bY$ with bounded overlap.
Step~8 defines a typical set for the host sequence $\bS$.
Step~9 upper bounds the conditional probability of correct decoding
in terms of a mutual information.
Step~10 derives an analogous result conditioned on the event that both colluders are
assigned codewords from the bad set.
Step~11 combines the bounds for good and bad codewords into a single bound.
Step~12 removes the conditioning on $\bS,V$ and upper bounds the unconditional probability
of correct decoding (\ref{eq:Pc}) in terms of a mutual information.
Step~13 derives an upper bound on that mutual information and shows that
any achievable rate $R$ must be less than half of the upper bound.
The proof is completed by letting $\delta \downarrow 0$.

\underline{Step~1}.
Denote by
\[ d_H(\bx,\bx') = \sum_{t=1}^N \mathds1\{\bx_t \ne \bx'_t \} \]
the Hamming distance between two sequences $\bx$ and $\bx'$ in $\calX^N$,
and by
\begin{eqnarray}
  \calM_j(\bs,v,\delta) & = & \{ k~\in\calM_N ~:~
	d_H(\bx_j(\bs,v), \bx_k(\bs,v)) \le N\delta \} , \nonumber \\
	& & \hspace*{1in} j \in \calM_N, \,\bs \in \calS^N, \,v \in \calV_N, \,0 \le \delta \le 1
\label{eq:Mj}
\end{eqnarray}
the set of indices $k$ for the codewords $\bx_k(\bs,v)$
that are within Hamming distance $N\delta$ of codeword $\bx_j(\bs,v)$, and by
$M_j(\bs,v,\delta) = |\calM_j(\bs,v,\delta)|$ the cardinality of this set.
The function $M_j(\bs,v,\cdot)-1$ is akin to a cumulative distance distribution.
It is nondecreasing, with $M(\bs,v,0) \ge 1$ and $M(\bs,v,1) = 2^{NR}$.
Note that for $S=\emptyset$ and random codes over $\calX=\{0,1\}$,
$M_j(V,\delta)-1$ is a random variable whose expectation vanishes as $N \to \infty$
for $\delta < \delta_{GV}(R)$, the Gilbert-Varshamov distance at rate $R$ \cite{Barg02}.

Denote by
\begin{equation}
   \calM_N^{\good}(\bs,v,\delta) = \{ j \in \calM_N
	~:~ |\calM_j(\bs,v,\delta)| \le 2^{N 3\sqrt{\delta} } \} ,
	\quad v \in \calV_N, \,0 \le \delta \le 1
\label{eq:M-good}
\end{equation}
a set of ``good'' indices $j$ (there are at most $2^{N 3\sqrt{\delta}}$
codewords within Hamming distance $N\delta$ of codeword $\bx_j(\bs,v)$), and by
\begin{eqnarray}
   \calM_N^{\bad}(\bs,v,\delta) 
	& = & \calM_N \setminus \calM_N^{\good}(\bs,v,\delta) \nonumber \\
	& = & \{ j \in \calM_N ~:~ |\calM_j(\bs,v,\delta)| > 2^{N 3\sqrt{\delta}} \}
\label{eq:M-bad}
\end{eqnarray}
the complementary set of ``bad'' indices. 

Note that any code with normalized minimum distance $\delta_{\min} > 0$
satifies $M_j(\bs,v,\delta) \equiv 1$ and thus
$\calM_N^{\good}(\bs,v,\delta) \equiv \calM_N$ for all $0 < \delta < \delta_{\min}$.
However the derivations in Steps 2---8 of the proof make no assumption on the size of the sets
$\calM_N^{\good}(\bs,v,\delta)$. 
Finally, for the RM codes considered here, the sets (\ref{eq:Mj}), (\ref{eq:M-good}), and 
(\ref{eq:M-bad}) depend on the host sequence $\bs$ only via its type $p_{\bs}$.

\underline{Step~2}.
Channels $p_{Y|X_1 X_2}$ that satisfy $p_{Y|X_1 X_2}(y|x_1,x_2)=0$ for some $y,x_1,x_2$ or
$p_{Y|X_1 X_2}(\cdot|x_1,x_2) \equiv p_{Y|X_1 X_2}(\cdot|x_1',x_2')$ for some $(x_1,x_2) \ne (x_1',x_2')$
require special handling. To this end, we define the following nested family of subsets
of $\scrW_K^{\fair}$, indexed by $0 < \delta \le 1/|\calY|$:
\begin{eqnarray}
   \lefteqn{\scrW_{K,\delta}^{\fair} = \left\{ p_{Y|X_1 X_2} \in \scrW_K^{\fair} ~:~
   p_{Y|X_1 X_2}(y|x_1,x_2) \ge \delta , \quad \forall y,x_1,x_2 , \right.} ,  \nonumber \\
    & & \hspace*{0.5in} \left. \delta \le D(p_{Y|X_1=x_1, X_2=x_2} \| p_{Y|X_1=x_1', X_2=x_2'}) 
	\le \log \delta^{-1} , \quad \forall (x_1,x_2) \ne (x_1',x_2') \right\} 
\label{eq:WK-delta}
\end{eqnarray}
where the upper bound on divergence is implied by the lower bound on $p_{Y|X_1 X_2}$.
By continuity of the correct-decoding probability functional (\ref{eq:Pc})
and by the definition (\ref{eq:WK-delta}), we have
\[ \widetilde{C}^{\one}(D_1, \scrW_{K,\delta}^{\fair}) \downarrow 
	\widetilde{C}^{\one}(D_1, \scrW_K^{\fair})  \quad \mathrm{as~} \delta \downarrow 0  . \]


Denote by
\begin{equation}
   D_{ijk} \triangleq \frac{1}{N} \sum_{t=1}^N D(p_{Y|X_1=x_{it}(\bs,v), X_2=x_{jt}(\bs,v)}
	\| p_{Y|X_1=x_{it}(\bs,v), X_2=x_{kt}(\bs,v)})
\label{eq:D-ijk}
\end{equation}
the normalized conditional Kullback-Leibler divergence (given $\bs,v$) between
the distributions on $\bY$ induced by codeword pairs $(i,j)$ and $(i,k)$, respectively.  It follows from 
(\ref{eq:WK-delta}) that for each $p_{Y|X_1 X_2} \in \scrW_{K,\delta}^{\fair} $,
\begin{equation}
   d_H(\bx_j(\bs,v),\bx_k(\bs,v)) \le N\delta \quad \Rightarrow \quad
		 D_{ijk} \le \delta \log \delta^{-1} .
\label{eq:dH-ineq1}
\end{equation}
and
\[ d_H(\bx_j(\bs,v),\bx_k(\bs,v)) > N\delta  \quad \Rightarrow \quad D_{ijk} > \delta^2 \]
Conversely,
\begin{equation}
   D_{ijk} \le \delta^2
		\quad \Rightarrow \quad d_H(\bx_j(\bs,v),\bx_k(\bs,v)) \le N\delta .
\label{eq:dH-ineq2}
\end{equation}
When $D_{ijk}$ is small, we say that
the codewords $\bx_j(\bs,v)$ and $\bx_k(\bs,v)$ are {\em nearly indistinguishable}
at the channel output. For any $(\bs,v,i,j,k)$ and
$p_{Y|X_1 X_2} \in \scrW_{K,\delta}^{\fair}$,
(\ref{eq:dH-ineq1}) and (\ref{eq:dH-ineq2}) describe an equivalence
between statistical distinguishability of two codewords and Hamming distance.

\underline{Step~3}.
To analyze the probability of correct decoding conditioned on the event
$\calK \in (\calM_N^{\good}(\bS,V,\delta))^2$ that both colluders are assigned good codewords,
we define the following random variables.
Define $Q_t = \{V, S_j, j \ne t\}$ over the alphabet
$\calQ_N \triangleq \calV_N \times \calS^{N-1}$.
We have $(S_t,Q_t) = (\bS,V)$ for each $1 \le t \le N$.
Since the host sequence type $p_{\bs}$ together with any $q_t, 1 \le t \le N$,
uniquely determines $s_t$ and thus the pair $(\bs,v)$ (and vice-versa),
we may also use $(p_{\bs},q)$ as an equivalent representation of the pair $(\bs,v)$. 
Define a time-sharing random variable $T$ uniformly distributed over
$\{1,2,\cdots,N\}$ and independent of the other random variables. Let
\[ S=S_T, \;Q=Q_T, \;Y=Y_T, \;\mathrm{and}\;
    X_i = x_{i,T}(\bS,V) ,\;\forall i \in \calM_N .
\]
Define the random variable $X$ drawn uniformly from
$\{X_i, \,i\in\calM_N^{\good}(\bS,V,\delta)\}$.
The conditional p.m.f of $X$ given $\bS,V,T$ is given by
\begin{equation}
   p_{X|\bS VT}(x|\bs,v,t) = \frac{1}{|\calM_N^{\good}(\bs,v,\delta)|} 
	\sum_{i\in\calM_N^{\good}(\bs,v,\delta)} \mathds1\{x_{it}(\bs,v)=x\} ,
	\quad \forall x, \bs, v, t .
\label{eq:pxsvt}
\end{equation}

Given $\bs,v,t$, the conditional distribution of $(X_i,X_j,Y)$ is
$p_{X_i X_j|\bS=\bs, V=v, T=t} \,p_{Y|X_1 X_2}$ where 
\begin{eqnarray}
   p_{X_i X_j|\bS VT}(x_1,x_2|\bs,v,t)
	& = & \mathds1 \{ x_{it}(\bs,v)=x_1, \,x_{jt}(\bs,v)=x_2\} , \nonumber \\
	& & \hspace*{0.5in} x_1,x_2\in\calX, \,i,j \in \calM_N^{\good}(\bs,v,\delta) , \, 1 \le t \le N .
\label{eq:pxixjsvt}
\end{eqnarray}
By (\ref{eq:pxsvt}), the average of (\ref{eq:pxixjsvt}) over $i,j \in \calM_N^{\good}(\bs,v,\delta)$
is the product conditional p.m.f
\begin{eqnarray}
   \lefteqn{\frac{1}{|\calM_N^{\good}(\bs,v,\delta)|^2} \sum_{i,j\in\calM_N^{\good}(\bs,v,\delta)} 
		p_{X_i X_j|\bS VT}(x_1,x_2|\bs,v,t) } \nonumber \\
	& = & p_{X|\bS VT}(x_1|\bs,v,t) \,p_{X|\bS VT}(x_2|\bs,v,t) .
\label{eq:p2XSQT}
\end{eqnarray}

\underline{Step~4}.
The conditional distribution of each $Y_t, \,1 \le t \le N$, given $(\bS,V)$
and $\calK \in (\calM_N^{\good}(\bS,V,\delta))^2$, is given by
\begin{eqnarray}
   p_{Y_t|\bS V}(y|\bs,v)
     & = & p_{Y|\bS VT}(y|\bs,v,t) \nonumber \\
    & = & \frac{1}{|\calM_N^{\good}(\bs,v,\delta)|^2} \sum_{i,j\in\calM_N^{\good}(\bs,v,\delta)} 
			p_{Y|X_1 X_2}(y|x_{it}(\bs,v),x_{jt}(\bs,v)) \label{eq:pYV-0} \\
    & = & \frac{1}{|\calM_N^{\good}(\bs,v,\delta)|^2} \sum_{i,j\in\calM_N^{\good}(\bs,v,\delta)} 
			\sum_{x_1,x_2 \in \calX} \,p_{X_i X_j|\bS VT}(x_1,x_2|\bs,v,t) \,p_{Y|X_1 X_2}(y|x_1,x_2) \nonumber \\
	& = & \sum_{x_1,x_2 \in \calX} p_{X|\bS VT}(x_1|\bs,v,t)
		\,p_{X|\bS VT}(x_2|\bs,v,t) \,p_{Y|X_1 X_2}(y|x_1,x_2) .
\label{eq:pYV}
\end{eqnarray}
For any permutation $\pi$ of $\{1,2,\cdots,N\}$ we have 
$p_{Y_{\pi(t)}|\bS V}(y|\pi(\bs),v) = p_{Y_t|\bS V}(y|\bs,v)$
for all  RM codes.
The product conditional distribution
\begin{equation}
   r(\by|\bs,v) \triangleq \prod_{t=1}^N p_{Y_t|\bS V}(y_t|\bs,v)
\label{eq:r}
\end{equation}
is strongly exchangeable for each $v\in\calV_N$ and
will be used as a {\em reference conditional p.m.f} for $\bY$ given $\bS, V$
in the sequel.
We also define the following conditional self-informations (i.e., mutual information
for coalition $(i,j)$ averaged over $Y_t$ (resp. $\bY$) and conditioned on $\bS,V$):
\begin{eqnarray}
   \theta_{ij,t}(\bs,v) & \triangleq & \sum_{y_t \in \calY}
		p_{Y|X_1 X_2}(y_t|x_{it}(\bs,v),x_{jt}(\bs,v))
		\log \frac{p_{Y|X_1 X_2}(y_t|x_{it}(\bs,v),x_{jt}(\bs,v))}
			{p_{Y_t|\bS V}(y_t|\bs,v)} ,							\nonumber \\
		& = & D(p_{Y|X_1=x_{it}(\bs,v), X_2=x_{jt}(\bs,v)} \| p_{Y_t|\bS=\bs,V=v})  \label{eq:Iijt} \\
	\theta_{ij}(\bs,v) & \triangleq & \frac{1}{N} \sum_{t=1}^N \theta_{ij,t}(\bs,v)
																	\nonumber \\
		& = & \frac{1}{N} \sum_{t=1}^N \sum_{y_t \in \calY}
			p_{Y|X_1 X_2}(y_t|x_{it}(\bs,v),x_{jt}(\bs,v))
			\log \frac{p_{Y|X_1 X_2}(y_t|x_{it}(\bs,v),x_{jt}(\bs,v))}
			{p_{Y_t|\bS V}(y_t|\bs,v)} 								\nonumber \\
		& = & \frac{1}{N} \sum_{t=1}^N \sum_{x_1,x_2,y}
			\,\mathds1 \{ x_{it}(\bs,v)=x_1, \,x_{jt}(\bs,v)=x_2\}
			\,p_{Y|X_1 X_2}(y|x_1,x_2) \log \frac{p_{Y|X_1 X_2}(y|x_1,x_2)}
			{p_{Y_t|\bS V}(y|\bs,v)} \nonumber \\
		& = & \sum_{t,x_1,x_2,y} p_T(t) \,p_{X_i X_j|\bS VT}(x_1,x_2|\bs,v,t)
			\,p_{Y|X_1 X_2}(y|x_1,x_2) \log \frac{p_{Y|X_1 X_2}(y|x_1,x_2)}
			{p_{Y|\bS VT}(y|\bs,v,t)}	.					\label{eq:Iij-sv}
\end{eqnarray}
Since $p_{Y|X_1 X_2}$ is symmetric, the expressions
(\ref{eq:Iijt}) and (\ref{eq:Iij-sv}) are symmetric in $i$ and $j$.
The average of $\theta_{ij}(\bs,v)$ over all
$(i,j) \in (\calM_N^{\good}(\bs,v,\delta))^2$ is the conditional mutual information
\begin{eqnarray}
   I(\bs,v) & \triangleq & \frac{1}{|\calM_N^{\good}(\bs,v,\delta)|^2}
			\sum_{i,j\in\calM_N^{\good}(\bs,v,\delta)} \theta_{ij}(\bs,v)	\label{eq:I-sv-def} \\
   & = & \sum_{t,x_1,x_2,y} p_T(t) \,p_{X|\bS VT}(x_1|\bs,v,t) \,p_{X|\bS VT}(x_2|\bs,v,t)
				\,p_{Y|X_1 X_2}(y|x_1,x_2) \log \frac{p_{Y|X_1 X_2}(y|x_1,x_2)}
				{p_{Y|\bS VT}(y|\bs,v,t)} \nonumber \\
	& = & I_{p_T p_{X|\bS VT}^2 p_{Y|X_1 X_2}}(X_1 X_2;Y|\bS=\bs,V=v,T) .
\label{eq:I-sv}
\end{eqnarray}
For RM codes, both $\theta_{ij}(\bs,v)$ and $I(\bs,v)$ depend on $\bs$ only via its type $p_{\bs}$.

Since the average value of $\theta_{ij}(\bs,v)$ is $I(\bs,v)$, there may not be
too many pairs $(i,j)$ for which $\theta_{ij}(\bs,v)$ is well above the mean.
More precisely, there exists a symmetric subset
$\tilde{\calA}(\bs,v,\delta) \subseteq (\calM_N^{\good}(\bs,v,\delta))^2$ of size
\[ |\tilde{\calA}(\bs,v,\delta)| \ge \frac{\delta^2}{\delta^2 + I(\bs,v)} \,|\calM_N^{\good}(\bs,v,\delta)|^2 
	\ge \frac{\delta^2}{\delta^2 + \log |\calY|} \,|\calM_N^{\good}(\bs,v,\delta)|^2 \]
such that $\tilde{\calA}(\bs,v,\delta)$ depends on $\bs$ only via $p_{\bs}$ and
\begin{equation}
  (i,j) \in \tilde{\calA}(\bs,v,\delta) \quad \Rightarrow \quad
	\theta_{ij}(\bs,v) \le I(\bs,v) + \delta^2 .
\label{eq:theta-max0}
\end{equation}
This claim is seen to hold by contrapositive.
If there existed a subset $\tilde{\calA}^c(\bs,v,\delta)$ of size
$\frac{I(\bs,v)}{\delta^2 + I(\bs,v)} \,|\calM_N^{\good}(\bs,v,\delta)|^2$ or larger such that
\[ \forall (i,j) \in \tilde{\calA}^c(\bs,v,\delta) :
	\quad \theta_{ij}(\bs,v) > I(\bs,v) + \delta^2 \]
we would have
\[ \sum_{i,j \in \calM_N} \theta_{ij}(\bs,v)
	> (I(\bs,v) + \delta^2) \,|\tilde{\calA}^c(\bs,v,\delta)| 
	\ge |\calM_N^{\good}(\bs,v,\delta)|^2 I(\bs,v) 
\]
which would contradict (\ref{eq:I-sv-def}). \footnote{
	As mentioned by a reviewer, the claim could alternatively be proven 
	by application of Markov's inequality.}

Moreover, the interval $[0, \log|\calY|]$ is covered by the finite
collection of intervals
\[ \Theta_l \triangleq \left[ l \,\frac{\delta^2}{2}, \,(l+1) \,\frac{\delta^2}{2} \right) ,
	\quad l=0,1, \cdots, \left\lfloor \frac{2\log|\calY|}{\delta^2} \right\rfloor
	\triangleq l_{\max} \]
of width $\delta^2/2$, and at least one of these intervals must contain
many $\theta_{ij}(\bs,v)$. Specifically, for some integer $0 \le l < l_{\max}$
there must exist a subset 
$\calA(\bs,v,\delta) \subseteq \tilde{\calA}(\bs,v,\delta)$ 
with the following properties:
\begin{eqnarray}
  (i,j) \in \calA(\bs,v,\delta)
	& & \Rightarrow \quad \theta_{ij}(\bs,v) \in \Theta_l \nonumber \\
	& & \Rightarrow \quad
	|\theta_{ij}(\bs,v) - \underline{I}(\bs,v)| \le \frac{\delta^2}{4} ,
\label{eq:theta-max}
\end{eqnarray}
\begin{equation}
   \underline{I}(\bs,v) \triangleq \left(l+\frac{1}{2}\right)
	\frac{\delta^2}{2} \le I(\bs,v) \le \log|\calY| ,
\label{eq:I_}
\end{equation}
$\calA(\bs,v,\delta)$ is symmetric with size at least equal to
\begin{eqnarray}
  |\calA(\bs,v,\delta)|
	\ge \frac{\delta^2}{2\log |\calY|} |\tilde{\calA}(\bs,v,\delta)| 
	& \ge & \frac{\delta^4}{2\log|\calY| \,(\delta^2 + \log |\calY|)} \,|\calM_N^{\good}(\bs,v,\delta)|^2 \nonumber \\
	& \ge & \frac{\delta^4}{4\log^2|\calY|} \,|\calM_N^{\good}(\bs,v,\delta)|^2 ,
\label{eq:setA-size}
\end{eqnarray}
and $\calA(\bs,v,\delta)$ depends on $\bs$ only via $p_{\bs}$.

To summarize, the subset $\calA(\bs,v,\delta) \subseteq (\calM_N^{\good}(\bs,v,\delta))^2$ has size
nearly equal to $|\calM_N^{\good}(\bs,v,\delta)|^2$ and consists of the indices of the codeword pairs
whose conditional self-information $\theta_{ij}(\bs,v)$ is close to some
$\underline{I}(\bs,v) \le I(\bs,v)$.

Recalling (\ref{eq:I-sv}) and the equivalence of the representations $(\bS,V)$
and $(S,Q)$, we define
\begin{equation}
   I(p_S',q) \triangleq I_{p_T p_S' p_{X|SQT}^2 p_{Y|X_1 X_2}}(X_1,X_2;Y|S,Q=q,T) 
	\quad \forall p_S' \in \scrP_S, \,q \in \calQ_N
\label{eq:Ipsw}
\end{equation}
which is a linear functional of $p_S'$ and coincides with $I(\bs,v)$ in (\ref{eq:I-sv})
when $p_S' = p_{\bs}$.

\underline{Step~5}.
Define the following subset of $\calY^N$:
\begin{eqnarray}
   \!\!\!\!\!\!\widetilde{\calT}_\delta(\bs,v,i,j) & \triangleq & \left\{ \by \in \calY^N
		~:~ \left| \frac{1}{N} \sum_{t=1}^N \log
			\frac{p_{Y|X_1 X_2}(y_t|x_{it}(\bs,v),x_{jt}(\bs,v))}
			{p_{Y_t|\bS V}(y_t|\bs,v)} - \theta_{ij}(\bs,v) \right|
			\le \frac{\delta^2}{8} \right\} 
\label{eq:typ-Y}
\end{eqnarray}
which satisfies the symmetry property $\widetilde{\calT}_\delta(\bs,v,i,j) = \widetilde{\calT}_\delta(\bs,v,j,i)$
and the letter permutation-invariance property (for RM codes)
\[ \by \in \widetilde{\calT}_\delta(\bs,v,i,j) \quad \Rightarrow \quad
	\pi(\by) \in \widetilde{\calT}_\delta(\pi(\bs),v,i,j) \]
for any permutation $\pi$ of $\{1,2,\cdots,N\}$.

We show that $\widetilde{\calT}_\delta(\bs,v,i,j)$ is a typical set for $\bY$ conditioned on
$\bS=\bs$, $\bV=v$, and $\calK=\{i,j\}$, in the following sense:
\begin{equation}
   Pr[\bY \notin \widetilde{\calT}_\delta(\bs,v,i,j)|\bS=\bs,V=v,\calK=\{i,j\}]
		\le \frac{64\log^2 \delta}{N\delta^4} , \quad \forall \bs, v,i,j
\label{eq:Pr-Aeps}
\end{equation}
vanishes as $N \to \infty$. Indeed we may rewrite (\ref{eq:Pr-Aeps}) as
\begin{eqnarray}
   \lefteqn{Pr[\bY \notin \widetilde{\calT}_\delta(\bs,v,i,j)|\bS=\bs,V=v,\calK=\{i,j\}]} \nonumber \\
	& = & Pr \left[ \left. |\hat{\theta}_{ij}(\bs,v) - \theta_{ij}(\bs,v)|
		\ge \frac{\delta^2}{8} ~\right| \bS=\bs, V=v, \calK = \{i,j\} \right]
\label{eq:Pr-Aeps-2}
\end{eqnarray}
where
\begin{equation}
   \hat{\theta}_{ij}(\bs,v) \triangleq \frac{1}{N} \sum_{t=1}^N \log
		\frac{p_{Y|X_1 X_2}(Y_t|x_{it}(\bs,v),x_{jt}(\bs,v))}{p_{Y_t|\bS V}(Y_t|\bs,v)} .
\label{eq:Ihat}
\end{equation}
Since $Y_t, 1 \le t \le N$, are conditionally independent given
$\bS,V,\calK$, $\hat{\theta}_{ij}(\bs,v)$ is the average of $N$ random variables
that are conditionally independent given $\bS=\bs, V=v, \calK=\{i,j\}$. 
Recalling (\ref{eq:Iijt}), the conditional expectation of these random variables is given by
\begin{equation}
  \eE_{Y_t|\bS V \calK} \left[ \log \frac{p_{Y|X_1 X_2}(Y_t|x_{it}(\bs,v),x_{jt}(\bs,v)}
			{p_{Y_t|\bS V}(Y_t|\bs, v)} \right] 
	= \theta_{ij,t}(\bs,v) , \quad 1 \le t \le N ,
\label{eq:E-thetahat}
\end{equation}
and averaging (\ref{eq:E-thetahat}) over $t$ yields
$\eE_{\bY|\bS V \calK} (\hat{\theta}_{ij}(\bs,v)) = \theta_{ij}(\bs,v)$.
The conditional variances of these random variables are
\begin{equation}
   \zeta_t(\bs,v,i,j) \triangleq \mathrm{var}_{Y_t|\bS V \calK} \left[
	\log \frac{p_{Y|X_1 X_2}(Y_t|x_{it}(\bs,v),x_{jt}(\bs,v)}
			{p_{Y_t|\bS V}(Y_t|\bs, v)} \right] , \quad 1 \le t \le N .
\label{eq:zeta}
\end{equation}
By our assumption (\ref{eq:WK-delta}) that $p_{Y|X_1 X_2}(y|x_1,x_2) \ge \delta$
for every $y,x_1,x_2$, the argument of the log above is in the range
$[1/\delta, \delta]$. Hence $\zeta_t(\bs,v,i,j) \le \log^2 \delta$, and
\[ \mathrm{var}_{\bY|\bS V \calK} (\hat{\theta}_{ij}(\bs,v))
	= \frac{1}{N^2} \sum_{t=1}^N \zeta_t(\bs,v,i,j) \le \frac{\log^2 \delta}{N} . \]

By Chebyshev's inequality, the probability of (\ref{eq:Pr-Aeps-2}) is upper-bounded by
\begin{eqnarray*}
   \frac{\eE_{\bY|\bS V \calK} [(\hat{\theta}_{ij}(\bs,v) - \theta_{ij}(\bs,v))^2]}
		{(\delta^2/8)^2} 
	=\frac{\mathrm{var}_{\bY|\bS V \calK} (\hat{\theta}_{ij}(\bs,v))}{(\delta^2/8)^2}
	\le \frac{64\log^2 \delta}{N\delta^4}
\end{eqnarray*}
which establishes (\ref{eq:Pr-Aeps}).

\underline{Step~6}.
Define the following subsets  of $\calM_N^{\good}(\bs,v,\delta)$, indexed by 
$i \in \calM_N^{\good}(\bs,v,\delta)$:
\begin{equation}
   \calM_N^{\calA-\good}(\bs,v,i,\delta) \triangleq
		\{ j \in \calM_N^{\good}(\bs,v,\delta) ~:~ (i,j) \in \calA(\bs,v,\delta) \}  
\label{eq:M-good*}
\end{equation}
which depend on $\bs$ only via $p_{\bs}$.

We show that the typical sets
$\widetilde{\calT}_\delta(\bs,v,i,j), \,j \in \calM_N^{\calA-\good}(\bs,v,i,\delta)$,
have weak overlap for any fixed $\bs,v,i$.
Define the overlap factor of the good sets at $\bY=\by$:
\begin{equation}
   M_\delta(\by,\bs,v,i) \triangleq
	\sum_{k \in \calM_N^{\calA-\good}(\bs,v,i,\delta)}
	\,\mathds1\{\by \in \widetilde{\calT}_\delta(\bs,v,i,k)\} .
\label{eq:My}
\end{equation}
We show there exists $\delta^* > 0$ such that
\begin{eqnarray}
 \lefteqn{Pr[M_\delta(\bY,\bs,v,i) > 2^{N 3\sqrt{\delta}} ~|~ \bS=\bs, V=v, \calK=\{i,j\},
		\bY \in \widetilde{\calT}_\delta(\bs,v,i,j)] < \frac{1}{N}} \nonumber \\
	& & \hspace*{4in} \forall N > \delta^{-8} , \,\delta < \delta^* .
\label{eq:Pr-Mdelta}
\end{eqnarray}

To do so, define the normalized loglikelihood ratio
\begin{equation}
   \hat{D}_{ijk}(\bY) = \frac{1}{N} \log \frac{p_{Y|X_1 X_2}^N(\bY|\bx_i(\bs,v),\bx_j(\bs,v))}
		{p_{Y|X_1 X_2}^N(\bY|\bx_i(\bs,v),\bx_k(\bs,v))} .
\label{eq:Dhat-ijk}
\end{equation}
If $\bY \in \widetilde{\calT}_\delta(\bs,v,i,j) \cap \widetilde{\calT}_\delta(\bs,v,i,k)$
for some $j,k \in \calM_N^{\calA-\good}(\bs,v,i,\delta)$, then 
\begin{eqnarray}
   \hat{D}_{ijk}(\bY) & \le & |\hat{D}_{ijk}(\bY)| \nonumber \\
	& \stackrel{(a)}{\le} & 
		|\theta_{ij}(\bs,v) - \theta_{ik}(\bs,v)| + 2 \times \frac{\delta^2}{8} \nonumber \\
	& \stackrel{(b)}{\le} & \frac{3\delta^2}{4}
\label{eq:|Dhat-ijk|}
\end{eqnarray}
where inequality (a) follows from (\ref{eq:typ-Y})
and (b) from (\ref{eq:theta-max}) and the fact that both $(i,j)$ and $(i,k)$
are in $\calA(\bs,v,\delta)$.

If $j \in \calM_N^{\calA-\good}(\bs,v,i,\delta)$
and $\bY \in \widetilde{\calT}_\delta(\bs,v,i,j)$, it follows from (\ref{eq:My}) and
(\ref{eq:|Dhat-ijk|}) that
\begin{eqnarray}
   M_\delta(\bY,\bs,v,i)
	& = & \sum_{k \in \calM_N^{\calA-\good}(\bs,v,i,\delta)}
		\mathds1\{\bY \in \widetilde{\calT}_\delta(\bs,v,i,j) \cap 
		\widetilde{\calT}_\delta(\bs,v,i,k) \} \nonumber \\
	& \le & \hat{\zeta}(\bY) \,\mathds1\{\bY \in \widetilde{\calT}_\delta(\bs,v,i,j)\}
\label{eq:Mdelta2}
\end{eqnarray}
where we have defined the random variable
\begin{equation}
   \hat{\zeta}(\bY) \triangleq \sum_{k \in \calM_N^{\calA-\good}(\bs,v,i,\delta)} 
		\,\mathds1 \left\{ \hat{D}_{ijk}(\bY) \le \frac{3\delta^2}{4} \right\} .
\label{eq:zeta-hat}
\end{equation}

Now recalling the definition of the normalized divergence $D_{ijk}$ in (\ref{eq:D-ijk}), define
\begin{eqnarray}
   \zeta & \triangleq & \sum_{k \in \calM_N^{\calA-\good}(\bs,v,i,\delta)} 
			\,\mathds1\{ D_{ijk} \le \delta^2 \} \nonumber \\
   & \stackrel{(a)}{\le} & \sum_{k \in \calM_N^{\calA-\good}(\bs,v,i,\delta)} 
			\,\mathds1\{ d_H(\bx_j(\bs,v),\bx_k(\bs,v)) \le N\delta \} \nonumber \\
   & \le & \sum_{k \in \calM_N} 
			\,\mathds1\{ d_H(\bx_j(\bs,v),\bx_k(\bs,v)) \le N\delta \} \nonumber \\   
   & \stackrel{(b)}{=} & |\calM_j(\bs,v,\delta)| \nonumber \\
   & \stackrel{(c)}{\le} & 2^{N 3\sqrt{\delta}}
\label{eq:zeta-UB}
\end{eqnarray}
where inequality (a) follows from (\ref{eq:dH-ineq2}), 
(b) from (\ref{eq:Mj}), and (c) from (\ref{eq:M-good}).

In Appendix~\ref{sec:zeta}, we show that $\hat{\zeta}(\bY) \le \zeta$ with probability approaching 1
as $N \to \infty$, and more specifically,
\begin{equation}
   Pr[\hat{\zeta}(\bY) > \zeta ~|~ \bS=\bs, V=v, \calK=\{i,j\}] 
		< |\calX|^3 \,|\calY| \,(N+1)^{|\calX|^3} \,2^{-N\delta^7}
\label{eq:zeta-zetahat}
\end{equation}
for all $\delta$ smaller than some $\delta^{**} > 0$.
Then there exists some $\delta^* \in (0,\delta^{**})$ such that
\begin{eqnarray}
  \lefteqn{Pr[M_\delta(\bY,\bs,v,i) > 2^{N 3\sqrt{\delta}} ~|~ \bS=\bs, V=v, \calK=\{i,j\},
		\bY \in \widetilde{\calT}_\delta(\bs,v,i,j)]} \nonumber \\
	& \stackrel{(a)}{\le} & Pr[ \hat{\zeta}(\bY) > 2^{N 3\sqrt{\delta}} ~|~ \bS=\bs, V=v, \calK=\{i,j\},
		\bY \in \widetilde{\calT}_\delta(\bs,v,i,j)] \nonumber \\
    & \le & \frac{Pr[\hat{\zeta}(\bY) > 2^{N 3\sqrt{\delta}} ~|~ \bS=\bs, V=v, \calK=\{i,j\} }
		{Pr [\bY \in \widetilde{\calT}_\delta(\bs,v,i,j) ~|~ \bS=\bs, V=v, \calK=\{i,j\} ]} \nonumber \\
    & \stackrel{(b)}{\le} & \frac{Pr[\hat{\zeta}(\bY) > 2^{N 3\sqrt{\delta}} ~|~ \bS=\bs, V=v, \calK=\{i,j\} }
		{1 - \frac{64 \log^2 \delta}{N\delta^4}} \nonumber \\
    & \stackrel{(c)}{\le} & \frac{Pr[\hat{\zeta}(\bY) > \zeta ~|~ \bS=\bs, V=v, \calK=\{i,j\} }
		{1 - \frac{64 \log^2 \delta}{N\delta^4}} \nonumber \\
	& \stackrel{(d)}{<} & \frac{1}{N} \qquad \forall N > \delta^{-8} , \,\delta < \delta^* \nonumber
\end{eqnarray}
where (a) follows from (\ref{eq:Mdelta2}),
(b) from (\ref{eq:Pr-Aeps}),
(c) from (\ref{eq:zeta-UB}), and (d) from (\ref{eq:zeta-zetahat}).
This establishes (\ref{eq:Pr-Mdelta}).


\underline{Step~7}.
We now prune the typical sets $\widetilde{\calT}_\delta(\bs,v,i,j)$ to exclude the points $\by$
that lie within more than $2^{N 3\sqrt{\delta}}$ of the typical sets.
For each $\bs,v,i,j$, define the pruned typical set
\begin{equation}
   \calT_\delta(\bs,v,i,j) \triangleq \{ \by \in \widetilde{\calT}_\delta(\bs,v,i,j) 
		~:~ M_\delta(\by,\bs,v,i) \le 2^{N 3\sqrt{\delta}} \} .
\label{eq:Teps-refined}
\end{equation}
It follows from (\ref{eq:Teps-refined}) and (\ref{eq:My}) that
\begin{eqnarray}
   \forall \by,\bs,v,i ~: \quad \sum_{j \in \calM_N^{\calA-\good}(\bs,v,i,\delta)}
		\,\mathds1\{\by \in \calT_\delta(\bs,v,i,j)\} & = & M_\delta(\by,\bs,v,i) \nonumber \\
	& \le & 2^{N 3\sqrt{\delta}} .
\label{eq:M-mu}
\end{eqnarray}
The pruned set $\calT_\delta(\bs,v,i,j)$ is still typical for $\bY$ conditioned on
$\bS=\bs$, $\bV=v$, and $\calK=\{i,j\}$ because
\begin{eqnarray}
   \lefteqn{Pr[\bY \notin \calT_\delta(\bs,v,i,j)|\bS=\bs,V=v,\calK=\{i,j\}]} \nonumber \\
	& \stackrel{(a)}{=} & Pr[\bY \notin \widetilde{\calT}_\delta(\bs,v,i,j)|\bS=\bs,V=v,\calK=\{i,j\}] \nonumber \\
	& & \quad + Pr[M_\delta(\bY,\bs,v,i) > 2^{N 3\sqrt{\delta}} ~|~ 
		\bY \in \widetilde{\calT}_\delta(\bs,v,i,j), \bS=\bs,V=v,\calK=\{i,j\}] \nonumber \\
	& \stackrel{(b)}{\le} & \frac{64\log^2 \delta}{N\delta^4} + \frac{1}{N} \nonumber \\
	& \stackrel{(c)}{\le} & \frac{72\log^2 \delta}{N\delta^4} , \quad 
		\forall \bs, v,i,j, \,N > \delta^{-8} , \,\delta < \delta^*
\label{eq:Pr-Teps}
\end{eqnarray}
where (a) follows from the definition (\ref{eq:Teps-refined}), (b) from the inequalities
(\ref{eq:Pr-Aeps}) and (\ref{eq:Pr-Mdelta}), and (c) holds because $\delta < \frac{1}{2}$.

\underline{Step~8}.
Define the typical set for $\bS$ in the variational-distance sense:
\begin{equation}
   \calT_\delta \triangleq \{ \bs ~:~ d_V(p_{\bs}, p_S) \le \delta \} .
\label{eq:typ-S}
\end{equation}
We have the inequality
\begin{eqnarray}
   Pr[\bS \notin \calT_\delta]
    & = & \sum_{T_{\bs} ~:~ d_V(p_{\bs},p_S) > \delta} P_S^N(T_{\bs}) \nonumber \\
    & \stackrel{(a)}{\le} & \sum_{p_{\bs} ~:~ d_V(p_{\bs},p_S) > \delta} 2^{-N D(p_{\bs}\|p_S)} \nonumber \\
    & \stackrel{(b)}{\le} & (N+1)^{|\calS|} \max_{p_{\bs} ~:~ d_V(p_{\bs},p_S) > \delta}
			2^{-N D(p_{\bs}\|p_S)} \nonumber \\
    & \stackrel{(c)}{\le} & (N+1)^{|\calS|} \max_{p_{\bs} ~:~ D(p_{\bs}\|p_S) > \delta^2/\ln 4}
			2^{-N D(p_{\bs}\|p_S)} \nonumber \\
	& \le & (N+1)^{|\calS|} \,2^{-N \delta^2 / \ln 4}
\label{eq:Pc'-split2-term1}
\end{eqnarray}
where in (a) we have used the upper bound of \cite[p.~32]{Csiszar81} on
the probability of a type class,
in (b) the fact that the number of type classes $T_{\bs}$ is at most $(N+1)^{|\calS|}$
\cite[p.~29]{Csiszar81} and in (c) Pinsker's inequality $D(p\|q) \ge d_V^2(p,q)/\ln 4$
\cite[p.~58]{Csiszar81}.

Applying successively (\ref{eq:Ipsw}) and (\ref{eq:typ-S}), we have
\begin{eqnarray}
   |I(p_{\bs},q) - I(p_S,q)|
		& = & \left| \sum_{s \in \calS} (p_{\bs}(s)-p_S(s)) \,I(X_1,X_2;Y|S=s,Q=q,T) \right| \nonumber \\
		& \le & \delta \,\max_{s \in \calS} I(X_1,X_2;Y|S=s,Q=q,T) \nonumber \\
		& \le & \delta \,\log|\calY| , \qquad \forall \bs \in \calT_\delta, \,q \in \calQ_N .
\label{eq:I-var}
\end{eqnarray}

\underline{Step~9}.
Given $f_N,g_N,p_{Y|X_1,X_2}, \bs,v$, we will be interested in several conditional
probabilities that correct decoding occurs in conjunction with the typical event
$\bY \in T_\delta(\bs,v,\calK)$.
Define the following short hands:
\begin{eqnarray}
   \underline{P}_c(i,j|\bs,v)
		& = & Pr[\mathrm{correct~decoding~and~} \bY \in T_\delta(\bs,v,i,j)
				\,|\,\bS=\bs,V=v,\calK=\{i,j\}] \nonumber \\
		& = & \sum_{\by \in T_\delta(\bs,v,i,j)
		\cap (\calD_i(\bs,v) \cup \calD_j(\bs,v))}
		\,p_{Y|X_1 X_2}^N(\by|\bx_i(\bs,v),\bx_j(\bs,v)) , \label{eq:Pc_ijsv} \\
    \underline{P}_c^{\good}(\bs,v) 
		& = & Pr[\mathrm{correct~decoding~and~} \bY \in T_\delta(\bs,v,\calK)
				\,|\bS=\bs,V=v, \calK \in (\calM_N^{\good}(\bs,v,\delta))^2] \nonumber \\
		& = & \frac{1}{|\calM_N^{\good}(\bs,v,\delta)|^2}
			\sum_{i,j\in\calM_N^{\good}(\bs,v,\delta)}  \underline{P}_c(i,j|\bs,v) . \label{eq:Pc_good_sv} 
\end{eqnarray}
Note that $\underline{P}_c(i,j|\bs,v)$ depends on $\bs$ only via its type $p_{\bs}$
(because RM codes are used). 

The conditional probability of correct decoding given $(\bs,v)$ and the event that both colluders
are assigned good codewords is
\begin{eqnarray}
   P_c^{\good}(\bs,v) 
		& = & Pr[\mathrm{correct~decoding}
				\,|\bS=\bs,V=v, \calK \in (\calM_N^{\good}(\bs,v,\delta))^2] \nonumber \\
		& \stackrel{(a)}{\le} & \underline{P}_c^{\good}(\bs,v) + Pr[\bY \notin T_\delta(\bs,v,\calK)
				\,|\bS=\bs,V=v, \calK \in (\calM_N^{\good}(\bs,v,\delta))^2] \nonumber \\
		& \stackrel{(b)}{\le} & \underline{P}_c^{\good}(\bs,v) + \frac{72 \log^2 \delta}{N\delta^4} ,
			\quad \forall N > \delta^{-8} , \,\delta < \delta^*
\label{eq:Pc-good-sv}
\end{eqnarray}
where (a) and (b) follow from (\ref{eq:Pc_good_sv}) and (\ref{eq:Pr-Teps}), respectively.
For any subset $\calB \subseteq (\calM_N^{\good}(\bs,v,\delta))^2$,
possibly dependent on $\bs,v$, we also define
\begin{equation}
   \underline{P}_c(\calB|\bs,v)
	\triangleq \frac{1}{|\calB|} \sum_{(i,j)\in\calB} \underline{P}_c(i,j|\bs,v)  .
\label{eq:Pc_B}
\end{equation}
Combining (\ref{eq:Pc_good_sv}) and (\ref{eq:Pc_B}), we have
\begin{eqnarray}
   \underline{P}_c^{\good}(\bs,v)
	& = & \frac{|\calB|}{|\calM_N^{\good}(\bs,v,\delta)|^2} \underline{P}_c(\calB|\bs,v)
		+ \left( 1 - \frac{|\calB|}{|\calM_N^{\good}(\bs,v,\delta)|^2} \right) 
		\underbrace{\underline{P}_c((\calM_N^{\good}(\bs,v,\delta))^2 \setminus \calB|\bs,v)}_{\le 1} \nonumber \\
	& \le & 1 - \frac{|\calB|}{|\calM_N^{\good}(\bs,v,\delta)|^2} (1 - \underline{P}_c(\calB|\bs,v)) .
\label{eq:Pc_good_sv_UB}
\end{eqnarray}
Applying this inequality to $\calB = \calA(\bs,v,\delta)$ and using the cardinality bound (\ref{eq:setA-size})
yields
\begin{equation}
   \underline{P}_c^{\good}(\bs,v) 
	\le 1 - \frac{\delta^4}{4 \log^2 |\calY|} (1 - \underline{P}_c(\calA(\bs,v,\delta)|\bs,v)) .
\label{eq:Pc_B_good}
\end{equation}


Until this point, no assumption has made on the size of the set $\calM_N^{\good}(\bs,v,\delta)$.
We now assume (this assumption will be relaxed in Steps~10 and 11 of the proof) that
\begin{equation}
   |\calM_N^{\good}(\bs,v,\delta)| \ge 2^{N[R-\delta^2/3]} .
\label{eq:large-MNgood}
\end{equation}
Hence (\ref{eq:setA-size}) implies $|\calA(\bs,v,\delta)| > 2^{N(2R - \delta^2)}$
for $N$ larger than some $N_0(\delta)$.
Then we obtain the sphere-packing inequality
\begin{eqnarray}
  \lefteqn{\underline{P}_c(\calA(\bs,v,\delta)|\bs,v)} \nonumber \\
	& \stackrel{(a)}{=} & \frac{1}{|\calA(\bs,v,\delta)|} \sum_{(i,j)\in \calA(\bs,v,\delta)}
		\,\underline{P}_c(i,j|\bs,v) \nonumber \\
	& \stackrel{(b)}{=} & \frac{1}{|\calA(\bs,v,\delta)|} \sum_{(i,j)\in \calA(\bs,v,\delta)}
			\,\sum_{\by \in \calT_\delta(\bs,v,i,j) \cap (\calD_i(\bs,v) \cup \calD_j(\bs,v))}
			p_{Y|X_1 X_2}^N(\by|\bx_i(\bs,v),\bx_j(\bs,v)) \nonumber \\
	& \stackrel{(c)}{=} & \frac{2}{|\calA(\bs,v,\delta)|}
			\sum_{(i,j)\in \calA(\bs,v,\delta)}
			\,\sum_{\by \in \calT_\delta(\bs,v,i,j) \cap \calD_i(\bs,v)}
			p_{Y|X_1 X_2}^N(\by|\bx_i(\bs,v),\bx_j(\bs,v)) \nonumber \\
	& < & 2^{-N(2R-\delta^2)+1} \,\sum_{(i,j)\in \calA(\bs,v,\delta)}
			\,\sum_{\by \in \calT_\delta(\bs,v,i,j) \cap \calD_i(\bs,v)}
			p_{Y|X_1 X_2}^N(\by|\bx_i(\bs,v),\bx_j(\bs,v))  \nonumber \\
	& \stackrel{(d)}{\le} & 2^{-N(2R-\delta^2)+1} \,\sum_{i=1}^{2^{NR}}
			\,\sum_{j \in \calM_N^{\calA-\good}(\bs,v,i,\delta)}
			\,\sum_{\by \in \calT_\delta(\bs,v,i,j) \cap \calD_i(\bs,v)}
			p_{Y|X_1 X_2}^N(\by|\bx_i(\bs,v),\bx_j(\bs,v))  \nonumber \\
	& \stackrel{(e)}{\le} & 2^{-N(2R-\delta^2)+1}
		\,\sum_{i=1}^{2^{NR}} \,\sum_{j \in \calM_N^{\calA-\good}(\bs,v,i,\delta)}
		\,2^{N[\theta_{ij}(\bs,v) + \delta^2/8]} 
			\,\sum_{\by \in \calT_\delta(\bs,v,i,j) \cap \calD_i(\bs,v)}
			\,r(\by|\bs,v) \nonumber \\
	& \stackrel{(f)}{\le} & 2^{-N(2R-\delta^2 - I(\bs,v) - \delta^2 - \delta^2/8)+1}
		\,\sum_{i=1}^{2^{NR}} \,\sum_{j \in \calM_N^{\calA-\good}(\bs,v,i,\delta)}
		\,\sum_{\by \in \calT_\delta(\bs,v,i,j) \cap \calD_i(\bs,v)}
									r(\by|\bs,v) \nonumber \\
	& = & 2^{-N(2R-I(\bs,v) - 17\delta^2/8)+1} 
		\,\sum_{i=1}^{2^{NR}} \,\,\sum_{j \in \calM_N^{\calA-\good}(\bs,v,i,\delta)}
		\sum_{\by \in \calD_i(\bs,v)}
		\,\mathds1\{\by \in \calT_\delta(\bs,v,i,j)\} \,r(\by|\bs,v) \nonumber \\
	& \stackrel{(g)}{\le} & 2^{-N(2R-I(\bs,v) - 17\delta^2/8)+1}
		\underbrace{\sum_{i=1}^{2^{NR}} \,\sum_{\by \in \calD_i(\bs,v)} r(\by|\bs,v)}_{=1}
		\,2^{N 3\sqrt{\delta}} \nonumber \\
    & < & 2^{-N(2R-I(\bs,v) - 3\delta^2 - 3\sqrt{\delta})}	\qquad \forall N > \frac{8}{7\delta^2}
\label{eq:Agood-UB0}
\end{eqnarray}
where (a) follows from (\ref{eq:Pc_ijsv}), (b) from (\ref{eq:Pc_ijsv}), 
(c) holds because the decoding sets $\calD_i(\bs,v)$ are disjoint,
and because of the symmetry of $p_{Y|X_1 X_2}$, $\calA(\bs,v,\delta)$, and
$\calT_\delta(\bs,v,i,j)$; (d) holds because 
$\calA(\bs,v,\delta) \subseteq \{(i,j)~:~ j \in \calM_N^{\calA-\good}(\bs,v,i,\delta)\}$;
(e) follows from (\ref{eq:typ-Y}) and (\ref{eq:r});
and (f) and (g) follow from (\ref{eq:theta-max0}) and (\ref{eq:M-mu}), respectively.

Combining (\ref{eq:Pc-good-sv}), (\ref{eq:Pc_B_good}), and (\ref{eq:Agood-UB0}) yields
\begin{equation}
   P_c^{\good}(\bs,v) \le
		1 - \frac{\delta^4}{4 \log^2 |\calY|} (1 - 2^{-N(2R-I(\bs,v) - 3\delta^2 - 3\sqrt{\delta})}) 
		+  \frac{72 \log^2 \delta}{N\delta^4} , \quad \forall N > \delta^{-8} , \,\delta < \delta^* .
\label{eq:Pc-sv-good-UB}
\end{equation}
Observe that for all $2R > I(\bs,v) + \delta \log |\calY| + 3\delta^2 + 3\sqrt{\delta}$,
the conditional correct-decoding probability $P_c^{\good}(\bs,v)  \lesssim 1 - \frac{\delta^4}{4 \log^2 |\calY|}$
is bounded away from 1 as $N \to \infty$, under the assumption that
$|\calM_N^{\good}(\bs,v,\delta)| \ge 2^{N[R-\delta^2/3]}$.

\underline{Step~10}.
We now relax the assumption (\ref{eq:large-MNgood}). If (\ref{eq:large-MNgood})
does not hold, then $|\calM_N^{\bad}(\bs,v,\delta)| \ge 2^{NR} (1- 2^{-N\delta^2/3})$.
Further assume both colluders are assigned bad codewords. (This last assumption
is relaxed in Step~11). Analogously to (\ref{eq:Pc-good-sv}), we define the conditional
probability of correct decoding,
\begin{equation}
    P_c^{\bad}(\bs,v) \triangleq Pr[\mathrm{correct~decoding} 
				\,|\bS=\bs,V=v, \calK \in (\calM_N^{\bad}(\bs,v,\delta))^2] .
\label{eq:Pc-bad-sv}
\end{equation}
We show that this probability vanishes as $N \to \infty$,
for any rate $R > 0$ and any channel $p_{Y|X_1 X_2}$. In particular,
for $\delta < \frac{1}{4000}$ we have
\begin{equation}
  P_c^{\bad}(\bs,v) \le \frac{2}{N\delta^2} \quad \forall \bs, v .
\label{eq:Pc-bad-sv-asym}
\end{equation}
The proof of (\ref{eq:Pc-bad-sv-asym}) uses the same techniques as in Steps 3, 4, 5, 9,
and is given in Appendix~\ref{sec:Pc-bad}.

\underline{Step~11}.
We finally consider the most general scenario in which a mix of good codewords and
bad codewords is used, and the mix depends on $(\bs,v)$. By application of the inequality
$Pr[A] \le Pr[A \cap B] + Pr[B^c]$ for any two events $A$ and $B$, we obtain 
the following two upper bounds on the correct-decoding probability,
conditioned on $\bS=\bs$ and $V=v$:
\begin{eqnarray}
   P_c(\bs,v) 
	& = & Pr[\mathrm{correct~decoding} | \bS=\bs, V=v] \nonumber \\
	& \le & \left\{ \begin{array}{l} Pr[\calK \notin (\calM_N^{\good}(\bs,v,\delta))^2] 
				+ P_c^{\good}(\bs,v)  \\
			Pr[\calK \notin (\calM_N^{\bad}(\bs,v,\delta))^2] + P_c^{\bad}(\bs,v) .
			\end{array} \right.
\label{eq:Pc-sv-UB}
\end{eqnarray} 
Let $\beta_N(\bs,v) \triangleq 2^{-NR} |\calM_N^{\bad}(\bs,v,\delta)| \in [0,1]$
be the fraction of bad codewords. 
Substituting (\ref{eq:Pc-bad-sv-asym}) into
(\ref{eq:Pc-sv-UB}) yields
\[ 
   P_c(\bs,v) \le \min \left\{ 1 - (1- \beta_N(\bs,v))^2 + P_c^{\good}(\bs,v),
			\;1- \beta_N^2(\bs,v) \right\} + \frac{2}{N\delta^2} .
\]
The first argument of $\min\{\cdot,\cdot\}$ increases with $\beta_N$
and the second decreases. The value of $\beta_N(\bs,v)$ that maximizes the expression above
is the equalizer, $\frac{1}{2} [1-P_c^{\good}(\bs,v)]$,
and thus we obtain 
\begin{equation}
   P_c(\bs,v) \le 1 - \frac{[1-P_c^{\good}(\bs,v)]^2}{4} + \frac{2}{N\delta^2} ,
	\quad \forall \beta_N(\bs,v) \in [0,1] .
\label{eq:Pc-sv-UB2}
\end{equation}
Hence if $P_c^{\good}(\bs,v)$ is bounded away from 1, so is $P_c(\bs,v)$.

In particular, if (\ref{eq:large-MNgood}) does not hold, then
$\beta_N(\bs,v) \ge 1- 2^{-N\delta^2/3}$, and $P_c(\bs,v) \le 1 - (1- 2^{-N\delta^2/3})^2
+ \frac{2}{N\delta^2}$ which vanishes as $N \to \infty$ for all $R > 0$.
Conversely, if (\ref{eq:large-MNgood}) holds, then so does the upper bound
(\ref{eq:Pc-sv-good-UB}) on $P_c^{\good}(\bs,v)$. Using this upper bound together
with the inequality $(b-a)^2 \ge b^2 - 2ac$ which is valid for $0 < a < b < c$, we obtain
\begin{eqnarray*}
    [1-P_c^{\good}(\bs,v)]^2  & \le & \left[ \underbrace{\underbrace{\frac{\delta^4}{4 \log^2 |\calY|}}_{=c}
		(1 - 2^{-N(2R-I(\bs,v) - 3\delta^2 - 3\sqrt{\delta})})}_{= b}
		- \underbrace{\frac{72 \log^2 \delta}{N\delta^4}}_{=a} \right]^2 \\
	& \ge & \frac{\delta^8}{16 \log^4 |\calY|} 
		(1 - 2^{-N(2R-I(\bs,v) - 3\delta^2 - 3\sqrt{\delta})})^2
		- \frac{36 \log^2 \delta}{N\log^2 |\calY|} , \quad \forall N > \delta^{-8} , \,\delta < \delta^* . 
\end{eqnarray*}
Combining this inequality with (\ref{eq:Pc-sv-UB2}) yields
\begin{equation}
   P_c(\bs,v) \le 1 - \frac{\delta^8}{64 \log^4 |\calY|} 
		(1 - 2^{-N(2R-I(\bs,v) - 3\delta^2 - 3\sqrt{\delta})})^2
		+ \frac{9 \log^2 \delta}{N\log^2 |\calY|} + \frac{2}{N\delta^2} 
		, \quad \forall N > \delta^{-8} , \,\delta < \delta^* . 
\label{eq:Pc-sv-UB3}
\end{equation}

\underline{Step~12.}
We shall maximize the upper bound of (\ref{eq:Pc-sv-UB3}) over $\bs \in T_\delta$ and $v \in \calV_N$,
which amounts to maximizing $I(\bs,v)$ in the exponent. In view of the equivalence
of the representations $(\bs,v)$ and $(p_{\bs},q)$, and recalling (\ref{eq:Ipsw}), we have
\begin{eqnarray}
   \max_{\bs \in T_\delta} \max_{v \in \calV_N} I(\bs,v)
	& = & \max_{p_{\bs}\,:\,d_V(p_{\bs},p_S) \le \delta} \,\max_{q \in \calQ_N} I(p_{\bs},q) \nonumber \\
	& \le & \max_{q \in \calQ_N} I(p_S,q) + \delta \log|\calY|
\label{eq:Isv-max}
\end{eqnarray}
where the inequality follows from (\ref{eq:I-var}).

The probability of correct decoding satisfies
\begin{eqnarray}
   \lefteqn{P_c(f_N,g_N,p_{Y|X_1 X_2})} \nonumber \\
	& = & \sum_{\bs \in \calS^N} p_S^N \,\sum_{v\in\calV_N} p_V(v) P_c(\bs,v) \nonumber \\
	& \le & Pr[\bS \notin \calT_\delta] + \max_{\bs \in \calT_\delta} \,\max_{v\in\calV_N} 
		P_c(\bs,v) \nonumber \\
	& \stackrel{(a)}{\le} & (N+1)^{|\calS|} \,2^{-N \delta^2 / \ln 4}
		+ \frac{9 \log^2 \delta}{N\log^2 |\calY|} + \frac{2}{N\delta^2} \nonumber \\
	& & + 1 - \frac{\delta^8}{64 \log^4 |\calY|} \,\left(
		1- \max_{\bs  \in \calT_\delta} \,\max_{v\in\calV_N} 
		\,2^{-N(2R- I(\bs,v) - 3\delta^2 - 3\sqrt{\delta})} \right) \nonumber \\
	& \stackrel{(b)}{\le} & (N+1)^{|\calS|} \,2^{-N \delta^2 / \ln 4}
		+ \frac{9 \log^2 \delta}{N\log^2 |\calY|} + \frac{2}{N\delta^2}  \nonumber \\
	& & + 1 - \frac{\delta^8}{64 \log^4 |\calY|} \,(1- \,2^{-N(2R- \max_q I(p_S,q) 
		- \delta \log |\calY| - 3\delta^2 - 3\sqrt{\delta})} )
		, \quad \forall N > \delta^{-8} , \,\delta < \delta^*
\label{eq:Pc-UB}
\end{eqnarray}
where (a) follows from  (\ref{eq:Pc'-split2-term1}) and (\ref{eq:Pc-sv-UB3})
and (b) from (\ref{eq:Isv-max}). Thus for all
$2R > \max_q I(p_S,q) + \delta \log |\calY| + 3\delta^2 + 3\sqrt{\delta}$,
\begin{equation}
   P_c(f_N,g_N,p_{Y|X_1 X_2}) \lesssim 1 - \frac{\delta^8}{64 \log^4 |\calY|} \quad 
	\mathrm{as~} N \to \infty 
\label{eq:Pc-UB-asym}
\end{equation}
is bounded away from 1.

\underline{Step~13}.
We now bound $\max_q I(p_S,q)$ in (\ref{eq:Pc-UB}) by a quantity that does
not depend on $N$. Since
\[ I_{p_Q p_T p_S p_{X|SQT}^2 p_{Y|X_1 X_2}}(X_1,X_2;Y|S,Q,T)
	= \sum_{q \in \calQ_N} p_Q(q) \,I(p_S,q) , \]
we have
\begin{eqnarray}
   \max_{q  \in \calQ_N} I(p_S,q)
	& = & \max_{p_Q \in \scrP_Q} I_{p_Q p_T p_S p_{X|SQT}^2 p_{Y|X_1 X_2}}(X_1,X_2;Y|S,Q,T) \nonumber \\
	& \stackrel{(a)}{\le} & \max_{p_{QT} \in \scrP_{QT}}
		I_{p_{QT} p_S p_{X|SQT}^2 p_{Y|X_1 X_2}}(X_1,X_2;Y|S,Q,T) \nonumber \\
	& \stackrel{(b)}{=} & \max_{p_W \in \scrP_W}
		I_{p_W p_S p_{X|SW}^2 p_{Y|X_1 X_2}}(X_1,X_2;Y|S,W)
\label{eq:max-Iv}
\end{eqnarray}
where (a) holds because the maximization is over a larger domain ($p_{QT}$ is now
unconstrained over $\calW_N \triangleq \calQ_N \times \{1,2,\cdots,N\}$),
and (b) is obtained  by defining the random variable $W =(Q,T) \in \calW_N$.
Moreover
\begin{eqnarray}
   \max_{p_W \in \scrP_W} I(X_1,X_2;Y|S,W)
	& \le & \sup_{L \to \infty} \max_{p_W \in \scrP_W}
		I(X_1,X_2;Y|S,W) \nonumber \\
	& = & \lim_{L \to \infty} \max_{p_W \in \scrP_W}
		I(X_1,X_2;Y|S,W)
\label{eq:lim-sup}
\end{eqnarray}
where the alphabet for $W$ in the right side is $\{1,2,\cdots,L\}$,
and the supremum and the limit are equal because the supremand
is nondecreasing in $L$.

Combining (\ref{eq:Pc-UB}), (\ref{eq:max-Iv}), and (\ref{eq:lim-sup}),
we conclude that
\begin{equation}
   P_c^*(f_N,g_N,\scrW_{K,\delta}^{\fair})
	\triangleq \min_{p_{Y|X_1 X_2} \in \scrW_{K,\delta}^{\fair}}
		\,P_c(f_N,g_N,p_{Y|X_1 X_2})
\label{eq:Pc*}
\end{equation}
is bounded away from 1 as $N \to \infty$ for all $\delta \in (0,\delta^*)$
and all sequences of codes $(f_N,g_N)$ of rate
\begin{eqnarray*}
  R & > & \frac{1}{2} \left[
	\min_{p_{Y|X_1 X_2} \in \scrW_{K,\delta}^{\fair}}
	\lim_{L \to \infty} \max_{p_W \in \scrP_W}
	I(X_1,X_2;Y|S,W) + \delta \log|\calY| + 3\delta^2 + 3\sqrt{\delta} \right] .
\end{eqnarray*}

Letting $\delta \downarrow 0$, we conclude that reliable decoding is possible only if
\begin{eqnarray*}
   R & \le & \min_{p_{Y|X_1 X_2} \in \scrW_K^{\fair}}
				\lim_{L \to \infty} \max_{p_W \in \scrP_W} 
				\frac{1}{2} I(X_1,X_2;Y|S,W) \\
	& = & \lim_{L \to \infty} \min_{p_{Y|X_1 X_2} \in \scrW_K^{\fair}}
				\max_{p_W \in \scrP_W}
				\frac{1}{2} I(X_1,X_2;Y|S,W) \\
	& = & \lim_{L \to \infty} \max_{p_W \in \scrP_W}
				\min_{p_{Y|X_1 X_2} \in \scrW_K^{\fair}} 
				\frac{1}{2} I(X_1,X_2;Y|S,W)
\end{eqnarray*}
where the second equality holds by application of the minimax theorem: the mutual
information functional is linear (hence concave) in $p_W$ and convex in
$p_{Y|X_1 X_2}$, and the domains of $p_W$ and $p_{Y|X_1 X_2}$ are convex.
Since the above inequality holds for all feasible $p_{X|SW}$, we obtain
\begin{eqnarray*}
   R & \le & \lim_{L \to \infty} \;\max_{p_{X_1 X_2 W|S} \in \scrP_{X_1 X_2 W|S}(p_S,L,D_1)}
				\;\min_{p_{Y|X_1 X_2} \in \scrW_K^{\fair}} 
				\frac{1}{2} I(X_1,X_2;Y|S,W) \\
	& = & \widetilde{C}^{\one}(D_1, \scrW_K^{\fair}) .
\end{eqnarray*}
This concludes the proof.
\footnote{The case of more than two colluders would be treated as follows. Say there are three colluders.
The definition of the restricted class of channels (\ref{eq:WK-delta}) would be extended as follows:
\begin{eqnarray*}
   \lefteqn{\scrW_{K,\delta}^{\fair} = \left\{ p_{Y|X_1 X_2 X_3} \in \scrW_K^{\fair} ~:~
   p_{Y|X_1 X_2 X_3}(y|x_1,x_2,x_3) \ge \delta , \quad \forall y,x_1,x_2,x_3 , \right.} ,  \nonumber \\
    & & \left. \delta \le D(p_{Y|X_1=x_1,X_2=x_2,X_3=x_3} \| p_{Y|X_1=x_1',X_2=x_2',X_3=x_3'})
	\le \log \delta^{-1} , \quad \forall (x_1,x_2,x_3) \ne (x_1',x_2',x_3') \right\} 
\end{eqnarray*}
  Then the notions of equivalence of Hamming distance and statistical
  indistinguishability of two codewords apply similarly to the case of
  two colluders, as does the key property of bounded overlap of the typical
  sets, and the derivation of the sphere-packing inequality in Step~9.}
\hfill $\Box$

\section{Proof of Theorem~\ref{thm:simple}}
\label{Sec:SimpleProof}
\setcounter{equation}{0}

We derive the error exponents for the threshold decision rule (\ref{eq:decision-simple}).
By symmetry of the codebook construction, the error probabilities will be independent
of $\calK$. Without loss of optimality, we assume that $\calK = \sfK = \{1,2,\cdots,K\}$.
Recalling that $\calW = \{1,2,\cdots,L\}$, denote by $\scrP_{XW}^{[N]}(L)$
the set of joint types over $\calX \times \calW$.
Define
\begin{eqnarray}
   \scrP_{YX_{\sfK}|W}^{[N]}(p_{\bx\bw}, \scrW_K, R,L, m)
	& = & \left\{ p_{\by\bx_{\sfK}|\bw} \,:~p_{\bx_{\sfK}|\bw} \in \scrM(p_{\bx|\bw}),
		~p_{\by|\bx_{\sfK}} \in \scrW_K(p_{\bx_{\sfK}}), ~I(\bx_m;\by|\bw) \le R \right\} \nonumber \\
    \tilde{E}_{\psp,m,N}(R,L,p_{\bx\bw},\scrW_K)
	& = & \min_{p_{\by\bx_{\sfK}|\bw} \in \scrP_{YX_{\sfK}|W}^{[N]}(p_{\bx\bw}, \scrW_K, R,L, m)}	
		\;D(p_{\by\bx_{\sfK}|\bw} \| p_{\by|\bx_{\sfK}} \,p_{\bx|\bw}^K \,|\,p_{\bw})
															\label{eq:Epsp-m-simple-N} \\
   \overline{\tilde{E}}_{\psp,N}(R,L,p_{\bx\bw},\scrW_K)
	& = & \max_{m\in\sfK} \tilde{E}_{\psp,m,N}(R,L,p_{\bx\bw},\scrW_K) ,
															\label{eq:Emax-tilde-simple-N} \\
   \underline{\tilde{E}}_{\psp,N}(R,L,p_{\bx\bw},\scrW_K)
	& = & \min_{m\in\sfK} \tilde{E}_{\psp,m,N}(R,L,p_{\bx\bw},\scrW_K)
															\label{eq:Emin-tilde-simple-N} 
\end{eqnarray}
and
\begin{equation}
   E_{\psp,N}(R,L,\scrW_K) = \max_{p_{\bx\bw} \in \scrP_{XW}^{[N]}(L)}
	\;\tilde{E}_{\psp,1,N}(R,L,p_{\bx\bw},\scrW_{K_{\nom}}^{\fair}) .
\label{eq:Epsp-simple-N}
\end{equation}
Denote by $p_{\bx\bw}^*$ the maximizer above (which implicitly depends on $R$)
and by $T_{\bx\bw}^*$ the corresponding type class. Let
\begin{eqnarray}
   \overline{E}_{\psp,N}(R,L,\scrW_K) & = & 
	\overline{\tilde{E}}_{\psp,N}(R,L,p_{\bx\bw}^*,\scrW_K) ,
															\label{eq:Emax-simple-N} \\
   \underline{E}_{\psp,N}(R,L,\scrW_K) & = & 
	\underline{\tilde{E}}_{\psp,N}(R,L,p_{\bx\bw}^*,\scrW_K) .
															\label{eq:Emin-simple-N}
\end{eqnarray}
The expressions (\ref{eq:Epsp-m-simple-N})---(\ref{eq:Emin-simple-N}) differ from
(\ref{eq:Epsp-m-simple})---(\ref{eq:Emin-simple}) in that the optimizations
are performed over types instead of general p.m.f.'s. We have
\begin{eqnarray}
 	\lim_{N \to \infty} \overline{E}_{\psp,N}(R,L,\scrW_K)
		& = & \overline{E}_{\psp}(R,L,\scrW_K)
														\label{eq:lim-Emax-simple-N} \\
 	\lim_{N \to \infty} \underline{E}_{\psp,N}(R,L,\scrW_K)
		& = & \underline{E}_{\psp}(R,L,\scrW_K)
														\label{eq:lim-Emin-simple-N}
\end{eqnarray}
by (\ref{eq:WK-continuous}) and continuity of the divergence and mutual-information functionals.

With the joint type class $T_{\bx\bw}^*$ specified below (\ref{eq:Epsp-simple-N}),
we now restate the coding and decoding scheme.

{\bf Codebook}.
A random constant-composition code $\calC(\bw) = \{\bx_m, \,1 \le m \le 2^{NR}\}$
is generated for each $\bw \in T_{\bw}^*$ by drawing $2^{NR}$ sequences independently
and uniformly from the conditional type class $T_{\bx|\bw}^*$. 

{\bf Encoder}.
A sequence $\bw$ is drawn uniformly from $T_{\bw}^*$ and shared
with the receiver. User $m$ is assigned codeword $\bx_m$ from $\calC(\bw)$,
for $1 \le m \le 2^{NR}$.

{\bf Decoder}. Given $(\by,\bw)$, the decoder places user $m$ on the guilty list
if $I(\bx_m;\by|\bw) > R+\Delta$.

{\bf Collusion Channel.}
The random code described above is a RM code.
By Prop.~\ref{prop:exch}, it is sufficient to restrict
our attention to strongly exchangeable collusion channels
for the error probability analysis.
Recall from (\ref{eq:pYXS}) and (\ref{eq:bound1}) that for such channels,
\begin{equation}
   p_{\bY|\bX_{\sfK}}(\tby|\bx_{\sfK})
	= \frac{Pr[T_{\by|\bx_{\sfK}}]}{|T_{\by|\bx_{\sfK}}|} 
	\le \frac{1}{|T_{\by|\bx_{\sfK}}|} \,\mathds1\{ p_{\by|\bx_{\sfK}} \in \scrW_K(p_{\bx_\sfK}) \} ,
		\quad \forall \,\tby \in T_{\by|\bx_{\sfK}} .
\label{eq:bound2}
\end{equation}

{\bf Error Exponents.}
The derivation is based on the following two asymptotic equalities which are
special cases of (\ref{eq:I-rc}) and (\ref{eq:I-sp}) proven later.

\noindent
1) Fix $\bw$ and $\by$ and draw $\bx$ uniformly from a fixed conditional type class
$T_{\bx|\bw}^*$, independently of $\by$. Then for any $\nu \ge 0$,
\vspace*{-0.1in}
\begin{equation}
   Pr[I(\bx;\by|\bw) \ge \nu] \dotle 2^{-N \nu} .
\label{eq:I-rc-simple}
\end{equation}
2) Fix $\bw$, draw $\bx_m, \,m\in\sfK$, i.i.d. uniformly from a fixed conditional type
class $T_{\bx|\bw}$, and then draw $\bY$ uniformly from the type class $T_{\by|\bx_{\sfK}}$.
For any strongly exchangeable collusion channel, for any $m \in \sfK$ and $\nu \ge 0$, we have
\begin{equation}
   Pr[I(\bx_m;\by|\bw) \le \nu] \doteq
	\exp_2\{ -N \tilde{E}_{\psp,m,N}(\nu,L,p_{\bx\bw}, \scrW_K) \} .
\label{eq:I-sp-simple}
\end{equation}

{\bf (i). False Positives.}
A false positive occurs if
\begin{equation}
   \exists m \notin \sfK \,:\quad I(\bx_m;\by|\bw) > R+\Delta .
\label{eq:FP-simple}
\end{equation}
By construction of the codebook, $\bx_m$ is conditionally independent of $\by$
given $\bw$, for each $m \notin \sfK$.
There are at most $2^{NR} -K$ possible values for $m$ in (\ref{eq:FP-simple}).
Hence the probability of false positives, conditioned on the joint type class
$T_{\by\bx_{\sfK}\bw}$, is
\begin{eqnarray}
   P_{\FP}(T_{\by\bx_{\sfK}\bw},\scrW_K)
	& = & Pr[\exists m \notin \sfK \,:\quad I(\bx_m;\by|\bw) > R+\Delta] \nonumber \\
	& \stackrel{(a)}{\le} & (2^{NR} -K) \,Pr_{\bX}[I(\bx;\by|\bw) > R+\Delta ] 	\nonumber \\
	& \stackrel{(b)}{\dotle} & 2^{NR} \,2^{-N (R+\Delta)} = 2^{-N \Delta}
\label{eq:P-FP-simple}
\end{eqnarray}
where (a) follows from the union bound, and (b) from (\ref{eq:I-rc-simple})
with $\nu=R+\Delta$.
Averaging over all type classes $T_{\by\bx_{\sfK}\bw}$, we obtain
$P_{\FP} \dotle 2^{-N \Delta}$, from which (\ref{eq:E-FP-simple}) follows.

{\bf (ii). Detect-One Error Criterion}. (Miss all colluders.)
We first derive the error exponent for the event that the decoder misses
a specific colluder $m \in \sfK$.
Any coalition $\hat{\calK}$ that contains $m$ fails the test (\ref{eq:decision-simple}),
i.e., for any such $\hat{\calK}$,
\vspace*{-0.1in}
\begin{equation}
   I(\bx_m;\by|\bw) \le R+\Delta .
\label{eq:miss-K*-simple}
\end{equation}
The probability of the miss-$m$ event, given the joint type $p_{\bx\bw}^*$,
is therefore upper-bounded by the probability of the event (\ref{eq:miss-K*-simple}).
From (\ref{eq:I-sp-simple}) we obtain
\begin{eqnarray}
  p_{\mathrm{miss}-m}(p_{\bx\bw}^*,\scrW_K)
   & \le & Pr \left[ I(\bx_m;\by|\bw) \le R+\Delta \right] \nonumber \\
	& \stackrel{(a)}{\dotle} & 
		\exp_2\{ -N \tilde{E}_{\psp,m,N}(R+\Delta,L,p_{\bx\bw}^*,\scrW_K) \} .
\label{eq:pmiss-m}
\end{eqnarray}

The miss-all event is the intersection of the miss-$m$ events over $m \in \sfK$.
Its probability is
\begin{eqnarray*}
  p_{\mathrm{miss-all}}(p_{\bx\bw}^*,\scrW_K)
	& = & Pr \left[ \bigcap_{m \in \sfK} \left\{ \mathrm{miss~} m ~|~p_{\bx\bw}^* \right\} \right] \\
    & \le & \min_{m \in \sfK} \,p_{\mathrm{miss}-m}(p_{\bx\bw}^*,\scrW_K) \\
	& \stackrel{(a)}{\doteq} & \min_{m \in \sfK} \exp_2\{
				-N \tilde{E}_{\psp,m,N}(p_{\bx\bw}^*,R+\Delta,L,\scrW_K) \} \\
	& \stackrel{(b)}{\doteq} & \exp_2\{ -N \overline{E}_{\psp,N}(R+\Delta,L,\scrW_K) \} \\
	& \stackrel{(c)}{\doteq} & \exp_2\{ -N \overline{E}_{\psp}(R+\Delta,L,\scrW_K) \}
\end{eqnarray*}
were (a) follows from (\ref{eq:pmiss-m}), (b) from (\ref{eq:Emax-tilde-simple-N})
and (\ref{eq:Emax-simple-N}), and (c) from (\ref{eq:lim-Emax-simple-N}).

{\bf (iii). Detect-All Error Criterion.} (Miss Some Colluders.)
%
The miss-some event is the union of the miss-$m$ events over $m \in \sfK$.
Its probability is
\begin{eqnarray*}
  p_{\mathrm{miss-some}}(p_{\bx\bw}^*,\scrW_K)
	& = & Pr \left[ \bigcup_{m \in \sfK} \left\{ \mathrm{miss~} m ~|~p_{\bx\bw}^* \right\} \right] \\
    & \le & \sum_{m \in \sfK} p_{\mathrm{miss}-m}(p_{\bx\bw}^*,\scrW_K) \\
	& \doteq & \max_{m \in \sfK} \exp_2\{ -N \tilde{E}_{\psp,m,N}
			(R+\Delta,L,p_{\bx\bw}^*,\scrW_K) \} \\
	& \stackrel{(a)}{=} & \exp_2\{ -N \underline{E}_{\psp,N}(R+\Delta,L,\scrW_K) \} \\
	& \stackrel{(b)}{\doteq} & \exp_2\{ -N \underline{E}_{\psp}(R+\Delta,L,\scrW_K) \}
\end{eqnarray*}
where (a) follows from (\ref{eq:Emin-tilde-simple-N})
and (\ref{eq:Emin-simple-N}), and (b) from (\ref{eq:lim-Emin-simple-N}).

{\bf (iv). Fair Collusion Channels.}
Recall (\ref{eq:set-m}), restated here for convenience:
\begin{eqnarray*}
  \scrP_{YX_{\sfK}|W}(p_{XW}, \scrW_K, R,L, m)
	& \triangleq  & \left\{ \tp_{YX_{\sfK}|W}\,: ~\tp_{X_{\sfK}|W} \in \scrM(p_{X|W}) ,
	\;\tp_{Y|X_{\sfK}} \in \scrW_K(\tp_{X_{\sfK}}) , \right. \nonumber \\
	& & \qquad \qquad \left. \;I_{\tp_{YX_{\sfK}|W} p_W}(X_m;Y|W) \le R \right\} ,
		\quad m \in \sfK .
\end{eqnarray*}
The union of these sets over $m$,
\begin{equation}
   \calP^*(\scrW_K) \triangleq \bigcap_{m\in\sfK} 
		\scrP_{YX_{\sfK}|W}(p_{XW}, \scrW_K, R,L, m)
\label{eq:set-P*}
\end{equation}
is convex and permutation-invariant because so is $\scrW_K$, by assumption.
%
Combining (\ref{eq:set-P*}), (\ref{eq:set-m}), and (\ref{eq:Epsp-m-simple}),
we may write (\ref{eq:Emax-tilde-simple}) as
\begin{eqnarray}
	\overline{\tilde{E}}_{\psp}(R,L,p_{XW},\scrW_K)
	= \min_{\tp_{YX_{\sfK}|W} \,\in \,\calP^*(\scrW_K)}
		D(\tp_{YX_{\sfK}|W} \| \tp_{Y|X_{\sfK}} \,p_{X|W}^K \,|\, p_W) .
\label{eq:Epsp-m-simple1}
\end{eqnarray}

For any $\tp_{YX_{\sfK}|W} \in \calP^*(\scrW_K)$ and permutation $\pi$ of $\sfK$,
define the permuted conditional p.m.f.
\[ \tp_{YX_{\sfK}|W}^\pi (y,x_{\sfK}|w) = \tp_{YX_{\sfK}|W} (y,x_{\pi(\sfK)}|w) \]
and the permutation-averaged p.m.f.
$\tp_{YX_{\sfK}|W}^{\fair} =  \frac{1}{K!} \sum_\pi \tp_{YX_{\sfK}|W}^\pi$
which also belongs to the convex set $\calP^*(\scrW_K)$.
We similarly define $\tp_{Y|X_{\sfK}}^\pi$ and $\tp_{Y|X_{\sfK}}^{\fair}$.
Observe that $D(\tp_{YX_{\sfK}|W}^\pi \| \tp_{Y|X_{\sfK}}^\pi \,p_{X|W}^K \,| p_W)$
is independent of $\pi$. By convexity of Kullback-Leibler divergence, this implies
\begin{eqnarray}
   D(\tp_{YX_{\sfK}|W}^{\fair} \| \tp_{Y|X_{\sfK}}^{\fair} \,p_{X|W}^K \,| p_W)
	& \le & \frac{1}{K!} \sum_\pi
		D(\tp_{YX_{\sfK}|W}^\pi \| \tp_{Y|X_{\sfK}}^\pi \,p_{X|W}^K \,| p_W) \nonumber \\
	& = & D(\tp_{YX_{\sfK}|W} \| \tp_{Y|X_{\sfK}} \,p_{X|W}^K \,| p_W) .
\label{eq:KL-convex-simple}
\end{eqnarray}
Therefore the minimum in (\ref{eq:Epsp-m-simple1}) is achieved by a permutation-invariant
$\tp_{YX_{\sfK}|W} = \tp_{YX_{\sfK}|W}^{\fair}$, and the same minimum would have been
obtained if $\scrW_K$ had been replaced with $\scrW_K^{\fair}$.
Hence
\[ \overline{\tilde{E}}_{\psp}(R,L,p_{XW},\scrW_K) 
	= \overline{\tilde{E}}_{\psp}(R,L,p_{XW},\scrW_K^{\fair}) .
\]
Substituting into (\ref{eq:Emax-simple}) and (\ref{eq:E-one-simple}), we obtain
\[ E^{\one}(R,L,\scrW_K,\Delta) = E^{\one}(R,L,\scrW_K^{\fair},\Delta) . \]

{\bf (v).} The equality
\[ E^{\one}(R,L,\scrW_K^{\fair},\Delta) = E^{\all}(R,L,\scrW_K^{\fair},\Delta) \]
is straightforward because $\tilde{E}_{\psp,m}(R,L,p_{XW},\scrW_K^{\fair})$
in (\ref{eq:Epsp-m-simple}) is the same for all $m \in \sfK$, and thus
$\overline{E}_{\psp}(R,L,\scrW_K^{\fair}) = \underline{E}_{\psp}(R,L,\scrW_K^{\fair})$.

{\bf (vi). Positive Error Exponents.}
From Part (v) above, we may restrict our attention to $\scrW_K = \scrW_K^{\fair}$.
Consider any $\calW = \{1,\cdots,L\}$ and $p_W$ that is positive over its support set
(if it is not, reduce the value of $L$ accordingly.)
For any $m \in \sfK$, the minimand in the expression (\ref{eq:Epsp-m-simple}) for
$\tilde{E}_{\psp,m}(R,L,p_{XW},\scrW_K^{\fair})$ is zero if and only if 
\[ \tp_{YX_{\sfK}|W} = \tp_{Y|X_{\sfK}} \,p_{X|W}^{\sfK} ,
	\quad \mathrm{with~} \tp_{Y|X_{\sfK}} \in \scrW_K^{\fair}(\tp_{X_{\sfK}}) . \]
Such $ \tp_{YX_{\sfK}|W}$ is feasible for (\ref{eq:set-m})
if and only if $(p_{XW},\tp_{Y|X_{\sfK}})$ is such that $I(X_m;Y|W) \le R$.
It is not feasible, and thus a positive exponent $E^{\one}$ is guaranteed,
if $R < I(X_1;Y|W)$. The supremum of all such $R$ is given by (\ref{eq:C1})
and is achieved by letting $\Delta \to 0$ and $L \to \infty$.
\hfill $\Box$

\section{Proof of Theorem~\ref{thm:joint}}
\label{Sec:JointProof}
\setcounter{equation}{0}

We derive the error exponents for the MPMI decision rule (\ref{eq:MPMI}).
Again by symmetry of the codebook construction, the error probabilities
will be independent of $\calK$. Without loss of optimality, we assume that
$\calK = \sfK = \{1,2,\cdots,K\}$. We have also defined $\calW = \{1,2,\cdots,L\}$.
Define for all $\sfA \subseteq \sfK$
\begin{eqnarray}
	\scrP_{YX_{\sfK}|SW}^{[N]}(p_{\bw}, \,p_{\bs|\bw}, \,p_{\bx|\bs\bw}, \scrW_K, R,L,\sfA)
	& = & \left\{ p_{\by\bx_{\sfK}|\bs\bw} \,:\,p_{\bx_{\sfK}|\bs\bw} \in \scrM(p_{\bx|\bs\bw}),
		\,p_{\by|\bx_{\sfK}} \in \scrW_K(p_{\bx_{\sfK}}) \right. , \nonumber \\
	& & \qquad \left. \oI(\bx_\sfA;\by\bx_{\sfK\setminus\sfA}|\bs\bw) \le |\sfA| R \right\}
																		\label{eq:set-A-N} \\
    \breve{E}_{\psp,\sfA,N}(R,L,p_{\bw}, p_{\bs|\bw}, p_{\bx|\bs\bw},\scrW_K)
	& = & \min_{p_{\by\bx_{\sfK}|\bs\bw} \in \scrP_{YX_{\sfK}|SW}^{[N]}
				(p_{\bw}, \,p_{\bs|\bw}, \,p_{\bx|\bs\bw}, \scrW_K, R,L,\sfA)} \nonumber \\
	&   & \qquad D(p_{\by\bx_{\sfK}|\bs\bw} \| p_{\by|\bx_{\sfK}} \,p_{\bx|\bs\bw}^K \,|\,p_{\bs\bw}) ,
																		\label{eq:Epsp-cond-N} \\
    \hat{E}_{\psp,\sfA,N}(R,L,p_{\bw}, p_{\bs|\bw}, p_{\bx|\bs\bw},\scrW_K)
	& = & D(p_{\bs|\bw}\|p_S\,|p_{\bw}) + 
		\breve{E}_{\psp,\sfA,N}(R,L,p_{\bw}, p_{\bs|\bw}, p_{\bx|\bs\bw},\scrW_K)
																		\nonumber \\
	& = & \min_{p_{\by\bx_{\sfK}|\bs\bw} \in \scrP_{YX_{\sfK}|SW}^{[N]}
				(p_{\bw}, \,p_{\bs|\bw}, \,p_{\bx|\bs\bw}, \scrW_K, R,L,\sfA)} \nonumber \\
	&   & \qquad D(p_{\by\bx_{\sfK}|\bs\bw} \,p_{\bs|\bw}
		\| p_{\by|\bx_{\sfK}} \,p_{\bx|\bs\bw}^K \,p_S \,|\,p_{\bw}) ,		\label{eq:Epsp-A-N} \\
	\overline{\hat{E}}_{\psp,N}(R,L,p_{\bw}, p_{\bs|\bw}, p_{\bx|\bs\bw},\scrW_K)
	& = & \hat{E}_{\psp,\sfK,N}(R,L,p_{\bw}, p_{\bs|\bw}, p_{\bx|\bs\bw},\scrW_K) ,
																		\label{eq:Emax-tilde-N} \\
   \underline{\hat{E}}_{\psp,N}(R,L,p_{\bw}, p_{\bs|\bw}, p_{\bx|\bs\bw},\scrW_K)
	& = & \min_{\sfA \subseteq \sfK}
			\hat{E}_{\psp,\sfA,N}(R,L,p_{\bw}, p_{\bs|\bw}, p_{\bx|\bs\bw},\scrW_K) ,
																		\label{eq:Emin-tilde-N} \\
   E_{\psp,N}(R,L,D_1,\scrW_K) & = & \max_{p_{\bw} \in \scrP_W^{[N]}}
	\;\min_{p_{\bs|\bw} \in \scrP_{S|W}^{[N]}}
	\max_{p_{\bx|\bs\bw}  \in \scrP_{X|SW}^{[N]}(p_{\bs\bw},L,D_1)} \nonumber \\
	& & \hat{E}_{\psp,\sfK,N}(R,L,p_{\bw}, p_{\bs|\bw}, p_{\bx|\bs\bw},\scrW_{K_{\nom}}^{\fair}) ,
																		\label{eq:Epsp-N}
\end{eqnarray}
where the second equality in (\ref{eq:Epsp-A-N}) is obtained
by application of the chain rule for divergence.

Denote by $p_{\bw}^*$ and $p_{\bx|\bs\bw}^*$ the maximizers in (\ref{eq:Epsp-N}), the latter
viewed as a function of $p_{\bs|\bw}$. Moreover, both $p_{\bw}^*$ and $p_{\bx|\bs\bw}^*$
implicitly depend on $R$ and $\scrW_{K_{\nom}}^{\fair}$. Denote by $T_{\bw}^*$ and
$T_{\bx|\bs\bw}^*$ the corresponding type and conditional type classes. Let
\begin{eqnarray}
   \overline{E}_{\psp,N}(R,L,D_1,\scrW_K)
	& = & \min_{p_{\bs|\bw}} \overline{\hat{E}}_{\psp,N}
			(R,L,p_{\bw}^*, p_{\bs|\bw}, p_{\bx|\bs\bw}^*,\scrW_K)		\label{eq:Emax-N} \\
   \underline{E}_{\psp,N}(R,L,D_1,\scrW_K)
	& = & \min_{p_{\bs|\bw}} \underline{\hat{E}}_{\psp,N}
			(R,L,p_{\bw}^*, p_{\bs|\bw}, p_{\bx|\bs\bw}^*,\scrW_K) .		\label{eq:Emin-N}
\end{eqnarray}
The exponents (\ref{eq:Epsp-A-N})---(\ref{eq:Emin-N}) differ from
(\ref{eq:Epsp-A})---(\ref{eq:Emin}) in that the optimizations are performed
over conditional types instead of general conditional p.m.f.'s.
We have
\begin{eqnarray}
 	\lim_{N \to \infty} \overline{E}_{\psp,N}(R,L,D_1,\scrW_K)
		& = & \overline{E}_{\psp}(R,L,D_1,\scrW_K)
														\label{eq:lim-Emax-N} \\
 	\lim_{N \to \infty} \underline{E}_{\psp,N}(R,L,D_1,\scrW_K)
		& = & \underline{E}_{\psp}(R,L,D_1,\scrW_K)
														\label{eq:lim-Emin-N}
\end{eqnarray}
by (\ref{eq:WK-continuous}) and continuity of the divergence and mutual-information functionals.

{\bf Codebook}.
For each $\bw \in T_{\bw}^*$ and $\bs \in \calS^N$, a codebook
$\calC(\bs,\bw) = \{ \bx_m, \,1 \le m \le 2^{NR} \}$
is generated by drawing $2^{NR}$ random vectors independently and uniformly
from $T_{\bx|\bs\bw}^*$.

{\bf Encoder}. 
A sequence $\bw$ is drawn uniformly from $T_{\bw}^*$ and shared with the decoder.
Given $\bs$ and $\bw$, user $m$ is assigned codeword $\bx_m \in \calC(\bs,\bw)$.

{\bf Decoder}. The decoding rule is the MPMI rule of (\ref{eq:MPMI}).

{\bf Collusion Channel.}
This random code is a RM code, hence by application of Prop.~\ref{prop:exch},
it is sufficient to restrict our attention to strongly exchangeable collusion channels.

{\bf Error Probability Analysis.}
To analyze the error probability for our random-coding scheme under strongly
exchangeable collusion channels, we will again use the bound (\ref{eq:bound2})
as well as the following three properties, which originate from the basic inequalities
(\ref{eq:type-size1}) and (\ref{eq:type-size2}).

\noindent
1) Fix $(\bs,\bw)$ and $\bz \in \calZ^N$, and draw $\bx_{\sfK} = \{\bx_m, \,m \in \sfK\}$
i.i.d. uniformly from a conditional type class $T_{\bx|\bs\bw}$, independently of $\bz$.
We have the asymptotic equality
\begin{equation}
   Pr[T_{\bx_{\sfK}|\bz\bs\bw}] = \frac{|T_{\bx_{\sfK}|\bz\bs\bw}|}{|T_{\bx|\bs\bw}|^K}
	\doteq 2^{-N [KH(\bx|\bs\bw)-H(\bx_{\sfK}|\bz\bs\bw)]} = 2^{-N \oI(\bx_{\sfK};\bz|\bs\bw)}
\label{eq:cond-type-prob}
\end{equation}
where the last equality is due to (\ref{eq:I-empirical}). Then
\begin{eqnarray}
   Pr[\oI(\bx_{\sfK};\bz|\bs\bw) \ge \nu] 
	& = & \sum_{T_{\bx_{\sfK}|\bz\bs\bw}} \,Pr[T_{\bx_{\sfK}|\bz\bs\bw}] 
			\,\mathds1\{\oI(\bx_{\sfK};\bz|\bs\bw) \ge \nu\} 		\nonumber \\ 
	& \doteq & \sum_{T_{\bx_{\sfK}|\bz\bs\bw}} \,2^{-N \oI(\bx_{\sfK};\bz|\bs\bw)}
			\,\mathds1\{\oI(\bx_{\sfK};\bz|\bs\bw) \ge \nu\} 		\nonumber \\ 
	& \doteq & \max_{T_{\bx_{\sfK}|\bz\bs\bw}} \,2^{-N \oI(\bx_{\sfK};\bz|\bs\bw)}
			\,\mathds1\{\oI(\bx_{\sfK};\bz|\bs\bw) \ge \nu\} 		\nonumber \\ 
	& \dotle & 2^{-N \nu} .
\label{eq:I-rc}
\end{eqnarray}
2) Fix $\bw$ and draw $\bs$ i.i.d. $p_S$. We have \cite{Csiszar81}
\begin{equation}
  Pr[T_{\bs|\bw}] \doteq 2^{-N D(p_{\bs|\bw}\|p_S|p_{\bw})} .
\label{eq:Pr-psw}
\end{equation}
3) Fix $(\bs,\bw)$, draw $\bx_k, \,k\in\sfK$, i.i.d. uniformly from
a conditional type class $T_{\bx|\bs\bw}$, and then draw $\bY$ uniformly from
a single conditional type class $T_{\by|\bx_{\sfK}}$. 
We have
\begin{eqnarray}
   Pr[T_{\by\bx_{\sfK}|\bs\bw}]
	& = & Pr[T_{\by|\bx_{\sfK} \bs\bw}] \,Pr[T_{\bx_{\sfK}|\bs\bw}] \nonumber \\
	& = & \frac{|T_{\by|\bx_{\sfK} \bs\bw}|}{|T_{\by|\bx_{\sfK}}|}
			\,\frac{|T_{\bx_{\sfK}|\bs\bw}|}{|T_{\bx|\bs\bw}|^K} \nonumber \\
	& \doteq & 2^{-N [H(\by|\bx_{\sfK}) - H(\by|\bx_{\sfK} \bs\bw)]}
				\;2^{-N[KH(\bx|\bs\bw) - H(\bx_{\sfK}|\bs\bw)]} \nonumber \\
	& = & \exp_2 \left\{ -N [I(\by;\bs\bw|\bx_{\sfK})
		+ \oI(\bx_1; \cdots; \bx_K |\bs\bw)] \right\} .
\label{eq:PrT-1}
\end{eqnarray}

Consider the two terms in brackets above.
The first one may be written as
\begin{eqnarray*}
	I(\by;\bs\bw|\bx_{\sfK}) & = & D(p_{\by\bs\bw|\bx_{\sfK}} 
			\| p_{\by|\bx_{\sfK}} \,p_{\bs\bw|\bx_{\sfK}} \,|\,p_{\bx_{\sfK}}) \\
	& = & D(p_{\by\bs\bw\bx_{\sfK}} \| p_{\by|\bx_{\sfK}} \,p_{\bs\bw\bx_{\sfK}}) \\
	& = & D(p_{\by\bx_{\sfK}|\bs\bw} \| p_{\by|\bx_{\sfK}} \,p_{\bx_{\sfK}|\bs\bw} \,|\,p_{\bs\bw})
\end{eqnarray*}
and the second one as
\[ \oI(\bx_1; \cdots; \bx_K |\bs\bw) = D(p_{\bx_{\sfK}|\bs\bw} \| p_{\bx|\bs\bw}^K \,|\,p_{\bs\bw}) . \]
By application of the chain rule for divergence, the sum of these two terms is
$D(p_{\by\bx_{\sfK}|\bs\bw} \| p_{\by|\bx_{\sfK}} \,p_{\bx|\bs\bw}^K \,|\,p_{\bs\bw})$.
Substituting into (\ref{eq:PrT-1}), we obtain
\begin{eqnarray}
   Pr[T_{\by\bx_{\sfK}|\bs\bw}] \doteq \exp_2 \left\{ -N
	D(p_{\by\bx_{\sfK}|\bs\bw} \| p_{\by|\bx_{\sfK}} \,p_{\bx|\bs\bw}^K \,|\,p_{\bs\bw}) \right\} .
\label{eq:PrT}
\end{eqnarray}
In the derivation below we use the shorthand $e(p_{\by\bx_{\sfK}|\bs\bw})$
to represent the exponential above, and fix $T_{\bx|\bs\bw} = T_{\bx|\bs\bw}^*$.

For any feasible, strongly exchangeable collusion channel, for any $\sfA \subseteq \sfK$ and $\nu > 0$,
conditioning on $\bw \in T_{\bw}^*$ and $\bs \in \calS^N$, we have
\begin{eqnarray}
   \lefteqn{Pr\left[\oI(\bx_{\sfA};\by\bx_{\sfK\setminus\sfA}|\bs\bw)
		\le |\sfA| \nu \right]} \nonumber \\
	& \stackrel{(a)}{\le} & \sum_{\mathrm{feasible~} T_{\by\bx_{\sfK}|\bs\bw}}
				Pr[T_{\by\bx_{\sfK}|\bs\bw}] 
				\,\mathds1\{\oI(\bx_{\sfA};\by\bx_{\sfK\setminus\sfA}|\bs\bw)
				\le |\sfA| \nu \} \nonumber \\
	& \stackrel{(b)}{\doteq} &
				\sum_{\mathrm{feasible~} p_{\by\bx_{\sfK}|\bs\bw}} \,e(p_{\by\bx_{\sfK}|\bs\bw})
				\,\mathds1\{\oI(\bx_{\sfA};\by\bx_{\sfK\setminus\sfA}|\bs\bw)
				\le |\sfA| \nu \} \nonumber \\
	& \stackrel{(c)}{\doteq} & \max_{\mathrm{feasible~} p_{\by\bx_{\sfK}|\bs\bw}}
				\,e(p_{\by\bx_{\sfK}|\bs\bw})
				\,\mathds1\{\oI(\bx_{\sfA};\by\bx_{\sfK\setminus\sfA}|\bs\bw)
				\le |\sfA| \nu \} \nonumber \\
	& = & \max_{p_{\by\bx_{\sfK}|\bs\bw} \,:\,p_{\bx_{\sfK}|\bs\bw} \in \scrM(p_{\bx|\bs\bw}^*),
				\,p_{\by|\bx_{\sfK}} \in \scrW_K} \,e(p_{\by\bx_{\sfK}|\bs\bw})
				\,\mathds1\{\oI(\bx_{\sfA};\by\bx_{\sfK\setminus\sfA}|\bs\bw)
				\le |\sfA| \nu \} \nonumber \\
	& = & \max_{p_{\by\bx_{\sfK}|\bs\bw} ~:~p_{\bx_{\sfK}|\bs\bw} \in \scrM(p_{\bx|\bs\bw}^*),
				\,p_{\by|\bx_{\sfK}} \in \scrW_K ,
				\,\oI(\bx_{\sfA};\by\bx_{\sfK\setminus\sfA}|\bs\bw) \le |\sfA| \nu } 
				e(p_{\by\bx_{\sfK}|\bs\bw})					\nonumber \\
	& \stackrel{(d)}{=} & \max_{p_{\by\bx_{\sfK}|\bs\bw} \in \scrP_{YX_{\sfK}|SW}^{[N]}
				(p_{\bw}^*, \,p_{\bs|\bw}, \,p_{\bx|\bs\bw}^*, \scrW_K, \nu,L,\sfA)} 
				e(p_{\by\bx_{\sfK}|\bs\bw})					\nonumber \\
	& \stackrel{(e)}{=} & \exp_2 \left\{ -N \breve{E}_{\psp,\sfA,N}(\nu,L,p_{\bw}^*,p_{\bs|\bw},
				p_{\bx|\bs\bw}^*,\scrW_K) \right\}
\label{eq:I-sp}
\end{eqnarray}
where (a) follows from (\ref{eq:bound2}), (b) from (\ref{eq:PrT}),
(c) from the fact that the number of conditional types is polynomial in $N$,
(d) from (\ref{eq:set-A-N}), and (e) from (\ref{eq:Epsp-cond-N}).

{\bf (i). False Positives.}
A false positive occurs if $\hat{\calK} \setminus \calK \ne \emptyset$.
By application of (\ref{eq:property1}), we have
\begin{equation}
   \forall \calA \subseteq \hat{\calK} ~:\quad
	\oI(\bx_{\calA};\by\bx_{\hat{\calK}\setminus\calA}\,|\bs\bw) > |\calA| (R+\Delta) .
\label{eq:FP-condition1}
\end{equation}
Denote by $\calB$ the set of colluder indices $m\in\sfK$ that are correctly identified
by the decoder, and by $\calA \triangleq \hat{\calK} \setminus \calB$ the complement set,
which is comprised of all incorrectly accused users and has cardinality $|\calA| \ge 1$. 
By construction of the codebook, $\bx_{\calA}$ is independent of $\by$ and $\bx_{\calB}$.
The probability of the event (\ref{eq:FP-condition1}) is upper-bounded by the probability
of the larger event
\begin{equation}
   \exists \calA \not\subseteq \sfK, \; \exists \calB \subseteq \sfK \;:\quad
   \oI(\bx_{\calA}; \by\bx_{\calB}\,|\bs\bw) > |\calA| (R+\Delta) .
\label{eq:event-FP}
\end{equation}

Hence the probability of false positives, conditioned on $T_{\by\bx_{\sfK}\bs\bw}$, satisfies
\begin{eqnarray}
   \lefteqn{P_{\FP}(T_{\by\bx_{\sfK}\bs\bw},\scrW_K)} \nonumber \\
	& \le & Pr \left[ \bigcup_{|\calA| \ge 1} \bigcup_{\calB \subseteq \sfK} \left\{
		\exists \calA \not\subseteq \sfK \,:\quad
		\oI(\bx_{\calA}; \by\bx_{\calB}\,|\bs\bw)
		> |\calA| (R+\Delta) \right\} \right] \nonumber \\
	& = & Pr \left[ \bigcup_{|\calA| \ge 1} \left\{
		\exists \calA \not\subseteq \sfK \,:\quad
		\max_{\calB \subseteq \sfK} \oI(\bx_{\calA}; \by\bx_{\calB}\,|\bs\bw)
		> |\calA| (R+\Delta) \right\} \right] \nonumber \\
	& = & Pr \left[ \bigcup_{|\calA| \ge 1} \left\{
		\exists \calA \not\subseteq \sfK \,:\quad
		\oI(\bx_{\calA}; \by\bx_{\sfK}\,|\bs\bw)
		> |\calA| (R+\Delta) \right\} \right] \nonumber \\
	& \stackrel{(a)}{\le} & \sum_{|\calA| \ge 1} 2^{N |\calA| R}
			\,Pr \left[ \oI(\bx_{\calA};\by\bx_{\sfK}\,|\bs\bw)
			> |\calA| (R+\Delta) \right] 										\nonumber \\
	& \stackrel{(b)}{\doteq} & \sum_{|\calA| \ge 1}
			2^{N |\calA| R} \,2^{-N |\calA|(R+\Delta)}	\nonumber \\
	& = & \sum_{|\calA| \ge 1} 2^{-N |\calA| \Delta}							\nonumber \\
	& \doteq & 2^{-N\Delta}
\label{eq:P-FP2}
\end{eqnarray}
where (a) follows from the union bound, and (b) from (\ref{eq:I-rc}) with
$\bx_{\calA}$ and $\by\bx_{\sfK}$ in place of $\bx_\sfK$ and $\bz$, respectively.
Averaging over all joint type classes $T_{\by\bx_{\sfK}\bs\bw}$,
we obtain $P_{\FP} \dotle 2^{-N \Delta}$, from which (\ref{eq:E-FP}) follows.

{\bf (ii). Detect-All Error Criterion.} (Miss Some Colluders.)
Under the detect-all error event, {\em any} coalition $\tilde{\calK}$ that {\em contains}
$\calK$ fails the test. By (\ref{eq:property1}), this implies that
\begin{equation}
   \exists \calA \subseteq \tilde{\calK} ~:\quad
		\oI(\bx_{\calA};\by\bx_{\tilde{\calK}\setminus\calA}\,|\bs\bw)
		\le |\calA| (R+\Delta) .
\label{eq:miss-K*}
\end{equation}
In particular, for $\tilde{\calK} = \calK = \sfK$ we have
\begin{equation}
   \exists \sfA \subseteq \sfK ~:\quad \oI(\bx_{\sfA};\by\bx_{\sfK\setminus\sfA}\,|\bs\bw)
		\le |\sfA| (R+\Delta) .
\label{eq:miss-K*-2}
\end{equation}
The probability of the miss-some event, conditioned on $(\bs,\bw)$,
is therefore upper bounded by the probability of the event (\ref{eq:miss-K*-2}):
\begin{eqnarray}
  \lefteqn{p_{\mathrm{miss-some}}(p_{\bw}^* \,p_{\bs|\bw},p_{\bx|\bs\bw}^*,\scrW_K)} \nonumber \\
	& \le & Pr \left[ \bigcup_{\sfA \subseteq \sfK} \left\{
		\oI(\bx_{\sfA};\by\bx_{\sfK\setminus\sfA}\,|\bs\bw)
		\le |\sfA| (R+\Delta) \right\} \right] \nonumber \\
   & \le & \sum_{\sfA \subseteq \sfK} Pr \left[ \oI(\bx_{\sfA};
		\by\bx_{\sfK\setminus\sfA}\,|\bs\bw)
		\le |\sfA| (R+\Delta) \right] \nonumber \\
	& \stackrel{(a)}{\dotle} & \sum_{\sfA \subseteq \sfK} \exp_2 \left\{
		-N \breve{E}_{\psp,\sfA,N}(R+\Delta,L,p_{\bw}^*,p_{\bs|\bw},p_{\bx|\bs\bw}^*,\scrW_K) \right\}
																			\nonumber \\
	& \doteq & \max_{\sfA \subseteq \sfK} \exp_2 \left\{
		-N \breve{E}_{\psp,\sfA,N}(R+\Delta,L,p_{\bw}^*,p_{\bs|\bw},p_{\bx|\bs\bw}^*,\scrW_K) \right\}
																			\nonumber \\
	& = & \exp_2 \left\{ - N \min_{\sfA \subseteq \sfK} 
		\breve{E}_{\psp,\sfA,N}(R+\Delta,L,p_{\bw}^*,p_{\bs|\bw},p_{\bx|\bs\bw}^*,\scrW_K) \right\}
\label{eq:miss-K*-UB}
\end{eqnarray}
where (a) follows from (\ref{eq:I-sp}) with $\nu=R+\Delta$.


Averaging over $\bS$, we obtain
\begin{eqnarray*}
\lefteqn{p_{\mathrm{miss-some}}(\scrW_K)} \nonumber \\
	& = & \sum_{p_{\bs|\bw}} \,Pr[T_{\bs|\bw}]
		\,p_{\mathrm{miss-some}}(p_{\bw}^* \,p_{\bs|\bw},p_{\bx|\bs\bw}^*,\scrW_K) \\
	& \stackrel{(a)}{\doteq} & \max_{p_{\bs|\bw}}
		\exp_2 \left\{ -N \left[ D(p_{\bs|\bw}\|p_S \,|\,p_{\bw}^*)
			+ \min_{\sfA \subseteq \sfK} \breve{E}_{\psp,\{m\},N}
			(R+\Delta,L,p_{\bw}^*,p_{\bs|\bw},p_{\bx|\bs\bw}^*,\scrW_K) \right] \right\} \\
	& \stackrel{(b)}{=}  & \max_{p_{\bs|\bw}}
		\exp_2 \left\{ -N \underline{\hat{E}}_{\psp,N}
			(R+\Delta,L,p_{\bw}^*,p_{\bs|\bw},p_{\bx|\bs\bw}^*,\scrW_K) \right\} \\
	& \stackrel{(c)}{=} & \exp_2 \left\{
			-N \underline{E}_{\psp,N}(R+\Delta,L,D_1,\scrW_K) \right\} \\	
	& \stackrel{(d)}{\doteq} & \exp_2 \left\{
			-N \underline{E}_{\psp}(R+\Delta,L,D_1,\scrW_K) \right\} 
\end{eqnarray*}
which proves (\ref{eq:E-all}).
Here (a) follows from (\ref{eq:Pr-psw}) and (\ref{eq:miss-K*-UB}),
(b) from the definitions (\ref{eq:Emin-tilde-N}) and (\ref{eq:Epsp-A-N}),
(c) from (\ref{eq:Emin-N}), and
(d) from the limit property (\ref{eq:lim-Emin-N}).

{\bf (iii). Detect-One Criterion.} (Miss All Colluders.)
Under the detect-one error event, either the estimated coalition $\hat{\calK}$
is empty, or it is a set $\calI$ of innocent users (disjoint with $\calK$).
Hence $P_e^{\one} \le Pr[\hat{\calK} = \emptyset] + Pr[\hat{\calK} = \calI]$.
The first probability, conditioned on $(\bs,\bw)$, is bounded as \footnote{
	Using the bound $\min_{\calK' \subseteq \calK} Pr[MPMI(\calK') \le 0]$
	would not strengthen the inequality in (\ref{eq:empty-Khat}).}
\begin{eqnarray}
   Pr[\hat{\calK} = \emptyset]
	& = & Pr[\forall \calK' ~:~MPMI(\calK') \le 0] \nonumber \\
	& \le & Pr[MPMI(\calK) \le 0] \nonumber \\
	& = & Pr[\oI(\bx_{\calK};\by|\bs\bw) \le K(R+\Delta)] \label{eq:empty-Khat} \\
	& \doteq & \exp_2 \left\{ -N \breve{E}_{\psp,\sfK,N}
			(R+\Delta,L,p_{\bw}^*,p_{\bs|\bw},p_{\bx|\bs\bw}^*,\scrW_K) \right\} . \nonumber
\end{eqnarray}
To bound the second probability, we use property (\ref{eq:property2})
with $\hat{\calK} = \calI$ and $\calA=\calK$. We obtain
\[ \oI(\bx_{\calK};\by\bx_{\calI}|\bs\bw) \le K(R+\Delta) \]
Since
\[ \oI(\bx_{\calK};\by\bx_{\calI}|\bs\bw)
	= \oI(\bx_{\calK};\by|\bs\bw) + I(\bx_{\calK};\bx_{\calI}|\by\bs\bw)
	\ge \oI(\bx_{\calK};\by|\bs\bw) \]
combining the two inequalities above yields
\[ \oI(\bx_{\calK};\by|\bs\bw) \le K(R+\Delta) . \]
The probability of this event is again given by (\ref{eq:empty-Khat});
we conclude that
\[ p_{\mathrm{miss-all}}(p_{\bw}^* \,p_{\bs|\bw},p_{\bx|\bs\bw}^*,\scrW_K)
	\doteq \exp_2 \left\{ -N \breve{E}_{\psp,\sfK,N}
			(R+\Delta,L,p_{\bw}^*,p_{\bs|\bw},p_{\bx|\bs\bw}^*,\scrW_K) \right\} .
\]

Averaging over $\bS$ and proceeding as in Part (ii) above, we obtain
\begin{eqnarray*}
   p_{\mathrm{miss-all}}(\scrW_K)
	& \le & \sum_{p_{\bs|\bw}} \,Pr[T_{\bs|\bw}]
		\,p_{\mathrm{miss-all}}(p_{\bw}^* \,p_{\bs|\bw},p_{\bx|\bs\bw}^*,\scrW_K) \\
	& \doteq & \exp_2 \left\{ -N \overline{E}_{\psp}(R+\Delta,L,D_1,\sfK,\scrW_K) \right\}
\end{eqnarray*}
which establishes (\ref{eq:E-one}).

{\bf (iv). Fair Collusion Channels.}
The proof parallels that of Theorem~\ref{thm:simple}, Part (iv).
Define
\begin{equation}
   \scrP^*(\scrW_K) \triangleq \scrP_{YX_{\sfK}|SW}(p_W, \tp_{S|W}, p_{X|SW}, \scrW_K, R,L, \sfK)
\label{eq:P*}
\end{equation}
which is convex and permutation-invariant.
Then write (\ref{eq:Emax-tilde}) as
\begin{eqnarray}
	\overline{\tilde{E}}_{\psp}(R,L,p_W, \tp_{S|W},p_{X|SW},\scrW_K)
	= \min_{\tp_{YX_{\sfK}|SW} \,\in \,\scrP^*(\scrW_K)}
		D(\tp_{YX_{\sfK}|SW} \| \tp_{Y|X_{\sfK}} \,p_{X|SW}^K \,|\, \tp_{S|W} \,p_W) .
\label{eq:Epsp-m-1}
\end{eqnarray}

For any $\tp_{YX_{\sfK}|SW} \in \scrP^*(\scrW_K)$ and permutation $\pi$ of $\sfK$, define
the permuted conditional p.m.f. $\tp_{YX_{\sfK}|SW}^\pi$ and the permutation-averaged p.m.f.
$\tp_{YX_{\sfK}|SW}^{\fair} =  \frac{1}{K!} \sum_\pi \tp_{YX_{\sfK}|SW}^\pi$,
which also belongs to the convex set $\scrP^*(\scrW_K)$.
We similarly define $\tp_{Y|X_{\sfK}}^\pi$ and $\tp_{Y|X_{\sfK}}^{\fair}$.
The conditional divergence
$D(\tp_{YX_{\sfK}|SW}^\pi \| \tp_{Y|X_{\sfK}}^\pi \,p_{X|SW}^K \,|$ $\,\tp_{S|W} \,p_W)$
is independent of $\pi$. By convexity, we obtain
\begin{eqnarray}
   D(\tp_{YX_{\sfK}|SW}^{\fair} \| \tp_{Y|X_{\sfK}}^{\fair} \,p_{X|SW}^K \,|\, \tp_{S|W} \,p_W)
	& \le & D(\tp_{YX_{\sfK}|SW} \| \tp_{Y|X_{\sfK}} \,p_{X|SW}^K \,|\, \tp_{S|W} \,p_W) .
\label{eq:KL-convex}
\end{eqnarray}
Therefore the minimum in (\ref{eq:Epsp-m-1}) is achieved by a permutation-invariant
$\tp_{YX_{\sfK}|SW} = \tp_{YX_{\sfK}|SW}^{\fair}$, and the same minimum would have been
obtained if $\scrW_K$ had been replaced with $\scrW_K^{\fair}$.
Hence
\[ \overline{\tilde{E}}_{\psp}(R,L,p_W, \tp_{S|W},p_{X|SW},\scrW_K) 
	= \overline{\tilde{E}}_{\psp}(R,L,p_W, \tp_{S|W},p_{X|SW},\scrW_K^{\fair}) .
\]
Substituting into (\ref{eq:Emax}) and (\ref{eq:E-one}), we obtain
\[ E^{\one}(R,L,D_1,\scrW_K,\Delta) = E^{\one}(R,L,D_1,\scrW_K^{\fair},\Delta) . \]

{\bf (v). Detect-All Error Exponent for Fair Collusion Channels.}
Using (\ref{eq:set-A}) and (\ref{eq:Epsp-A}), observe that $\underline{\tilde{E}}_{\psp}$
in (\ref{eq:Emin-tilde}) may be written as
\begin{eqnarray}
   \underline{\tilde{E}}_{\psp}(R,L,p_W, p_{S|W}, p_{X|SW},\scrW_K)
	= \min_{\tp_{YX_{\sfK}|SW} \in \overline{\scrP}^*(\scrW_K)}
		D(\tp_{YX_{\sfK}|SW} \,\tp_{S|W} \| \tp_{Y|X_{\sfK}} \,p_{X|SW}^K \,p_S \,|\,p_W)
\label{eq:Epsp-m-2}
\end{eqnarray}
where
\begin{eqnarray*}
   \overline{\scrP}^*(\scrW_K) \triangleq \left\{ \tp_{YX_{\sfK}|SW} \right.
		& : & \tp_{X_{\sfK}|SW} \in \scrM(p_{X|SW}),
		\;\tp_{Y|X_{\sfK}} \in \scrW_K(\tp_{X_{\sfK}}) , \nonumber \\
		& & \left. \min_{\sfA \subseteq \sfK} \frac{1}{|\sfA|}
		\oI(X_\sfA;YX_{\sfK\setminus\sfA}|SW) \le R \right\} .
\end{eqnarray*}
Similarly to the discussion below (\ref{eq:Epsp-m-1}), when $\scrW_K = \scrW_K^{\fair}$
the minimum over $\tp_{YX_{\sfK}|SW}$ in (\ref{eq:Epsp-m-2}) is achieved by
a permutation-invariant conditional p.m.f.

Next we show that $\sfK$ minimizes
$\frac{1}{|\sfA|} \oI(X_\sfA;YX_{\sfK\setminus\sfA}|SW)$
over $\sfA \subseteq \sfK$. Indeed
\begin{eqnarray}
   \frac{1}{|\sfA|} \oI(X_\sfA;YX_{\sfK\setminus\sfA}|SW)
	& = & \frac{1}{|\sfA|} \left[ \sum_{m\in\sfA} H(X_m|SW)
			+ H(YX_{\sfK\setminus\sfA}|SW) - H(YX_{\sfK}|SW) \right] \nonumber \\
	& = & H(X|SW) - \frac{1}{|\sfA|} H(X_{\sfA}|YX_{\sfK\setminus\sfA} SW) \nonumber \\
	& \stackrel{(a)}{\ge} & H(X|SW) - \frac{1}{|\sfK|} H(X_{\sfK}|Y SW) \nonumber \\
	& = & \frac{1}{|\sfK|} \oI(X_\sfK;Y|SW)
\label{eq:A-min}
\end{eqnarray}
where (a) follows from (\ref{eq:HUS-fair}) with $Z=(Y,S,W)$.

Using (\ref{eq:A-min}) and (\ref{eq:P*}), we obtain
$\overline{\scrP}^*(\scrW_K^{\fair}) = \scrP^*(\scrW_K^{\fair})$. Hence
\begin{eqnarray*}
   \underline{\tilde{E}}_{\psp}(R,L,p_W, \tp_{S|W}, p_{X|SW},\scrW_K^{\fair})
	& = & \min_{\tp_{YX_{\sfK}|SW} \in \scrP^*(\scrW_K^{\fair})}
		D(\tp_{YX_{\sfK}|SW} \,\tp_{S|W}
		\| \tp_{Y|X_{\sfK}} \,p_{X|SW}^K \,p_S \,|\,p_W) \\
	& = & \overline{\tilde{E}}_{\psp}(R,L,p_W, \tp_{S|W}, p_{X|SW},\scrW_K^{\fair})
\end{eqnarray*}
and therefore
\[ E^{\all}(R,L,D_1,\scrW_K^{\fair},\Delta) = E^{\one}(R,L,D_1,\scrW_K^{\fair},\Delta) . \]

{\bf (vi). Positive Error Exponents.}
Consider any $\calW = \{1,\cdots,L\}$ and $p_W$ that is positive over its support set
(if it is not, reduce the value of $L$ accordingly.)
For any $\sfA \subseteq \sfK$, the divergence to be minimized in the expression
(\ref{eq:Epsp-A}) for
$\tilde{E}_{\psp,\sfA}(R,L,p_W,\tp_{S|W},p_{X|SW},\scrW_K)$ is zero if and only if 
\[ \tp_{YX_{\sfK}|SW} = \tp_{Y|X_{\sfK}} \,p_{X|SW}^{\sfK} 
	\quad \mathrm{and~} \tp_{S|W} = p_S . \]
These p.m.f.'s are feasible for (\ref{eq:set-A}) if and only if 
the resulting $I(X_{\sfA};YX_{\sfK\setminus\sfA}|SW) \le |\sfA|\,R$.
They are infeasible, and thus positive error exponents are guaranteed, if
\[ R < \min_{\sfA \subseteq \sfK}
	\frac{1}{|\sfA|} I(X_{\sfA};YX_{\sfK\setminus\sfA}|SW) .\]

From Part (iv) above, we may restrict our attention to $\scrW_K = \scrW_K^{\fair}$
under the detect-one criterion. Since the p.m.f. of $(S,W,X_{\sfK},Y)$ is permutation-invariant,
by application of (\ref{eq:I-fair}) we have
\begin{equation}
   \min_{\sfA \subseteq \sfK} \frac{1}{|\sfA|} I(X_{\sfA};YX_{\sfK\setminus\sfA}|SW)
	 = \frac{1}{K} I(X_{\sfK};Y|SW) .
\label{eq:I-min}
\end{equation}
Hence the supremum of all $R$ for error exponents are positive
is given by $\widetilde{C}^{\one}(D_1,\scrW_K)$ in (\ref{eq:C-one}) and is obtained
by letting $\Delta \to 0$ and $L \to \infty$.

For any $\scrW_K$, under the detect-all criterion,
the supremum of all $R$ for which error exponents are positive
is given by $\widetilde{C}^{\all}(D_1,\scrW_K)$ in (\ref{eq:C-all}) and
is obtained by letting $\Delta \to 0$ and $L \to \infty$. Since the optimal
p.m.f. is not necessarily permutation-invariant, (\ref{eq:I-min}) does not hold
in general. However, if $\scrW_K = \scrW_K^{\fair}$,
the same capacity is obtained for the detect-one and detect-all problems.
\hfill $\Box$


\section{Conclusion}
\label{sec:conclusion}

We have derived exact fingerprinting capacity formulas as opposed to bounds
derived in recent papers \cite{Somekh05,Somekh07,Barg08}, and constructed 
a universal fingerprinting scheme. A distinguishing feature of this new scheme
is the use of an auxiliary ``time-sharing'' randomized sequence $\bW$.
The analysis shows that optimal coalitions are fair and that capacity and
random-coding exponents are the same whether the problem is formulated as catching
one colluder or all of them.

Our study also allows us to reexamine previous fingerprinting system designs from a new angle.
First, randomization of the encoder via $\bW$ is generally needed because
the payoff function in the mutual-information game is
nonconcave with respect to $p_{X|S}$.
Thus capacity is obtained as the value of a mutual-information game with $p_{XW|S}$
as the maximizing variable. This has motivated the construction of our randomized fingerprinting
scheme, which may also be thought of as a generalization of Tardos' design \cite{Tardos03}.
Two other randomization methods are also fundamental: randomized permutation of user
indices to ensure that maximum error probability (over all possible coalitions) equals
average error probability; and randomized permutation of the letters $\{1,2,\cdots,N\}$
to cope with collusion channels with arbitrary memory.

Second, single-user decoders are simple but suboptimal. Such decoders
assign a score to each user based on his individual fingerprint and the received data,
and declare guilty those users whose score exceeds some threshold. While this is
a reasonable approach, performance can be improved by making joint decisions
about the coalition. Similarly, the fingerprinting schemes proposed in \cite{Tardos03}
and in much of the signal processing literature might be improved by adopting
a joint-decision principle, at the expense of increased decoding complexity.

Finally, several information-theoretic approaches to fingerprinting have been
studied in the two years since this paper was submitted for publication, including
work on spherical fingerprinting by the author \cite{Moulin09-ICASSP} and his coworkers
Wang \cite{Moulin09-ITA} and Jourdas \cite{Jourdas09}, on blind fingerprinting
\cite{Wang06,Wang08}, on binary fingerprinting under the Boneh-Shaw model
by Amiri and Tardos \cite{Amiri09},
Huang and Moulin \cite{Huang09,Huang09b}, and Furon and P\'{e}rez-Freire \cite{Furon09},
as well as research on two-level fingerprinting codes
by Anthapadmanabhan and Barg \cite{Barg09}.

Particularly noteworthy is \cite{Amiri09}, which presents a random coding scheme
closely related to ours, with a joint decoder
(improving on Tardos' earlier work \cite{Tardos03}) that maximizes a penalized empirical
mutual information criterion, similarly to Plotnik and Satt's universal decoder for
the random MAC \cite{Plotnik91}. Amiri and Tardos use ordinary empirical mutual information
instead of our empirical mutual information $\oI$ of $K$ variables. While both
choices are capacity-achieving, ours is geared towards obtaining better error exponents,
as is the case for the classical MAC decoding problem \cite{Liu96}. The paper
\cite{Amiri09} also outlines the proof of a converse theorem for the so-called
{\em weak fingerprinting model}, in which a helper discloses all colluders
except one to the decoder.

{\bf Acknowledgments.}
The author is very grateful to Dr.~Ying Wang for reading several drafts of this paper
and making comments and suggestions that have improved it. He also thanks 
Yen-Wei Huang, Dr.~Prasanth Anthapadmanabhan, Profs.~Barg and Tardos, and the anonymous reviewers
for helpful comments and corrections; and Prof.~Raymond Yeung and an anonymous reviewer of \cite{Wang08}
for bringing references \cite{Han78} and \cite{Plotnik91}, respectively, to our attention.

\newpage

\appendix
\renewcommand{\theequation}{\Alph{section}.\arabic{equation}}

\section{Proof of Lemma~\ref{lem:H-fair}}
\label{Sec:Proof-Lemma-Fair}
\setcounter{equation}{0}

Due to the permutation-invariant assumption
on the joint p.m.f. of $(X_{\sfK}, Z)$, it suffices the establish
(\ref{eq:HUY-fair}) for $\sfA = \{ 1, \cdots, k-1\}$ and $\sfB = \{ 1, \cdots, k\}$,
where $2 \le k \le K$. The claim then follows by induction over $k$.
Let $Z_k = (Z, X_{k+1}^N)$, hence $Z_{k-1} = (Z_k, X_k)$.
Then (\ref{eq:HUY-fair}) takes the form
\[ \frac{1}{k-1} H(X_1^{k-1}|Z, X_k^N) \le \frac{1}{k} H(X_1^k|Z, X_{k+1}^N) \]
or equivalently
\begin{equation}
   (k-1) H(X_1^k|Z_k) \ge k H(X_1^{k-1}|Z_k X_k) , \quad 2 \le k \le K.
\label{eq:H-recursive}
\end{equation}
And indeed the difference between left and right sides of (\ref{eq:H-recursive}) satisfies
\begin{eqnarray*}
   \lefteqn{(k-1) H(X_1^k|Z_k) - k H(X_1^{k-1}|Z_k X_k)} \\
	& = & (k-1) [ H(X_k|Z_k) + H(X_1^{k-1}|Z_k X_k)] - k H(X_1^{k-1}|Z_k X_k) \\
	& = & (k-1) H(X_k|Z_k) - H(X_1^{k-1}|Z_k X_k) \\
	& \stackrel{(a)}{=} & \sum_{i=1}^{k-1} H(X_i|Z_k) - H(X_1^{k-1}|Z_k X_k) \\
	& \stackrel{(b)}{\ge} & H(X_1^{k-1}|Z_k) - H(X_1^{k-1}|Z_k X_k) \\
	& = & I(X_1^{k-1};X_k|Z_k) \\
	& \stackrel{(c)}{\ge} & 0
\end{eqnarray*}
where (a) holds because the conditional p.m.f.'s $p_{X_i|Z_k}, \,1 \le i \le k$,
are identical due to the permutation invariance assumption. Inequalities (b)
and (c) hold with equality when $X_i , \,1 \le i \le k$, are conditionally
independent given $Z_k$.

Similarly, to establish (\ref{eq:HUS-fair}), it suffices to prove that
\begin{equation}
   (k-1) H(X_1^k|Z) \le k H(X_1^{k-1}|Z) .
\end{equation}
We have
\begin{eqnarray*}
   \lefteqn{(k-1) H(X_1^k|Z) - k H(X_1^{k-1}|Z)} \\
	& = & (k-1) [ H(X_1^{k-1}|Z) + H(X_k|Z,X_1^{k-1})] - k H(X_1^{k-1}|Z) \\
	& = & (k-1) H(X_k|Z,X_1^{k-1}) - H(X_1^{k-1}|Z) \\
	& \stackrel{(a)}{=} & \sum_{i=1}^{k-1} H(X_i|Z,X_1^{i-1},X_{i+1}^k) - H(X_1^{k-1}|Z) \\
	& \stackrel{(b)}{=} & \sum_{i=1}^{k-1} H(X_i|Z,X_1^{i-1},X_{i+1}^k)
									- \sum_{i=1}^{k-1} H(X_i|Z,X_1^{i-1}) \\
	& = & - \sum_{i=1}^{k-1} I(X_i;X_{i+1}^k|Z,X_1^{i-1}) \\
	& \le & 0
\end{eqnarray*}
where in (a) we have used the permutation invariance of the distribution of $X_1^k$,
and in (b) the chain rule for entropy.
\hfill $\Box$

\section{Proof of Lemma~\ref{lem:memoryless}}
\label{sec:Proof-Lemma-memoryless}
\setcounter{equation}{0}

The derivation below is given in terms of the detect-one criterion
but applies straightforwardly to the detect-all criterion as well.
Denote by $C_{\memoryless}^{\one}(D_1,\scrW_K)$ the compound capacity
under the detect-one criterion. 
To prove the claim
\begin{equation}
  C^{\one}(D_1,\scrW_K) \le C_{\memoryless}^{\one}(D_1,\scrW_K) ,
\label{eq:memoryless-LB}
\end{equation}
it suffices to identify a family of collusion channels satisfying 
the almost-sure fidelity constraint (\ref{eq:WK})
and for which reliable decoding is impossible at rates above
$C_{\memoryless}^{\one}(D_1,\scrW_K)$.
For any $\bx_\calK$, consider the class
\begin{equation}
   \scrW_K^{\epsilon}(p_{\bx_\calK}) \triangleq \left\{ \tp_{Y|X_\sfK} \in \scrP_{Y|X_\sfK}
		~:~ \min_{p_{Y|X_\sfK} \in \scrW_K(p_{\bx_\calK})} \max_{x_\sfK,y}
		\;|\tp_{Y|X_\sfK}(y|x_\sfK) - p_{Y|X_\sfK}(y|x_\sfK)| \le \epsilon \right\} ,
			\quad \epsilon \ge 0 ,
\label{eq:WK-epsilon}
\end{equation}
which is slightly larger than $\scrW_K(p_{\bx_\calK})$ but shrinks towards
$\scrW_K(p_{\bx_\calK})$ as $\epsilon \downarrow 0$.  Continuity of mutual
information and the mapping $\scrW_K(\cdot)$ with respect to variational distance
(per (\ref{eq:WK-continuous}) implies that
\begin{equation}
  C^{\one}(D_1,\scrW_K^{\epsilon}) \uparrow C^{\one}(D_1,\scrW_K) 
	\quad \mathrm{as~} \epsilon \downarrow 0 .
\label{eq:WKeps}
\end{equation}

We now claim that if the coalition selects a {\em memoryless channel}
$p_{Y|X_\sfK} \in \scrW_K(p_{\bx_\calK})$, the constraint
$p_{\by|\bx_\calK} \in \scrW_K^{\epsilon}(p_{\bx_\calK})$
is satisfied with probability approaching 1 as $N \to \infty$: 
\begin{equation}
   \forall \epsilon > 0 \;\exists N_0(\epsilon) ~:\quad 
	Pr[p_{\by|\bx_\calK} \in \scrW_K^{\epsilon}(p_{\bx_\calK})] \ge 1-\epsilon 
	\quad \forall N > N_0(\epsilon) .
\label{eq:p-ptilde}
\end{equation}
To show this, define the set
\[ \calE = \left\{ \bx_\sfK ~:~ \min_{x_\sfK \in \calX^K} \,p_{\bx_\sfK}(x_\sfK)
	\ge \epsilon |\calX|^{-K} \right\} . \]
Without loss of generality \footnote{
  One may always ``fill in'' each codeword $\bx_m$ with $2\epsilon |\calX|^{-K} N$
  dummy symbols drawn from the uniform p.m.f. on $\calX$ to ensure that (\ref{eq:good-xK})
  holds. The rate loss due to the ``fill-in'' symbols vanishes as $\epsilon \to 0$.
  }
assume $f_N$ is such that
\begin{equation}
  Pr[\bx_\calK \in \calE] \ge 1-\epsilon/2
\label{eq:good-xK}
\end{equation}
where the probability is taken with respect to $M_\sfK, \bS, V$.
For any $\bx_\sfK \in \calE$, $x_\sfK \in \calX^K$, $y \in \calY$, if $\by$ is generated
conditionally i.i.d. $p_{Y|X_\sfK}$, the random variable $p_{\by|\bx_\sfK}(y|x_\sfK)$  
converges in probability to $p_{Y|X_\sfK}(y|x^K)$ as $N \to \infty$. Hence
\begin{equation}
   P_{\bY|\bX_\sfK=\bx_\sfK} \left[\,\max_{x_\sfK,y}
	\,|p_{\by|\bx_\sfK}(y|x_\sfK) - p_{Y|X_\sfK}(y|x_\sfK)|
	\,\le \epsilon \right] \ge 1-\epsilon/2 , \quad \forall \bx_\sfK \in \calE
\label{eq:good-yxK}
\end{equation}
for any $N > N_0(\epsilon)$. Combining (\ref{eq:good-xK})
and (\ref{eq:good-yxK}), we obtain (\ref{eq:p-ptilde}).

A lower bound on error probability is obtained when a helper provides some
information to the decoder. Assume the constraint on the coalition is slightly
relaxed so that they are allowed to produce pirated copies that violate
the constraint $p_{\by|\bx_\sfK} \in \scrW_K^{\epsilon}(p_{\bx_\calK})$
with probability at most $\epsilon$, as in (\ref{eq:p-ptilde}).
In this event, the helper reveals the entire coalition to the decoder.
This contributes at most $\epsilon KNR$ bits of information
to the decoder and does not increase the decoder's error probability. 
Hence
\[ C^{\one}(D_1,\scrW_K^\epsilon) + \epsilon K \le C_{\memoryless}^{\one}(D_1,\scrW_K) . \]
Combining this inequality with (\ref{eq:WKeps}) establishes (\ref{eq:memoryless-LB}).
\hfill $\Box$

\section{Proof of (\ref{eq:zeta-zetahat})}
\label{sec:zeta}
\setcounter{equation}{0}

The quantities $\hat{\zeta}(\bY)$ and $\zeta$ are defined in (\ref{eq:zeta-hat}) and (\ref{eq:zeta-UB}),
respectively. We first analyze
\begin{eqnarray}
   \lefteqn{Pr \left[ \mathds1 \left\{ \hat{D}_{ijk}(\bY) \le \frac{3\delta^2}{4} \right\} > \mathds1\{ D_{ijk} \le \delta^2 \} 
			~|~ \bS=\bs, V=v, \calK=\{i,j\} \right]} \nonumber \\
	& = & Pr \left[ \hat{D}_{ijk}(\bY) \le \frac{3\delta^2}{4} \;\mathrm{and}\; D_{ijk} > \delta^2 
			~|~ \bS=\bs, V=v, \calK=\{i,j\} \right] \nonumber \\
	& \le & Pr \left[ \hat{D}_{ijk}(\bY) < D_{ijk} - \frac{\delta^2}{4} ~|~ \bS=\bs, V=v, \calK=\{i,j\} \right]
\label{eq:D-Dhat}
\end{eqnarray}
for any $k \in \calM_N^{\calA-\good}(\bs,v,i,\delta)$. The shorthand $Pr$ denotes the probability
distribution on $\hat{D}_{ijk}(\bY)$ induced by the conditional distribution 
$p_{Y|X_1 X_2}^N(\bY | \bx_i(\bs,v), \bx_j(\bs,v))$.
Conditioned on $\bS=\bs, V=v, \calK=\{i,j\}$, the normalized loglikelihood $\hat{D}_{ijk}(\bY)$
of (\ref{eq:Dhat-ijk}) is the average of $N$ independent random variables. We show that
$\hat{D}_{ijk}(\bY)$ converges in probability (and exponentially with $N$) to its expectation $D_{ijk}$
of (\ref{eq:D-ijk}).

We may write $\hat{D}_{ijk}(\bY)$ as a function of the joint type $p_{\bY\bx_i\bx_j\bx_k}$
of the quadruple $(\bY, \bx_i(\bs,v), \bx_j(\bs,v), \bx_k(\bs,v))$:
\begin{equation}
   \hat{D}_{ijk}(\bY) = D(p_{\bY\bx_i\bx_j\bx_k}) \triangleq
	\sum_{y,x_1,x_2,x_2'} p_{\bY\bx_i\bx_j\bx_k}(y,x_1,x_2,x_2') 
	\log \frac{p_{Y|X_1 X_2}(y|x_1,x_2)}{p_{Y|X_1 X_2}(y|x_1,x_2')} .
\label{eq:Dhat-ijk-app}
\end{equation}
Similarly, from (\ref{eq:D-ijk}) we obtain
\begin{equation}
   D_{ijk} = D(p_{\bx_i\bx_j\bx_k}) \triangleq
	\sum_{y,x_1,x_2,x_2'} p_{\bx_i\bx_j\bx_k}(x_1,x_2,x_2') \,p_{Y|X_1 X_2}(y|x_1, x_2)
	\log \frac{p_{Y|X_1 X_2}(y|x_1,x_2)}{p_{Y|X_1 X_2}(y|x_1,x_2')} .
\label{eq:Dijk-app}
\end{equation}
Subtracting (\ref{eq:Dijk-app}) from (\ref{eq:Dhat-ijk-app}) yields
\begin{eqnarray*}
   \hat{D}_{ijk}(\bY) - D_{ijk}
	& = & \sum_{y,x_1,x_2,x_2'} p_{\bx_i\bx_j\bx_k}(x_1,x_2,x_2')
		[p_{\bY|\bx_i\bx_j\bx_k}(y|x_1,x_2,x_2') - p_{Y|X_1 X_2}(y|x_1,x_2)] \\
	& & \qquad \times \log \frac{p_{Y|X_1 X_2}(y|x_1,x_2)}{p_{Y|X_1 X_2}(y|x_1,x_2')} \\
	& = & \sum_{y,x_1,x_2,x_2'} U(y,x_1,x_2,x_2')
		\,p_{Y|X_1 X_2}(y|x_1,x_2) \,\log \frac{p_{Y|X_1 X_2}(y|x_1,x_2)}{p_{Y|X_1 X_2}(y|x_1,x_2')}
\end{eqnarray*}
which is a linear combination of the random variables 
\[ U(y,x_1,x_2,x_2') \triangleq p_{\bx_i\bx_j\bx_k}(x_1,x_2,x_2')
		\,\left[\frac{p_{\bY|\bx_i\bx_j\bx_k}(y|x_1,x_2,x_2')}{p_{Y|X_1 X_2}(y|x_1,x_2)} - 1 \right] . \]
Note that for each $x_1,x_2,x_2'$, the minimum of $U(y,x_1,x_2,x_2')$ over $y\in\calY$ is nonpositive.

Owing to (\ref{eq:WK-delta}), we also have
\begin{eqnarray*}
   \sum_y p_{Y|X_1 X_2}(y|x_1,x_2) \,\log \frac{p_{Y|X_1 X_2}(y|x_1,x_2)}{p_{Y|X_1 X_2}(y|x_1,x_2')} 
	& = & D(p_{Y|X_1=x_1, X_2=x_2} \| p_{Y|X_1=x_1, X_2=x_2'}) \\
	& \in & [\delta, \;\log \delta^{-1} ] . 
\end{eqnarray*}
Hence
\begin{equation}
   \hat{D}_{ijk}(\bY) - D_{ijk} \ge (\log \delta^{-1}) \min_{y,x_1,x_2,x_2'} U(y,x_1,x_2,x_2') .
\label{eq:D-Dhat-UB}
\end{equation}
In the sequel we omit the conditioning on $\bs,v,i,j$, for conciseness of notation.
We bound (\ref{eq:D-Dhat}) by
\begin{eqnarray}
   Pr \left[ \hat{D}_{ijk}(\bY) < D_{ijk} - \frac{\delta^2}{4} \right]
	& \stackrel{(a)}{\le} &  Pr \left[ \min_{y,x_1,x_2,x_2'} U(y,x_1,x_2,x_2') 
			< - \frac{\delta^2}{4\log \delta^{-1}} \right] \nonumber \\
	& \stackrel{(b)}{\le} & |\calX|^3 |\calY| \,\max_{y,x_1,x_2,x_2'} Pr \left[ U(y,x_1,x_2,x_2')
			< - \epsilon \right]
\label{eq:Dhat-D-UB}
\end{eqnarray}
where (a) follows from (\ref{eq:D-Dhat-UB}); and in (b) we have used the union bound and the shorthand
$\epsilon = \frac{\delta^2}{4\log \delta^{-1}}$.
Denote by $D_b(\alpha\|p) = \alpha \ln \frac{\alpha}{p} +  (1-\alpha) \ln \frac{1-\alpha}{1-p},\;0 < \alpha < 1$,
the large-deviations function for the Bernoulli random variable with probability $p$.
Note that $D_b((1-\epsilon)p\|p) \sim \frac{p^2 \epsilon^2}{2(1-p)}$ as $\epsilon \downarrow 0$
and that $\frac{p^2}{1-p} > \frac{\delta^2}{1-\delta}$ for all $\delta < p < 1-\delta$.
Define $f(\delta) = \min_{\delta \le p \le 1-\delta} D_b((1-\epsilon)p\|p)$.
We have
\begin{equation}
   f(\delta) \sim D_b((1-\epsilon)\delta\|\delta ) \sim \frac{\epsilon^2 \delta^2}{2(1-\delta)} 
	= \frac{\delta^6}{32(1-\delta)\log^2 \delta^{-1}} \gg \delta^7 \quad \mathrm{as~} \delta \downarrow 0
\label{eq:f-delta}
\end{equation}
hence there exists $\delta^* > 0$ such that $f(\delta) > \delta^7$ for all $0 < \delta < \delta^*$.

Define the shorthand $\beta = p_{\bx_i\bx_j\bx_k}(x_1,x_2,x_2') \in [0,1]$.
For each $i,j,k$ and each $y,x_1,x_2,x_2'$, the count
\[  \beta N p_{\bY|\bx_i\bx_j\bx_k}(y|x_1,x_2,x_2') = N p_{\bY\bx_i\bx_j\bx_k}(y,x_1,x_2,x_2') 
	= \sum_{t=1}^N \mathds1\{Y_t=y, x_{it}=x_1,x_{jt}=x_2,x_{kt}=x_3\} \]
is a binomial random variable  with $\beta N$ trials and probability
$p \triangleq p_{Y|X_1 X_2}(y|x_1,x_2) \in [\delta, \,1-\delta]$.
By (\ref{eq:WK-delta}), we have $\delta \le p \le 1-\delta$.
Next
\begin{eqnarray*}
  Pr[U(y,x_1,x_2,x_2') < - \epsilon]
	& = & Pr \left[ \beta \,\left( \frac{p_{\bY|\bx_i\bx_j\bx_k}(y|x_1,x_2,x_2')}{p_{Y|X_1 X_2}(y|x_1,x_2)} - 1 \right)
		< - \epsilon \right] \\
	& = & Pr\left[\frac{\mathrm{Bi}(\beta N,p)}{Np} - \beta < -\epsilon \right] .
\end{eqnarray*}
For $\beta \le \epsilon$, this probability is zero. For $\epsilon < \beta \le 1$, we have
\begin{eqnarray}
  Pr[U(y,x_1,x_2,x_2') < - \epsilon]
	& = & Pr \left[ \beta \,\left( \frac{p_{\bY|\bx_i\bx_j\bx_k}(y|x_1,x_2,x_2')}{p_{Y|X_1 X_2}(y|x_1,x_2)} - 1 \right)
		< - \epsilon \right] \\
	& = & Pr[\mathrm{Bi}(\beta N,p) < \beta N(1-\epsilon/\beta) p] \nonumber \\
	& \stackrel{(a)}{\le}  & 2^{-N \beta D_b((1-\epsilon/\beta) p \| p) } \nonumber \\
	& \stackrel{(b)}{\le} & 2^{-N D_b((1-\epsilon) p \| p) } \nonumber \\
	& \stackrel{(c)}{\le} & 2^{-N f(\delta)} \nonumber \\
	& \stackrel{(d)}{<} & 2^{-N \delta^7} \quad \forall \delta < \delta^*
\label{eq:py-ratio-bounds}
\end{eqnarray}
where (a) holds by definition of the large-deviations function $D_b$; (b) holds by convexity
of the function $D_b(\cdot\|p)$: for all $\epsilon' = \epsilon/\beta \in [\epsilon, 1)$,
we have $D_b((1-\epsilon')p\|p) \ge (\epsilon'/\epsilon) D_b((1-\epsilon)p\|p)$ with equality if $\epsilon'=\epsilon$,
i.e., $\beta=1$;(c) holds by (\ref{eq:f-delta}); and (d) holds by
the lower bound on $f(\delta)$. Combining (\ref{eq:Dhat-D-UB}) and (\ref{eq:py-ratio-bounds}), we conclude
that (\ref{eq:D-Dhat}) is upper-bounded by an exponentially vanishing function of $N$
for each $\delta < \delta^*$:
\begin{eqnarray}
   \forall i,j,k :\;\; Pr \left[ \mathds1 \left\{ \hat{D}_{ijk}(\bY) \le \frac{3\delta^2}{4} \right\} 
			> \mathds1\{ D_{ijk} \le \delta^2 \} ~|~ \bS=\bs, V=v, \calK=\{i,j\} \right] 
	\le \,2^{-N\delta^7} .
\label{eq:C1-bound}
\end{eqnarray}
\vspace*{-0.2in}

This does not immediately imply that $\hat{\zeta}(\bY) \le \zeta$ with probability approaching 1 
because the definition of $\hat{\zeta}(\bY)$ in (\ref{eq:zeta-hat}) involves potentially exponentially
many terms $\hat{D}_{ijk}(\bY)$. However 
\begin{eqnarray*}
   \lefteqn{Pr[\hat{\zeta}(\bY) > \zeta ~|~ \bS=\bs, V=v, \calK=\{i,j\}]} \nonumber \\
    & = & Pr \left[ \sum_k \mathds1 \left\{ \hat{D}_{ijk}(\bY) \le \frac{3\delta^2}{4} \right\} 
			> \sum_k \mathds1\{ D_{ijk} \le \delta^2 \} ~|~ \bS=\bs, V=v, \calK=\{i,j\} \right] \\
	& \le & Pr \left[ \exists p_{\bx_i\bx_j\bx_k} ~:~ \mathds1 \left\{ \hat{D}_{ijk}(\bY) \le \frac{3\delta^2}{4} \right\} 
			> \mathds1\{ D_{ijk} \le \delta^2 \} ~|~ \bS=\bs, V=v, \calK=\{i,j\} \right] \\
	& \stackrel{(a)}{\le} & (N+1)^{|\calX|^3} \,\max_{p_{\bx_i\bx_j\bx_k}}
			Pr \left[ \mathds1 \left\{ \hat{D}_{ijk}(\bY) \le \frac{3\delta^2}{4} \right\} 
			> \mathds1\{ D_{ijk} \le \delta^2 \} ~|~ \bS=\bs, V=v, \calK=\{i,j\} \right] \\
	& \stackrel{(b)}{\le} & |\calX|^3 |\calY| \,(N+1)^{|\calX|^3} 
			\,2^{-N\delta^7} \rightarrow 0 \quad \mathrm{~as~} N \to \infty
\end{eqnarray*}
where (a) follows from the union bound and the fact that the number of joint types $p_{\bx_i\bx_j\bx_k}$
is at most $(N+1)^{|\calX|^3}$, and (b) from (\ref{eq:D-Dhat}) and (\ref{eq:C1-bound}).
This establishes (\ref{eq:zeta-zetahat}).
\hfill $\Box$

\section{Proof of (\ref{eq:Pc-bad-sv-asym})}
\label{sec:Pc-bad}
\setcounter{equation}{0}


\begin{lemma}
There exists a partition $\{\widetilde{\calM}_i\}_{i\in\calI}$ of $\calM_N^{\bad}(\bs,v,\delta)$
with the following properties:
\begin{description}
\item[(P1)] $\forall i\in\calI, \;\forall j \in \widetilde{\calM}_i :\; d_H(\bx_i(\bs,v), \bx_j(\bs,v)) \le 2N\delta$;
\item[(P2)] $\forall i\in\calI :\;|\widetilde{\calM}_i| \ge 2^{N3\sqrt{\delta}}$.
\end{description}
\label{lem:genie}
\end{lemma}
{\em Proof}. By assumption, $|\calM_N^{\bad}(\bs,v,\delta)| \ge 2^{NR} (1- 2^{-N\delta^2/3})$.
The index set $\calI$ and the sets $\{\widetilde{\calM}_i\}_{i\in\calI}$ 
are constructed iteratively as follows. Denote by $i$ the smallest index
in $\calM_N^{\bad}(\bs,v,\delta)$ and initialize $\calI = \{i\}$ and
$\widetilde{\calM}_i = \calM_i(\bs,v,\delta)$.
By the definition (\ref{eq:Mj}), $\widetilde{\calM}_i$
satisfies $d_H(\bx_i(\bs,v), \bx_j(\bs,v)) \le N\delta$ for all $j \in \widetilde{\calM}_i$,
hence Property (P1) holds. Also, owing to (\ref{eq:M-bad}), Property (P2) holds as well.
Next, find the smallest $i \in \calM_N^{\bad}(\bs,v,\delta)$ such that
$d_H(\bx_j(\bs,v), \bx_i(\bs,v)) > 2N\delta$ for all $j\in\calI$, and update $\calI \leftarrow \calI \cup \{i\}$.
By the triangle inequality, the sets $\{\widetilde{\calM}_i\}_{i\in\calI}$ are disjoint.
Repeat this operation till no such $i$ can be found. At this point, the set $\calI$ is fixed,
and each remaining codeword index $j \notin \cup_{i\in\calI} \widetilde{\calM}_i$, satisfies
$d_H(\bx_j(\bs,v), \bx_i(\bs,v)) \le 2N\delta$ for some $i\in\calI$.
Assign the index $j$ of this codeword to $\widetilde{\calM}_i$; ties can be broken arbitrarily.
Properties (P1) and (P2)  of the set $\widetilde{\calM}_i$ are preserved. Repeat this operation till
all the codeword indices  in $\calM_N^{\bad}(\bs,v,\delta)$ are exhausted. Upon completion of this process,
the sets $\{\widetilde{\calM}_i\}_{i\in\calI}$ form a partition of $\calM_N^{\bad}(\bs,v,\delta)$
and satisfy (P1) and (P2).
\hfill $\Box$

Assume that $\calK = \{i,j\} \in (\calM_N^{\bad}(\bs,v,\delta))^2$.
Consider the partition of Lemma~\ref{lem:genie} and a genie (helper) that reveals
the two ``clusters'' of indices $\widetilde{\calM}_{i^*}$ and $\widetilde{\calM}_{j^*}$ 
($i^*,j^*\in\calI$) to which $i$ and $j$ respectively belong. 
Thanks to the genie, we can enlarge the decoding regions, obtaining
$\{ \calD_m'(\bs,v), \,m \in \widetilde{\calM}_{i^*} \cup \widetilde{\calM}_{j^*} \}$
that are at least as large as the original decoding regions ---
$\calD_m(\bs,v) \subseteq \calD_m'(\bs,v)$ --- and form a partition of $\calY^N$. 
The conditional probability that $\bY$ is typical and that correct decoding occurs
(given $\bs,v, i\in\widetilde{\calM}_{i^*}, j\in\widetilde{\calM}_{j^*}$)
for the original decoder and for the genie-aided decoder are respectively given by
$\underline{P}_c(i,j|\bs,v)$ in (\ref{eq:Pc_ijsv}) and by
\[ \underline{P}_c'(i,j|\bs,v,i^*,j^*) \triangleq 
	\sum_{\by \in T_\delta(\bs,v,i,j) \cap (\calD_i'(\bs,v) \cup \calD_j'(\bs,v))} 
	p_{Y|X_1 X_2}^N (\by|\bx_i(\bs,v),\bx_j(\bs,v)) . 
\]
Since $\calD_m(\bs,v) \subseteq \calD_m'(\bs,v)$ for all $m$, we have
\begin{equation}
   \underline{P}_c(i,j|\bs,v) \le  \underline{P}_c'(i,j|\bs,v,i^*,j^*) ,
	\quad \forall i\in\widetilde{\calM}_{i^*}, \;j\in\widetilde{\calM}_{j^*} .
\label{eq:Pc'vsPc}
\end{equation}
The average of the right side over $i \in \widetilde{\calM}_{i^*}$ and $j \in \widetilde{\calM}_{j^*}$
is denoted by
\begin{eqnarray} 
   \underline{P}_c'(\bs,v,i^*,j^*) & \triangleq &  \frac{1}{|\widetilde{\calM}_{i^*}| \,|\widetilde{\calM}_{j^*}|} 
		\sum_{i\in\widetilde{\calM}_{i^*}} \sum_{j\in\widetilde{\calM}_{j^*}} \underline{P}_c'(i,j|\bs,v,i^*,j^*) .
\label{eq:Pc'-i*j*}
\end{eqnarray}
Let $(i^{**}, j^{**})$ achieve the maximum of $\underline{P}_c'(\bs,v,i^*,j^*)$ over $(i^*,j^*)$, and denote by
$\calC_1 = \widetilde{\calM}_{i^{**}}$ and $\calC_2 = \widetilde{\calM}_{j^{**}}$
the corresponding clusters of indices.

Analogously to Step~3, define the random variables $X_i = x_{iT}(\bS,V), \,i\in\calM_N$,
where $T$ is uniformly distributed over $\{1,2,\cdots,N\}$ and independent of all other random
variables. Define $X$ and $X'$ drawn uniformly and independently from the sets $\{ X_i, \,i\in\calC_1\}$ and
$\{ X_j, \,j\in\calC_2\}$ respectively.
The definitions and derivations in Steps~3 and 4 carry, with $\calC_1 \times \calC_2$
in place of $(\calM_N^{\good}(\bs,v,\delta))^2$. In particular, (\ref{eq:pYV-0}) becomes
\begin{equation}
   p_{Y_t|\bS V}(y|\bs,v) = \frac{1}{|\calC_1| \,|\calC_2|} \sum_{i\in\calC_1} \sum_{j\in\calC_2}
	p_{Y|X_1 X_2} (y|x_{it}(\bs,v),x_{jt}(\bs,v)) .
\label{eq:pYt-bad}
\end{equation}
We again use the reference conditional distribution (\ref{eq:r}), repeated below for convenience:
\begin{equation}
   r(\by|\bs,v) \triangleq \prod_{t=1}^N p_{Y_t|\bS V}(y_t|\bs,v) .
\label{eq:r-bad}
\end{equation}

For each $i,k \in \calC_1$ and $j,l \in \calC_2$,
it follows from the triangle inequality that
\[ d_H(\bx_i(\bs,v), \bx_k(\bs,v)) \le 4N\delta \quad \mathrm{and} \quad
	d_H(\bx_j(\bs,v), \bx_l(\bs,v)) \le 4N\delta . \]
Hence there are at most $8N\delta$ positions $t$ at which 
$(x_{it}(\bs,v),x_{jt}(\bs,v)) \ne (x_{kt}(\bs,v),x_{lt}(\bs,v))$. Therefore,
owing to (\ref{eq:WK-delta}), the Kullback-Leibler divergence between the distributions of $\bY$
conditioned on codeword pairs $(i,j)$ and $(k,l)$ respectively, satisfies
\begin{equation}
   D_{ijkl} \triangleq \frac{1}{N} \sum_{t=1}^N D(p_{Y|X_1=x_{it}(\bs,v),X_2=x_{jt}(\bs,v)} 
	\| p_{Y|X_1=x_{kt}(\bs,v),X_2=x_{lt}(\bs,v)}) 
	\le 8\delta \log \delta^{-1} .
\label{eq:Dijkl}
\end{equation}
Hence the conditional self-information of (\ref{eq:Iij-sv}) is
\begin{eqnarray*}
   \theta_{ij}(\bs,v)
   & \triangleq & \frac{1}{N} \sum_{t=1}^N D(p_{Y|X_1=x_{it}(\bs,v),X_2=x_{jt}(\bs,v)} \| p_{Y_t|\bS=\bs, V=v}) \\
   & \stackrel{(a)}{\le} & \frac{1}{|\calC_1| \,|\calC_2|} \sum_{(k,l) \in \calC_1 \times \calC_2} D_{ijkl} \\
   & \stackrel{(b)}{\le} & 8\delta \log \delta^{-1}
\end{eqnarray*}
where (a) holds by (\ref{eq:pYt-bad}) and convexity of the Kullback-Leibler divergence,
and (b) from (\ref{eq:Dijkl}). The average of the self-information $\theta_{ij}(\bs,v)$
over all $(i,j) \in \calC_1 \times \calC_2$ is the conditional mutual information
\begin{eqnarray*}
   \lefteqn{I_{p_T \,p_{X_1|\bS VT} \,p_{X_2|\bS VT} \,p_{Y|X_1 X_2}} (X_1 X_2;Y|\bS=\bs,V=v,T)} \\
	& = & I(\bs,v) \triangleq \frac{1}{|\calC_1| \,|\calC_2|} \sum_{(i,j) \in \calC_1 \times \calC_2} \theta_{ij}(\bs,v) 
	\le 8\delta \log \delta^{-1} .
\end{eqnarray*}

Analogously to Step~5, define the typical sets
\begin{equation}
   T_\delta(\bs,v,i,j) \triangleq \left\{ \by \in \calY^N ~:~ \underbrace{\frac{1}{N} \sum_{t=1}^N \log 
	\frac{p_{Y|X_1 X_2}(y_t|x_{it}(\bs,v),x_{jt}(\bs,v))}{p_{Y_t|\bS V}(y_t|\bs,v)} }_{\hat{\theta}_{ij}(\bs,v)}
	< 9\delta \log \delta^{-1} \right\} .
\label{eq:Ueps}
\end{equation}
The random variable $\hat{\theta}_{ij}(\bs,v)$ above is the average of $N$ conditionally independent
random variables (given $\bs,v$) and converges in probability to its mean $\theta_{ij}(\bs,v)
\le 8\delta \log \delta^{-1}$. Similarly to (\ref{eq:Pr-Aeps}), we have
\begin{equation}
   Pr[\bY \notin T_\delta(\bs,v,i,j)|\bS=\bs,V=v,\calK=\{i,j\}]
		\le \frac{1}{N\delta^2} , \quad \forall \bs, v,i,j
\label{eq:Pr-Ueps}
\end{equation}
which vanishes as $N \to \infty$.

Analogously to  (\ref{eq:Pc_good_sv}), we define
\begin{equation}
    \underline{P}_c^{\bad}(\bs,v) 
		\triangleq Pr[\mathrm{correct~decoding~and~} \bY \in T_\delta(\bs,v,\calK)
				\,|\bS=\bs,V=v, \calK \in (\calM_N^{\bad}(\bs,v,\delta))^2] . 
\label{eq:Pc_bad_sv}
\end{equation}
We have
\vspace*{-0.1in}
\begin{eqnarray}
   \underline{P}_c^{\bad}(\bs,v) 
    & = & \frac{1}{|\calM_N^{\bad}(\bs,v,\delta)|^2} \sum_{i,j\in\calM_N^{\bad}(\bs,v,\delta)} 
		\underline{P}_c(i,j|\bs,v) \nonumber \\
    & \stackrel{(a)}{=} & \frac{1}{|\calM_N^{\bad}(\bs,v,\delta)|}
		\sum_{i^*,j^*\in\calI} \sum_{i\in\widetilde{\calM}_{i^*}} 
		\sum_{j\in\widetilde{\calM}_{j^*}} \underline{P}_c(i,j|\bs,v) \nonumber \\
    & \stackrel{(b)}{\le} & \frac{1}{|\calM_N^{\bad}(\bs,v,\delta)|^2} 
		\sum_{i^*,j^*\in\calI} \sum_{i\in\widetilde{\calM}_{i^*}} 
		\sum_{j\in\widetilde{\calM}_{j^*}} \underline{P}_c'(i,j|\bs,v,i^*,j^*) \nonumber \\
	& \stackrel{(c)}{=} & \frac{1}{|\calM_N^{\bad}(\bs,v,\delta)|^2} \sum_{i^*,j^*\in\calI} |\widetilde{\calM}_{i^*}|
		\,|\widetilde{\calM}_{j^*}| \,\underline{P}_c'(\bs,v,i^*,j^*) \nonumber \\
	& \stackrel{(d)}{\le} & \max_{i^*,j^* \in \calI} \underline{P}_c'(\bs,v,i^*,j^*) \nonumber \\
	& = & \frac{1}{|\calC_1| \,|\calC_2|} \sum_{i\in\calC_1} \sum_{j\in\calC_2} 
		\sum_{\by \in T_\delta(\bs,v,i,j) \cap (\calD_i'(\bs,v) \cup \calD_j'(\bs,v))} 
		p_{Y|X_1 X_2}^N (\by|\bx_i(\bs,v),\bx_j(\bs,v)) \nonumber \\
  & \stackrel{(e)}{\le} & \frac{2^{N 9\delta \log \delta^{-1}}}{|\calC_1| \,|\calC_2|} 
	\sum_{i\in\calC_1} \sum_{j\in\calC_2}
	\sum_{\by \in T_\delta(\bs,v,i,j) \cap (\calD_i'(\bs,v) \cup \calD_j'(\bs,v))} r(\by|\bs,v) \nonumber \\
   & \stackrel{(f)}{\le} & \frac{2^{N 9\delta \log \delta^{-1}}}{|\calC_1| \,|\calC_2|} 
	\sum_{i\in\calC_1} \sum_{j\in\calC_2} \,\sum_{\by \in \calD_i'(\bs,v) \cup \calD_j'(\bs,v)} r(\by|\bs,v) \nonumber \\
   & \stackrel{(g)}{=} & \frac{2^{N 9\delta \log \delta^{-1}}}{|\calC_1| \,|\calC_2|} 
	\sum_{i\in\calC_1} \sum_{j\in\calC_2} \left[ \sum_{\by \in \calD_i'(\bs,v)} r(\by|\bs,v) 
	+ \sum_{\by \in \calD_j'(\bs,v)} r(\by|\bs,v) \right] \nonumber \\
   & = & \frac{2^{N 9\delta \log \delta^{-1}}}{|\calC_1| \,|\calC_2|}
	(|\calC_2| + |\calC_1|) \nonumber \\
   & \stackrel{(h)}{\le} & 2^{N [9\delta \log \delta^{-1} - 3\sqrt{\delta}] + 1} \nonumber \\
   & \le & 2^{-N\sqrt{\delta} \,+1} , \quad \forall \delta < \frac{1}{4000}
\label{eq:Pcsv-bad-UB}
\end{eqnarray}
where (a) and (d) hold because $\{\widetilde{\calM}_i\}_{i\in\calI}$ form a partition of 
$\calM_N^{\bad}(\bs,v,\delta)$,
(b) because of (\ref{eq:Pc'vsPc}),
(c) because of (\ref{eq:Pc'-i*j*}),
(e) follows from (\ref{eq:r-bad}) and (\ref{eq:Ueps}), 
(f) is obtained by dropping the restriction $\by \in T_\delta(\bs,v,i,j)$,
(g) holds because the decoding regions $\calD_i'(\bs,v)$ and $\calD_j'(\bs,v)$ are disjoint,
and (h) because $|\calC_1| , \,|\calC_2| \ge 2^{N3\sqrt{\delta}}$. 
Thus
\vspace*{-0.15in}
\begin{eqnarray}
  P_c^{\bad}(\bs,v) & \stackrel{(a)}{\le} & Pr[\bY \notin T_\delta(\bs,v,i,j)|\bS=\bs,V=v,\calK=\{i,j\}] 
		+ \underline{P}_c^{\bad}(\bs,v) \nonumber \\ 
	& \stackrel{(b)}{\le} & \frac{1}{N\delta^2} + 2^{-N\sqrt{\delta} \,+1} 
\label{eq:Pc_bad_sv_2}
\end{eqnarray}
where (a) follows from (\ref{eq:Pc-bad-sv}) and (\ref{eq:Pc_bad_sv}),
and (b) from (\ref{eq:Pr-Ueps}) and (\ref{eq:Pcsv-bad-UB}).
Hence $P_c^{\bad}(\bs,v)$ vanishes for all $R > 0$, all $(\bs,v)$, and all $p_{Y|X_1 X_2}$.
Moreover 
\vspace*{-0.1in}
\[ P_c^{\bad}(\bs,v) < \frac{2}{N\delta^2} \]
for all $\delta < 1/4000$, and this establishes (\ref{eq:Pc-bad-sv-asym}).
\hfill $\Box$


\newpage

\begin{thebibliography}{99}

\bibitem{Moulin02} P.~Moulin and A.~Briassouli, ``The Gaussian Fingerprinting Game,''
	{\em Proc. Conf. Information Sciences and Systems}, Princeton, NJ, March~2002.
	
\bibitem{Moulin02b} P.~Moulin and J.~A.~O'Sullivan, ``Optimal Key Design for
	Information-Embedding Systems,''
	{\em Proc. Conf. Information Sciences and Systems}, Princeton, NJ, March~2002.

\bibitem{Moulin03} P. Moulin and J. A. O'Sullivan,
	``Information-theoretic analysis of information hiding,'' 
	{\em IEEE Trans. on Information Theory},
	Vol.~49, No.~3, pp. 563---593, March 2003.
	
\bibitem{Somekh05} A. Somekh-Baruch and N. Merhav,
	``On the capacity game of private
	fingerprinting systems under collusion attacks,'' {\em IEEE Trans.
	Information Theory}, vol.~51, no.~3, pp.~884---899, Mar.~2005.

\bibitem{Somekh07} A.~Somekh-Baruch and N. Merhav,
	``Achievable error exponents for the private fingerprinting game,''
	{\em IEEE Trans. Information Theory}, Vol.~53, No.~5, pp.~1827---1838, May~2007.
	
\bibitem{Wang06} Y.~Wang and P.~Moulin, ``Capacity and Random-Coding Error
    Exponent for Public Fingerprinting Game,''
    {\em Proc. Int. Symp. on Information Theory}, Seattle, WA, July 2006.
    
\bibitem{Moulin07b} P. Moulin and N. Kiyavash, ``Expurgated Gaussian Fingerprinting Codes,''
	{\em Proc. IEEE Int. Symp. on Information Theory}, Nice, France, June~2007.
   
\bibitem{Boneh95} D. Boneh and J. Shaw, ``Collusion--Secure Fingerprinting 
	for Digital Data,'' in {\em Advances in Cryptology: 
	Proc.~CRYPTO'95}, Springer--Verlag, New York, 1995. 
	
\bibitem{Tardos03} G.~Tardos, ``Optimal Probabilistic Fingerprinting Codes,''
	{\em ACM Symp. on Theory of Computing}, San Diego, CA, 2003.

\bibitem{Barg08} N.~P.~Anthapadmanabhan, A.~Barg and I.~Dumer,
	``On the Fingerprinting Capacity Under the Marking Assumption,''
	{\em IEEE Trans. Information Theory}, Vol.~54, No.~6, pp.~2678---2689, June 2008.

\bibitem{Plotnik91} E.~Plotnik and A.~Satt, ``Decoding Rule and Error Exponent
	for the Random Multiple-Access Channel,'' {\em Proc.~Int. Symp. Information
	Theory}, p.~216, Budapest, Hungary, 1991.

    
\bibitem{Csiszar81} I. Csisz\'{a}r and J. K\"{o}rner,
	{\em Information Theory: Coding Theory for Discrete Memoryless Systems},
    Academic Press, NY, 1981.
    
\bibitem{Csiszar98} I. Csisz\'{a}r, `The Method of Types,''
	{\em IEEE Trans. on Information Theory}, Vol.~44, No.~6, pp.~2505---2523, Oct.~1998.
	
\bibitem{Moulin07} P. Moulin and Y. Wang,
	``Capacity and Random-Coding Exponents for Channel Coding with Side Information,''
	{\em IEEE Trans. on Information Theory}, Vol.~53, No.~4, pp.~1326---1347, Apr.~2007.

\bibitem{Forney68} G.~D.~Forney, Jr.,
	``Exponential Error Bounds for Erasure, List,
	and Decision Feedback Schemes,'' {\em IEEE Trans. Information Theory},
	Vol.~14, No.~2, pp.~206---220, 1968.

\bibitem{Gallager68} R.~G.~Gallager, {\em Information Theory and Reliable Communication},
	Wiley, New York, 1968.

\bibitem{Ahlswede71} R.~Ahlswede, ``Multiway Communication Channels,''
	{\em Proc. IEEE Int. Symp. on Information Theory}, pp.~23---52, Tsahkadsor, Armenia, 1971.

\bibitem{Liao72} H.~Liao, ``Multiple Access Channels,''
	{\em Ph.~D. dissertation}, EE Department, U. of Hawaii, 1972.
	
\bibitem{Das02} A.~Das and P.~Narayan,
	``Capacities of Time-Varying Multiple-Access Channels With Side Information,''
	{\em IEEE Trans. Information Theory}, Vol.~48, No.~1, pp.~4---25, Jan.~2002.
	
\bibitem{Barg08a} A.~Barg, {\em personal communication}, Jan.~2008.

\bibitem{Ahlswede82} R.~Ahlswede, ``An Elementary Proof of the Strong Converse
	Theorem for the Multiple-Access Channel,'' {\em J. Combinatorics,
	Information and System Sci.}, Vol.~7, No.~3, pp.~216---230, 1982.
	
\bibitem{Han78} T.~S.~Han, ``Nonnegative Entropy Measures of Multivariate Symmetric
	Correlations,'' {\em Information and Control}, Vol.~36, No.~2, pp.~133---156, 1978.

\bibitem{Liu96} Y.-S. Liu and B. L. Hughes,
	``A new universal random coding bound for the multiple-access channel,''
	{\em IEEE Trans. Information Theory}, vol.~42, no.~2, pp.~376–--386, Mar.~1996.

\bibitem{Barg02} A.~Barg and G.~D.~Forney, ``Random Codes: Minimum Distances
	and Error Exponents,'' {\em IEEE Trans. Information Theory}, Vol.~48, No.~9,
	pp.~2568---2573, Sep.~2002.

\bibitem{Moulin09-ICASSP} P.~Moulin, ``Optimal Gaussian Fingerprint Decoders,''
	{\em Proc. IEEE Int. Conf. Acoustics, Speech and Signal Processing},
	Taipei, Taiwan, Apr.~2009.
	
\bibitem{Moulin09-ITA} P.~Moulin and Y.~Wang, ``Information-Theoretic Analysis
	of Spherical Fingerprinting,'' {\em Proc. Symp. on Information Theory and Applications},
	San Diego, CA, Feb.~2009.
	
\bibitem{Jourdas09} J.-F.~Jourdas and P.~Moulin, ``High-Rate Random-Like Spherical
	Fingerprinting Codes with Linear Decoding Complexity,'' {\em IEEE Transactions
	on Information Forensics and Security}, Vol.~4, No.~4, pp.~768---780, Dec.~2009.
	
\bibitem{Wang08} Y. Wang and P. Moulin, ``Blind Fingerprinting,''
	submitted to {\em IEEE Trans. Information Theory}, Feb. 2008.
	Available from {\tt arXiv:0803.0265 [cs.IT]}


\bibitem{Amiri09} E.~Amiri and G.~Tardos, ``High Rate Fingerprinting Codes
	and the Fingerprinting Capacity,'' {\em Proc. 20th Annual ACM-SIAM Symposium on
	Discrete Algorithms}, New York, NY, Jan.~2009.

\bibitem{Huang09} Y.-W. Huang and P. Moulin, ``Saddle-Point Solution of the Fingerprinting
	Capacity Game Under the Marking Assumption,''
	{\em Proc. IEEE Int. Symp. on Information Theory}, Seoul, Korea, July 2009.
	
\bibitem{Huang09b} Y.-W. Huang and P. Moulin, ``Capacity-Achieving Fingerprint Decoding,''
	{\em Proc. 1st IEEE Workshop on Information Forensics and Security}, London, UK, Dec.~2009.
	
\bibitem{Furon09} T.~Furon and L.~P\'{e}rez-Freire, ``Worst Case Attacks Against
	Binary Probabilistic Traitor Tracing Codes,'' 
	{\em Proc. 1st IEEE Workshop on Information Forensics and Security}, London, UK, Dec.~2009.

\bibitem{Barg09} N.~P.~Anthapadmanabhan and A.~Barg, ``Two-Level Fingerprinting Codes,''
	{\em Proc. IEEE Int. Symp. on Information Theory}, Seoul, Korea, July 2009.

 		
\end{thebibliography}
\end{document}